\documentclass[fleqn,usenatbib]{aa}

\DeclareRobustCommand{\VAN}[3]{#2}
\let\VANthebibliography\thebibliography
\def\thebibliography{\DeclareRobustCommand{\VAN}[3]{##3}\VANthebibliography}

\usepackage{graphicx} 
\usepackage{amsmath}  
\usepackage{bm}  
\usepackage{float}
\usepackage{xfrac}
\usepackage{txfonts}
 
\usepackage[colorlinks=true, linkcolor=blue, citecolor=blue, urlcolor=blue]{hyperref}  

\begin{document}

    \title{Studying the absorption signatures of H\,{\sc i} Ly$\alpha$ in the warm-hot circumgalactic medium with TNG50}

   \author{
          P. Aparicio Marcos\inst{1},
          P. Richter  \inst{1},
          M. Sparre \inst{1},
          F. R\"unger \inst{1},
          \and
          D. Nelson \inst{2}
          }

   \institute{Institut f\"ur Physik und Astronomie, Universit\"at Potsdam,
             Karl-Liebknecht-Str.\,24/25, 14476 Golm, Germany
   \and          
   Universit\"at Heidelberg, Zentrum für Astronomie,
   ITA, Albert-Ueberle-Str. 2, 69120 Heidelberg, Germany}

\date{Received xxx; yyy}

\abstract
{In this study, we investigate the spectral signatures of neutral hydrogen Lyman-$\alpha$ absorption arising from the warm-hot gas component of the circumgalactic medium (CGM) around $z=0$ Milky Way (MW)-like galaxies using the high-resolution TNG50 cosmological simulation. We used synthetic absorption spectra to identify and characterise coronal broad Ly$\alpha$ absorbers (CBLAs), which represent 
\ion{H}{I} absorption features produced by the warm-hot CGM at temperatures above $10^5$ K. Our study implies that CBLAs have a significant absorption
cross-section, $f_c$, around MW-like galaxies. Based on an 
analysis of 75 sightlines intersecting the CGM of 15 galaxies in the mass range $10^{11.7}M_{\odot} \leq M_{200} \leq 10^{12.3}M_{\odot}$, we find $f_c\approx 0.8$ for $\log N_{\text{HI}}\geq 13$, where CBLAs span a total column-density range $\log N_{\text{HI}}=11.6-15.4$. Therefore, CBLAs provide a significant contribution to the overall H\,{\sc i} 
optical depth in the CGM with $\sim 50 \%$ of the CGM absorbers being dominated by CBLA absorption (in terms of H\,{\sc i} column density). Furthermore, we find that CBLAs trace warm-hot gas in a temperature range $T=10^{5.2-6.4}$ K,
which accounts for $\sim 7\%$ (median value) of the overall baryon budget in the TNG50 galaxies and $\sim 25 \%$ of the total CGM mass. Finally, we identify a population of strong CBLAs that exhibit substantial H\,{\sc i} column densities up to $\log N_{\text{HI}}=14.9$. This population represents a new absorber class that traces massive, extended circumgalactic structures composed of warm-hot gas even at large
radial distances. They arise in regions of increased gas density and long path lengths, and their radial velocities indicate contributions from both inflowing and outflowing warm-hot gas, consistent with warm-hot intergalactic medium-like CGM components. In conclusion, our study demonstrates that CBLAs represent an important absorber class that needs to be considered when interpreting the H\,{\sc i} absorption signatures from the multi-phase CGM of MW-like galaxies at low redshift.}


  \titlerunning{CBLAs in TNG50}

   \maketitle

\section{Introduction}

Galaxies grow and evolve through a complex interplay of inflow, outflow, and redistribution of gas within their halos. Feedback processes from active galactic nuclei (AGNs), supernovae (SNe), and stellar winds \citep{2013MNRAS.436.3031V,2023MNRAS.519.3154H}, together with events such as galaxy mergers \citep[e.g,][]
{Richter_2008,10.1093/mnras/stx3252,2018AAS...23144002D,2022MNRAS.509.2720S,refId0coldgas}, inject large amounts of kinetic energy and metal-enriched material, while also transporting cold and warm gas into the surrounding diffuse halo, also known as the circumgalactic medium (CGM), shaping its physical and chemical properties and evolution \citep[e.g,][]{refId0primera,10.1093/mnras/stae1248,Schellenberger_2024}. As a result of these processes, the CGM is a dynamically complex and multi-phase medium in which different gas phases exist simultaneously. These phases include: 1) cold gas ($T \leq 10^4$ K), characterised by neutral and weakly ionised metal species (e.g. O\,{\sc i}, N\,{\sc i}, and Ca\,{\sc ii}), as well as neutral atomic and molecular hydrogen \citep[e.g.][]{benbekhti2012,Zheng_2020,10.1093/mnras/stab1434,2022MNRAS.512.3717D}, 2) cool gas ($ T \sim 10^{4 -5}$ K) traced by low and intermediate ions such as 
Mg\,{\sc ii}, Si\,{\sc ii}, and Si\,{\sc iii}
\citep[e.g,][]{Werk_2014,10.1093/mnras/stab1673,richter2016,DeFelippis_2021}; 3) warm gas ($ T\sim 10^{5 -6}$ K), detectable through intermediate to high ions such as 
C\,{\sc iv}, Si\,{\sc iv}, N\,{\sc v} and O\,{\sc vi}
 \citep[e.g,][]{Tumlinson_2011, Werk_2016,sembach2003}; and 4) hot, shock-heated, and highly ionised gas ($ T> 10^6$ K), which is traced mainly through X-ray observations \citep[e.g.][]{Anderson2010}.
 
One of the key roles of the CGM is to sustain star formation in galaxies \citep{Carr_2023}. Because star formation requires a steady inflow of gas, galaxies must continuously replenish their fuel. However, observations of nearby disk galaxies have shown that they can exhaust their internal gas reservoirs within just a few billion years \citep{1994ApJ...435...22K,10.1093/mnras/stz2894}, making the CGM a crucial source of fresh material to support ongoing star formation. Additionally, the CGM might also serve as a significant cosmological reservoir for both baryons and metals, holding a substantial fraction of the Universe's matter \citep{2012MNRAS.425.1270S}. 
Indeed, observational studies of $L^{*}$ or Milky Way (MW)-like galaxies, such as those conducted by \cite{Werk_2014}, indicate that a significant portion of the galaxy's baryons reside in the CGM
($\geq 35$\% of the total baryonic mass in nearby $L^{*}$ galaxies).

Despite the important role that the CGM plays in addressing unresolved questions about galaxy evolution, detecting its different phases is challenging due to its low densities. Therefore, the best observational approach is to combine observations from both ground-based and space-based telescopes. While the CGM has been observed in both emission \citep[e.g.][]{2012MNRAS.420.1731F,2016ApJ...827..148C,2019astro2020T.403T} and absorption \citep[e.g.][]{2014MNRAS.445..794T,2019MNRAS.485.1595P,10.1093/mnras/stab871,Mathur_2023}, current emission measurements are limited to dense gas regions close to galaxies or halos with bright quasars (QSOs), which can enhance CGM emission to greater distances. This limitation arises from the intrinsic faintness of the circumgalactic gas and the limited sensitivity of current observational instruments \citep{10.1093/mnras/stz2238,Corlies_2020}. 

Therefore, many studies of the CGM have used absorption techniques, particularly QSO absorption-line spectroscopy. This method uses a bright background source, such as a quasar, to identify intervening gas systems along the line of sight (LOS). As light from the background source passes through gas, it leaves its absorption characteristics imprinted on the background source's spectrum \citep[e.g.][]{Tumlinson_2013,2015ApJ...813...46B,2016MNRAS.455.2662T,annurev:/content/journals/10.1146/annurev-astro-021820-120014}. For instance, ultraviolet (UV) absorption spectroscopy probes cold and hot gas in the CGM, particularly through \ion{H}{I} and metal-ion absorption from foreground galaxies \citep[e.g.][]{2013MNRAS.432...89F,2016MNRAS.460.2157O,2017MNRAS.465.2966S}. The UV range is rich in transitions ranging from the neutral hydrogen's Lyman series to different ionisation states of heavy elements \citep{1986A&A...155L...8B, 1995ApJ...442..538L, Ho_2020,DeFelippis_2021,refId0metalcontent}. A key advantage of this method is that the sensitivity of the observations does not depend on redshift, thus allowing for the detection of low-density CGM gas over a wide range of cosmic epochs, including the early Universe \citep{Lopez_2008,Rudie_2013, annurev:/content/journals/10.1146/annurev-astro-021820-120014}. Instruments such as the Space Telescope Imaging Spectrograph (STIS) and the Cosmic Origins Spectrograph (COS) on the Hubble Space Telescope (HST) have, over the years, greatly expanded our knowledge of the CGM through such absorption-line studies \citep[e.g,][]{Penton_2004,Prochaska_2011,Werk_2016,Bowen_2016, Prochaska_2017, Keeney_2017, 2017A&A...607A..48R}. 

However, the hot CGM, which is at temperatures close to the virial temperatures of the galaxies, cannot be observed directly in the UV range due to a lack of high-ion transitions at UV wavelengths that would trace highly ionised, million-degree gas in galaxy halos at low redshift. Absorption spectroscopy in the X-ray regime, on the other hand, provides access to high ions such as O\,{\sc vii} and O\,{\sc viii} \citep[e.g.][]{Miller_2013}, but such observations are restricted to the Milky Way CGM and very few individual galaxies due to the limited amount and quality of spectral data of bright X-ray background sources.  

Notably, only neutral hydrogen (H\,{\sc i}) is (in principle) detectable in all of the CGM phases, owing to its large cosmic abundance and the fact that the amount of H\,{\sc i} is non-zero even in highly ionised regions at million-degree temperatures due to the balance between ionisation and recombination. Indeed, H\,{\sc i} Lyman-$\alpha$ (hereafter H\,{\sc i} Ly$\alpha$) absorption signatures from the warm-hot ($T>10^5$ K) CGM of nearby galaxies have been predicted \citep[][]{liang2018} and observed \citep[][]{Richter_2020_hot_halos} in the spectra of HST/COS and HST/STIS along sightlines that pass the halos of foreground galaxies.

In this study, we follow up on this idea and use the TNG50 simulation from the IllustrisTNG simulation suite to systematically investigate the spatial distribution, kinematics, and physical properties of warm-hot gas surrounding MW-like galaxies, and to constrain its spectral signatures in H\,{\sc i} Ly$\alpha$ absorption. Specifically, we decomposed the H\,{\sc i} Ly$\alpha$ absorption into contributions from the different CGM phases, from cool to warm-hot, and determined the shape and strength of the H\,{\sc i} Ly$\alpha$ absorption arising from million-degree gas as a function of virial radius.

The paper is organized as follows: In Sect.\ \ref{hot_cgm}, we discuss in more detail the properties of the warm-hot CGM and its absorption signatures in H\,{\sc i} Ly$\alpha$ absorption. In Sect.\ \ref{set_up}, we describe the cosmological simulations used in this study and the setup of mock spectra. In Sect.\ \ref{properties_section}, we analyse and discuss the H\,{\sc i} absorber population in the (simulated) warm-hot CGM. Additionally, in Sect.\ \ref{CBLA_deep}, we introduce and study a newly identified class of hot absorbers. We discuss our results in Sect.\, \ref{discussion}, where we also provide a comparison with recent observations. A summary and conclusions are given in Sect.\ \ref{conclusions}.

\section{The warm-hot CGM and its H\,{\sc i} absorption signatures}\label{hot_cgm}

In this work, we will focus on exploring the shock-heated gas surrounding galaxies, commonly referred to as coronal gas - in analogy to the Sun's hot corona \citep{1956ApJ...124...20S,1977ApJ...215..483B,10.1093/mnras/183.3.341}. In the high-temperature regime of this warm-hot phase, the gas is expected to be close to the virial temperature of the halos [$\sim10^6$ K for MW type galaxies; see \citep{ 1991ApJ...379...52W,Anderson2010, Miller_2013, McQuinn_2018,10.1093/mnras/stz1859,2020ApJ...893...82F}], or even beyond in the super-virial regime \citep[e.g.][]{bisht2025}. In the lower-temperature regime, the gas resides at sub-virial temperatures in the range $T=10^5-10^6$ K, with part of it representing material that has cooled from the hotter regime.  

At $T>2\times 10^5$ K, the gas is nearly fully collisionally ionised, with neutral hydrogen fractions dropping below $f_{HI} \leq 10^{-5}$. In this temperature regime, there are mainly two processes that determine the ionisation state of hydrogen gas: collisional ionisation due to the high thermal energy and photoionisation by the cosmic far ultraviolet (FUV) background. At $T \gg 10^{5}$ K, collisional ionisation dominates, reducing the neutral hydrogen fraction to $f_{\mathrm{HI}} \lesssim 10^{-6}$ by $T \sim 10^6$ K. On the other hand, near the lower end of this regime ($T \sim 10^{5}$ K), photoionisation and collisional ionisation contribute comparably to the ionisation balance (by $T \sim 10^{5}$ K, $f_{\mathrm{HI,phot}} \sim f_{\mathrm{HI,coll}}$; \cite{1993ApJS...88..253S,Richter_2008}). Above this threshold, collisions rapidly take over, and hydrogen becomes almost fully collisionally ionised. However, as mentioned earlier, direct observations of the shock-heated phase of the CGM pose significant challenges due to its extremely low density (ranging from $10^{-2}$ to $10^{-5}$ particles per cubic centimetre) and its high ionisation state.
As an alternative to UV and X-ray absorption measurements, diffuse soft X-ray emission offers direct evidence of gas at these high temperatures \citep{Anderson2010, Anderson2013, Khabibullin_2018}, but observational data are limited to only a few cases.

Despite the low gas density in the warm-hot CGM, the presence of sufficient neutral hydrogen atoms along sightlines traversing the hot halo of a MW-like galaxy can produce a detectable Ly$\alpha$ absorption signal
\citep{2006A&A...451..767R,lehner2007,refId0whim,2010ApJ...712.1443N,Danforth_2016,Richter_2020_hot_halos}. This absorption, known as a coronal broad Lyman-$\alpha$ absorber (CBLA) \citep{Richter_2020_hot_halos}, is distinguished by its broad profile, due to the high temperature of the gas, and its relatively shallow depth, due to the tiny amount of neutral hydrogen. CBLAs represent a sub-class of broad Lyman-$\alpha$ absorbers 
\citep[BLAs;][]{richter2004} that have been observed to study the warm-hot, shock-heated component of the intergalactic medium (IGM), the so-called warm-hot intergalactic medium
\citep[WHIM;][]{Richter_2006,lehner2007,tepper2012}. While for the WHIM, the BLA line profile and depth are expected to be governed by the warm-hot gas, this is not the case for CBLAs: The complex and multi-phase nature of the CGM means that cooler gas phases tend to prevail in the \ion{H}{I} optical depth of most absorbers. Hence, typical CBLAs are likely embedded in the multi-component \ion{H}{I} Ly$\alpha$ absorption systems, which makes their detection particularly difficult and limits our ability to accurately study their physical properties and distribution using current observational instruments \citep[see][]{Richter_2020_hot_halos}. To address these challenges, it is essential to use simulations in the study of CBLAs. Current cosmological magnetohydrodynamic simulations of galaxy formation provide a valuable framework for addressing these observational limitations, allowing us to model and analyse the complex environments in which CBLAs form and evolve, thereby improving our understanding of their potential role in galaxy evolution. 

\begin{figure}
\centering
   \includegraphics[width=0.5\textwidth]{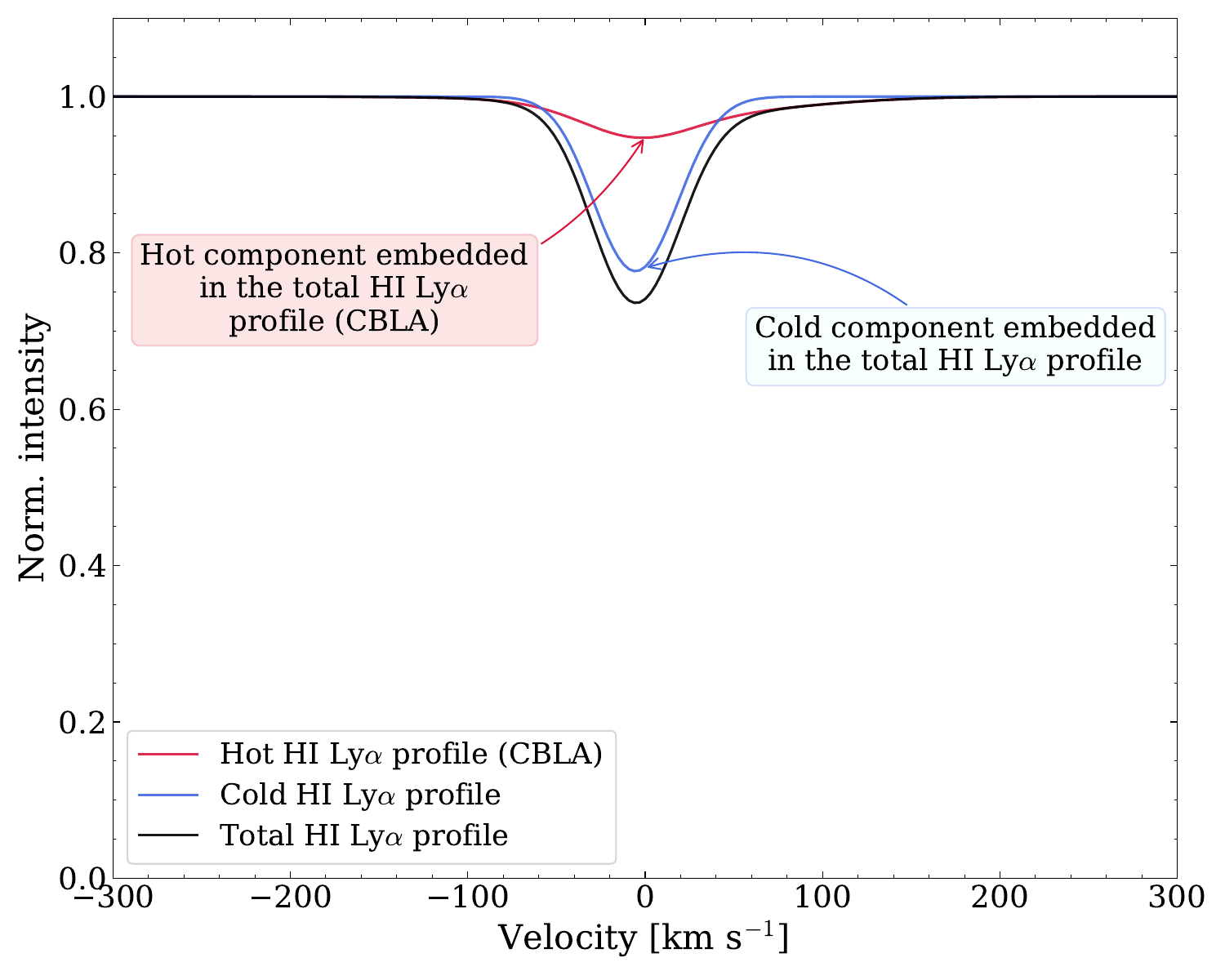}
    \caption{Example of the total \ion{H}{I} Ly$\alpha$ profile from our TNG50 sample. The normalized intensity is plotted as a function of the LOS velocity for a sightline passing within $R_{200}$ of a MW-like galaxy from TNG50 at $z=0$. The profile corresponds to the total observed \ion{H}{I} Ly$\alpha$ absorption and is decomposed into two components: a cold and a hot \ion{H}{I} phase. The hot contribution embedded in the total \ion{H}{I} absorption profile corresponds to a CBLA candidate.
    \label{figure1}}
\end{figure}

Figure \ref{figure1} shows a typical example of a CBLA in a MW-like galaxy with $R_{\mathrm{200}} = 209$ kpc and a halo mass of $M_{\mathrm{200}} = 9.75 \times 10^{11} \mathrm{M}_{\odot}$. The total \ion{H}{I} Ly$\alpha$ absorption profile is shown in black as it would be seen by an external observer, along with its hot and cold components. The cold \ion{H}{I} absorption stands out in blue as a deep and narrow feature that dominates the profile, while the warm-hot \ion{H}{I} gas, in red, adds a broader and shallower signal. In this case, the CBLA signal is blended with the stronger, narrower absorption from the warm/cool CGM, highlighting how difficult it can be to identify these warm-hot components in real observations. The overall shape and decomposition of cold and hot gas components mimic the component-decomposition applied to real CBLA absorbers seen in HST/STIS data \citep{Richter_2020_hot_halos}.
In the following, we study the typical properties of CBLAs in multi-phase CGM absorbers around galaxies spanning a range of masses in the TNG50 simulation and evaluate the role of the hot halo gas for our understanding of the overall \ion{H}{I} Ly$\alpha$ absorption properties in the CGM.     

\section{Cosmological simulations}\label{set_up}

\subsection{The TNG50 simulation}

The TNG50 simulation \citep{Nelson2021, 10.1007/978-3-030-66792-4_1} is the highest-resolution simulation of the IllustrisTNG (hereafter, TNG) simulation suite \citep{Nelson2017,Pillepich_2017,Springel_2017, Marinacci_2018, Naiman_2018}. This suite consists of three cosmological magneto-hydrodynamical simulations performed at different scales: TNG50, TNG100, and TNG300, and at different resolutions. Specifically, there are four resolution levels, labelled TNG50-1 to TNG50-4, with TNG50-1 being the highest. Expanding on the original Illustris simulation \citep{2013MNRAS.436.3031V, 10.1093/mnras/stu1536,10.1093/mnras/stu1654,2015MNRAS.452..575S}, the TNG project uses an improved version of the galaxy formation model, which integrates magneto-hydrodynamics \citep{10.1111/j.1365-2966.2011.19591.x, 2014ApJ...783L..20P} and updated feedback processes. In particular, TNG50 resolves individual galaxies down to the scale of dwarf galaxies, capturing the complex physics governing galaxy formation, such as gas accretion, star formation, and feedback from supernovae and black holes within a $(50 \ \text{Mpc})^3$ cubic volume sampled by $2160^3$ gas cells, and an average baryonic mass resolution of $8.5 \times 10^{4} M_\odot$. These features enable detailed studies of galaxy morphology and evolution across cosmic time, while also offering a solid framework to study the complex interplay of baryonic processes in the CGM. We refer readers to the TNG methods papers for details \cite{Springel_2017,Pillepich_2017,Nelson2017,Marinacci_2018,Naiman_2018,Nelson2021, 10.1007/978-3-030-66792-4_1}, and here briefly summarize the main characteristics of the simulation.

The TNG simulations are run with the AREPO code \citep{Springel_2010}, based on a dynamic, moving Voronoi mesh \citep{Springel_2010,10.1111/j.1365-2966.2011.19591.x} to solve the hydrodynamical equations, enabling natural adaptivity in both space and time, coupled to a Tree-PM algorithm for self-gravity  \citep{Bagla_2002, Bode_2003}. The TNG galaxy formation model includes a suite of physical processes essential for simulating the evolution of galaxies within a cosmological framework \citep{2018MNRAS.473.4077P, Weinberger_2016}. Gas physics is modelled through radiative mechanisms, including both primordial and metal-line cooling and heating \citep{2013MNRAS.436.3031V}, as well as an evolving background radiation field \citep{Faucher-Giguere_2009}. Star-formation is modelled by allowing stars to form in dense regions of the interstellar medium (ISM) once the gas exceeds a critical density threshold \citep{10.1046/j.1365-8711.2003.06206.x}. Therefore, the star formation rate is determined by the local properties of the gas, specifically its density, temperature, and metallicity.
The model also tracks the evolution of the stellar population and chemical enrichment from the asymptotic giant branch (AGB), as well as the Type II and Ia supernovae (SNe). These processes result in both the production and dispersion of heavy elements in the surrounding medium \citep{2018MNRAS.473.4077P,Emami_2021}. 

Additionally, stellar feedback is modelled using a kinetic wind scheme driven by Type II SNe, which generates galactic-scale flows that regulate star formation and enrich the ISM with metals \citep{2019MNRAS.490.3196P,Nelson2019}. Supermassive black holes (SMBHs) are seeded in Friends-of-Friends (FoF) halos once the halo mass exceeds $7 \times 10^{10}M_\odot$. Their subsequent growth occurs through mergers and gas accretion, the latter being governed by Bondi and Eddington accretion rates \citep{10.1046/j.1365-8711.2001.04912.x,Weinberger_2016}. The feedback from SMBHs is modelled in two modes, thermal (quasar) and kinetic (wind), depending on the accretion rate relative to the Eddington limit \citep{Weinberger_2016}. The model also accounts for the amplification of cosmic magnetic fields from primordial seed fields, which play a role in shaping galaxy morphology and influencing gas dynamics \citep{Marinacci_2018}. Finally, the TNG simulations adopt a flat $\Lambda$CDM cosmology consistent with the Planck 2015 results \citep{refId0}, with the set of cosmological parameters ($\Omega_{\mathrm{m,0}}$, $\Omega_{\mathrm{b,0}}$, $\Omega_{\mathrm{\Lambda,0}}$, $h$, $\sigma_{8}$, $n_{s}$) = (0.3089, 0.0486, 0.6911, 0.6774, 0.8159, 0.9667).

\subsection{The galaxy sample}\label{galaxy_sample}

In this work, we focus on characterising CBLAs in the halos of low-redshift Milky Way (MW) type galaxy halos. Our aim is to understand both their frequency and their physical properties, and to explore how these absorbers can serve as probes of the hot phase of the CGM. To this end, we selected a sample of MW-like galaxy halos at $z = 0$ from TNG50-1, using a halo-mass-based approach. This involved selecting central galaxies that reside in halos within a MW-like mass range and that exhibit disk-like stellar kinematics, as described below.
This approach differs from previous studies of MW/M31 analogues in TNG50, such as \cite{10.1093/mnras/stae2165}, both in how the sample is defined and in its motivation. While that work selects galaxies based on observable properties such as stellar mass, morphology, and environment to construct a representative sample of MW-like galaxies, our selection is instead based on halo mass (see Fig.~3 of \cite{10.1093/mnras/stae2165}, bottom left panel). This choice is motivated by our focus on the CGM, whose overall properties are closely tied to the host halo mass, which governs its thermodynamic state and kinematic structure.

\subsubsection{Halo mass range}\label{halo_mass_range}

In TNG50, FoF groups are identified with a linking length of $b = 0.2$ \citep{1985ApJ...292..371D} and are referred to as halos. Within these halos, subhalos (i.e. galaxies) are identified using the SUBFIND algorithm \citep{10.1046/j.1365-8711.2001.04912.x}. Each halo hosts a central galaxy, defined as the subhalo located at the minimum of the gravitational potential, together with its baryonic component. Additional galaxies within the same halo are classified as satellites, along with their associated baryons. For our analysis, we selected central galaxies with halo masses in the range $11.78 \leq \log M_{200}/M_\odot < 12.30$, where $M_{200}$ is defined as the total mass enclosed within a radius, $R_{200}$, for which the mean density is 200 times the critical density of the Universe.

\subsubsection{Morphology: Selecting disk galaxies in TNG50}\label{disk_selection}

Due to the absence of pre-computed morphological measurements in TNG50, we analysed the morphology using bulge-to-disk kinematic decomposition. To identify galaxy types (e.g. spiral or elliptical), we followed the methodology of \cite{2003ApJ...591..499A} and \cite{2014MNRAS.437.1750M} based on the circularity parameter. This parameter, defined as $\epsilon = J_z / J(E)$, is calculated for each star particle with specific angular momentum $J_{z}$ around the symmetry axis and $J(E)$ is the maximum angular momentum of the star particles at a given binding energy. In our study, we computed these values after aligning with the angular momentum vector of its stellar component, limiting our analysis to star particles within twice the stellar half-mass radius. The circularity distribution is centred symmetrically around $0.0$ for bulge particles, whereas for disk particles, it exhibits clustering around $1.0$. Circularities equal to or greater than $0.7$ are categorised as disk particles, while those below $0.7$ are classified as bulge particles. This classification allowed us to derive the disc-to-total ($D/T$) ratios, defined as $D/(D+B)$ where $D$ and $B$ are the stellar masses in disk and bulge particles, respectively \citep{2003ApJ...591..499A,Aumer2013,2014MNRAS.437.1750M}. We only included galaxies where the mass fraction of particles with $\epsilon > 0.7$ exceeds $0.4$.

\begin{figure*}
\centering
   \includegraphics[width=0.9\textwidth]{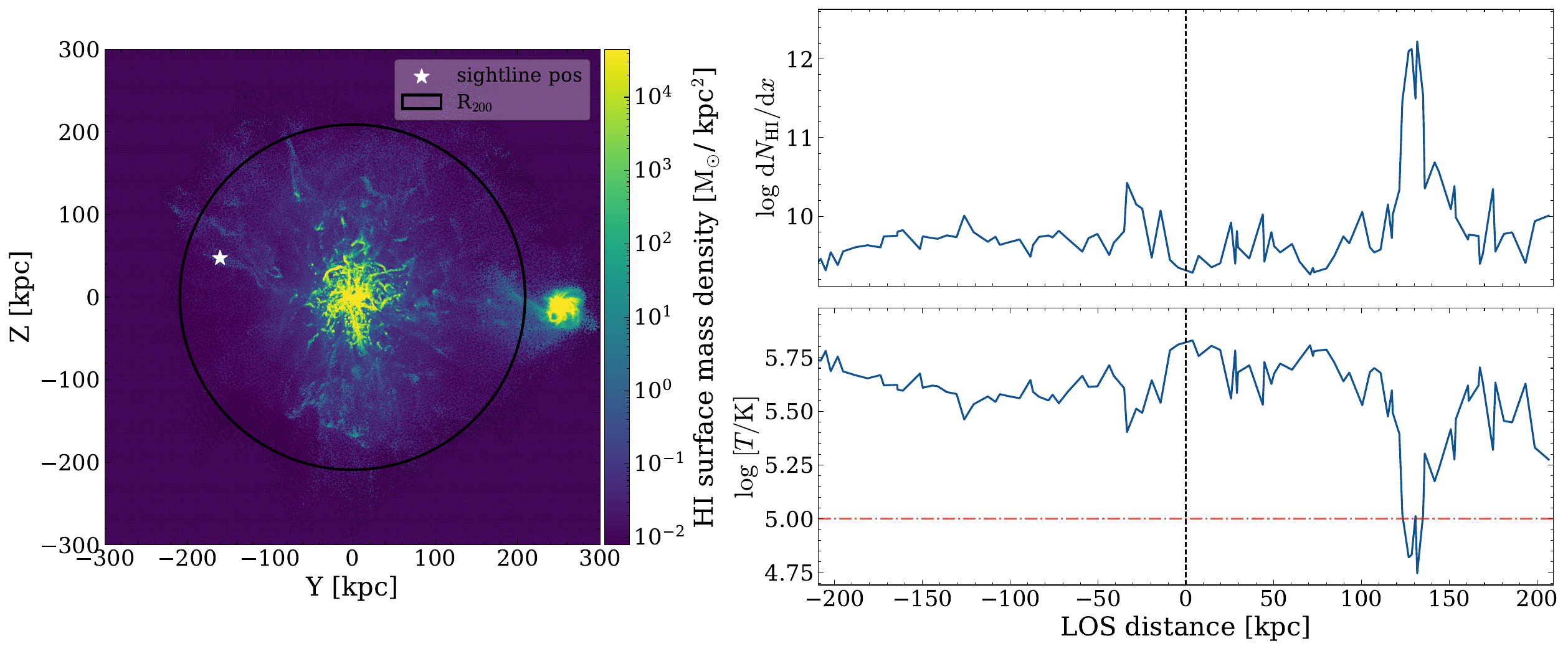}
    \caption{Illustration of gas properties along a representative sightline placed within $R_{\mathrm{200}}$ in one of our TNG50 MW-like galaxy halos at $z=0$. Left panel: Projected \ion{H}{I} surface mass density in the plane perpendicular to the sightline direction. The black circle shows $R_{200}$, and the white star marks the position of the sightline (in this example, the sightline runs along the x-axis). Right panel: Variation in column density per cell [$\mathrm{cm}^{-2}/\mathrm{kpc}$] and temperature along the sightline up to $\pm R_{200}$, which is the region we considered as the CGM in our analysis (see Sect. \ref{sim_hot_CGM}). The dashed red line indicates the temperature threshold of $10^5$ K used in this work for selecting warm-hot gas. The vertical dashed black line in both panels marks the location of the galaxy along the LOS in the simulation.
\label{figure2}}
\end{figure*}

\subsection{The simulated CGM and synthetic QSO spectra}\label{synthetic_spectra}

\subsubsection{Generating sightlines through TNG50 simulation box}

After selecting our galaxy sample, we generated sightlines around each galaxy halo following the Synthetic Absorption Line Spectral Almanac (SALSA) \citep{nelson2025}. The sightlines were generated in a large grid by ray-tracing through the Voronoi mesh. Specifically, as the sightlines passed through the simulation box in a specific direction aligned with one of the axes, they intersected multiple Voronoi cells. For every intersection, the local gas properties were sampled according to the characteristics of the cell, under the assumption that these quantities remain constant within each cell. By integrating these contributions along the path length of the ray, we obtained LOS profiles that provide a cumulative view of the physical conditions across the entire sightline. Since our goal is to study the CGM, we used a localized sampling approach, where sightlines were concentrated around specific galaxies of interest. Additionally, when placing sightlines, we also accounted for the orientation of the galaxy disk. This step is important because the hot halo may co-rotate with the disk, and if it does, the orientation directly influences how the resulting absorption profile is broadened and shifted due to the galaxy halo's rotation. For simplicity, we therefore selected galaxies whose rotational axis was aligned within 25 degrees of one of the simulation box axes. This criterion keeps the disk plane nearly parallel to the sightline, which helps reduce projection effects and makes it easier to interpret rotational signatures in the CGM. After applying these selection criteria, our final sample consisted of 15 MW-like galaxy halos.

\subsubsection{Characterising the hot CGM in our TNG50 sample}\label{sim_hot_CGM}

We characterised the CGM volume using a cylindrical region around each galaxy, i.e. all gas cells within a projected radius of $R_{200}$ (in the plane perpendicular to the sightline) and within $\pm R_{200}$ along the sightline, including gas bound to the central subhalo and its host FoF halo. To specifically isolate the warm-hot gas, we filtered the gas based on its temperature, selecting those with $T \ge 10^5\,\mathrm{K}$, as $T = 10^5\, \mathrm{K}$ marks the lower boundary of the warm-hot phase of the CGM, where hydrogen becomes highly ionised.

Figure \ref{figure2} shows the properties of a representative sightline through one of our MW-like galaxy halos from TNG50 at $z = 0$. The left panel shows the projected \ion{H}{i} surface mass density distribution in the plane perpendicular to the sightline, with the sightline position marked by a white star and the galaxy’s $R_{200}$ radius indicated by a black circle. On the other hand, the right panels show the corresponding profiles of column density (per gas cell) and temperature along the sightline direction. We also show the threshold temperature of $10^5$ K for identifying warm-hot gas used in our analysis, marked with a red dashed line. Therefore, for each sightline in each MW-galaxy halo of our sample, the warm–hot gas within $\pm\ R_{200}$ along the LOS is what produces, for instance, the CBLA shown in Fig. \ref{figure1}, as well as all CBLAs identified throughout this paper (see more examples in Appendix \ref{appendix:CBLA_sample}).

\subsubsection{Spectra generation}\label{spectra_generation}

In this study, we generated synthetic absorption spectra of \ion{H}{I} Ly$\alpha$ by integrating the physical properties of gas cells along the LOS. For each intersected cell, we extracted its path length, position, neutral hydrogen density, temperature and LOS velocity.
In TNG50 \citep{Nelson2021}, the neutral hydrogen fraction was computed self-consistently during the simulation using the ionisation model of \citet{2013MNRAS.436.3031V}, which includes primordial chemistry \cite{1996ApJS..105...19K}, metal-line cooling, and photoionisation from the \cite{Faucher-Giguere_2009} UV background. From these quantities, we computed the main observables as follows:

\begin{enumerate}

    \item Neutral hydrogen column density, $N_{\mathrm{HI},i}$: The contribution of each cell to the neutral hydrogen column density $N_{HI,i}$ was given by $N_{\mathrm{HI},i}= n_{\rm HI,i} \Delta l_{i}$ where $\Delta l_{i}$ is the physical LOS path length through the cell.
    \\
    \item Doppler broadening parameter, $b_{i}$: The Doppler broadening $b_{i}$ for each cell was calculated as $b_{i} = \sqrt{2k_{B}T_{i}/m_{p}}$ where $k_{B}$ is the Boltzmann constant, and $m_{p}$ is the proton mass. It represents the thermal broadening due to the cell's temperature $T_{i}$. In general, the total Doppler broadening was defined as $b_{\rm t}^2 = b^2_{\rm th} + b^2_{\rm nt}$ where $b_{\rm th}$ represents the thermal component and $b_{\rm nt}$ accounts for additional broadening due to bulk motions or turbulence. In warm-hot CGM gas, however, the line broadening is dominated by thermal motions. At $ \log T \sim 5.5-6$, the low atomic mass of hydrogen produces a thermal Doppler width of $\sim 70-120$ km s$^{-1}$, well above the typical non-thermal broadening, which is usually only a few tens of km s$^{-1}$ \citep[e.g.][]{Richter_2006, Tumlinson_2017}. Since each cell along the LOS contributes a different thermal width, we defined an effective Doppler parameter, $b_{\rm eff}$, that averages these contributions weighted by their column densities (since stronger absorbers dominate the line profile). This $b_{\rm eff}$ is given by $b_{\rm eff} = \sqrt{(\sum b_{i}^2N_\mathrm{HI,i})/(\sum N_\mathrm{HI,i})}$ and provides a $N_{\mathrm{HI}}$-weighted estimate of the characteristic line width. In our synthetic spectra, the line profile was calculated using a Voigt function, which combines both thermal Doppler broadening and natural broadening resulting from the atomic transition. Because every gas cell has its own LOS velocity, its absorption profile is centred at a different velocity. Therefore, when these are combined, the resulting line may appear shifted relative to the galaxy systemic velocity if there is a bulk offset, and also broadened if there is a spread of velocities along the LOS.
    \\
    \item Absorption velocities, $v_{\rm abs}$: We defined the absorption velocities in the galaxy rest-frame as $v_{\rm abs,i} = v_{\rm LOS,i} - v_{\rm gal}$ where $v_{\rm gal}$ is the systemic velocity of the galaxy and $v_{\rm LOS,i}$ is the cell LOS velocity. Since each gas cell contributes a Voigt profile centred on $v_{\rm abs,i}$, the position of the absorption directly reflects the velocity distribution of the gas. 
    \\
    \item Optical depth, $\tau$: For each gas cell, the optical depth $\tau$ was calculated from its column density $N_{\mathrm{HI}}$, Doppler parameter $b_{i}$, and absorption velocity $v_{\rm abs,i}$. The contribution of a single cell was modelled with a Voigt profile centred at its $v_{abs,i}$ as described before, with the Gaussian width set by $b_{i}$. The total optical depth $\tau$ along the LOS was calculated by summing the contributions from all gas cells. Finally, we computed the normalized transmitted flux as $F(\nu) = \text{exp}[-\tau(\nu)]$.
    \\
    \item Equivalent width, EW: The equivalent width was calculated by integrating the absorbed fraction of the transmitted flux along the line profile: $\mathrm{EW} = \int (1 - e^{-\tau}) \,d\lambda$. In the synthetic spectra, this integral was evaluated numerically in our wavelength grid using the trapezoidal rule.  

\end{enumerate}

As we were interested in the intrinsic properties of the CBLA absorption, the spectra were kept at full numerical resolution without applying any instrumental convolution. From this point onward, $N_\mathrm{HI}$, $b_{\mathrm{eff,HI}}$, and $\mathrm{EW}_{\mathrm{HI}}$ refer to quantities measured in the warm-hot gas phase, unless explicitly stated otherwise.

\footnotetext[2]{\url{https://www.tng-project.org/}}

\section{Properties of simulated CBLAs in TNG50}\label{properties_section}

In this section, we describe the approach used to identify CBLAs using all the parameters explained in Section \ref{synthetic_spectra}. Our setup consisted of nine sightlines placed at different impact parameters, extending out to $R_{200}$, and focused on edge-on MW galaxies where the sightlines run parallel to the disk plane. We initially restricted the CGM to gas with temperatures above $10^5$ K, in order to isolate the shock-heated, warm-hot phase, and selected halos within the mass range $10^{11.7}M_{\odot} \leq M_{200} \leq 10^{12.3}M_{\odot}$ (see Sec. \ref{set_up}). With these choices, we systematically explored how each parameter shaped the properties of CBLAs in our analysis.

\begin{figure}
\centering
\includegraphics[width= 0.45\textwidth]{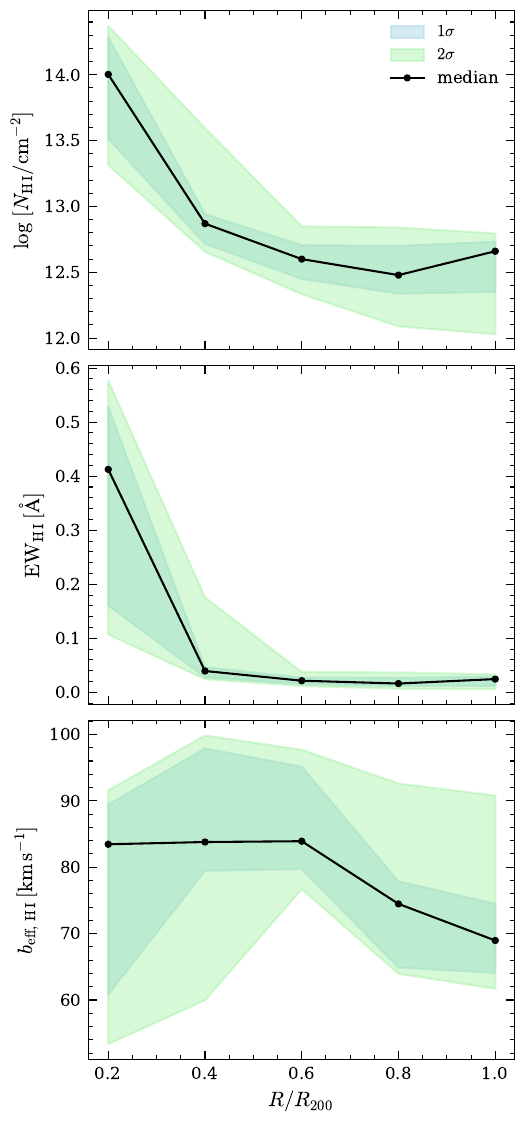}
\caption{Statistical radial dependence of the \ion{H}{I} column density ($\log N_{\mathrm{HI}}$, upper panel), \ion{H}{I} equivalent width ($\mathrm{EW}_{\mathrm{HI}}$, middle panel), and \ion{H}{I} effective Doppler width ($b_{\rm eff,HI}$, bottom panel) for CBLAs along nine sightlines randomly arranged in a ring pattern around a MW-like galaxy halo from TNG50 at $z=0$ with a mass of $M_{200} = 9.75 \times 10^{11}\,M_{\odot}$ (see Sect.~\ref{galaxy_sample}). The shaded regions represent the $1$-$2\,\sigma$ confidence intervals, while the black points indicate the median values at each radial distance. The trends highlight the transition from the dynamically active inner CGM to the more homogeneous and nearly isothermal outer CGM at larger radii.\label{figure3}}
\end{figure}

\subsection{CBLAs around MW-like galaxies: A representative case}

To illustrate our approach to studying the properties of CBLAs in TNG50, we discuss in Fig. \ref{figure3} a typical case of CBLA absorption identified along nine sightlines at different impact parameters up to $R_{200}$ of the hot halo of a MW-like galaxy with a mass of $M_{200} = 9.75 \times 10^{11} M_{\odot}$. 

In this figure, we present the radial trends of the neutral hydrogen column density $N_{\mathrm{HI}}$, equivalent width (EW), and effective Doppler parameter $b_{\rm eff}$ as a function of the normalized distance $R/R_{200}$ for our representative model halo. For this, we have generated 45 sightlines around that galaxy, arranged in a ring pattern with rings of radii $0.2, 0.4, 0.6, 0.8$ and $1.0\times R_{200}$ (i.e., 5 rings, each containing 9 sightlines randomly placed in that ring). In each panel, the shaded regions represent the $1-2\sigma$ intervals from the variations of these parameters at a given radius, while the black points indicate the median values at each radius (see also \ref{spectra_generation}).

The top panel shows the trend of \ion{H}{I} column density with distance, spanning from $\log N_\mathrm{HI} = 12.0 - 14.0$. The median values decline with radius: near the galaxy centre, $\sim 0.2R_{200}$, the \ion{H}{I} column densities reach $\log N_\mathrm{HI} = 13.5- 14.0$, and by $\sim R_{200}$ they drop closer to $\log N_\mathrm{HI} = 12.3- 12.5$. This behaviour is expected, as the gas density falls off in the outer halo, leading to smaller neutral hydrogen fractions. Likewise, the absorption-path length through the halo declines with increasing impact parameter, both effects leading to smaller \ion{H}{I} column densities in the CBLAs along sightlines that pass the outer halos. The $1-2\sigma$ deviations show substantial scatter, which reflects the clumpiness and multi-phase structure in the hot gas.

The middle panel illustrates the radial variation of the $\mathrm{EW}_\mathrm{HI}$ in the model galaxy halo. It spans a range from $0.01$ to $0.45$~\AA, reaching a constant value of $\sim 0.01$ at larger radii. Similar to the trend in $\log N_\mathrm{HI}$, the median $\mathrm{EW}_\mathrm{HI}$ declines with increasing distance for the same reasons as described above.

The bottom panel shows the radial trend of the effective Doppler parameter, $b_\mathrm{eff, HI}$. The $b_\mathrm{eff, HI}$ median values range from $\sim 90$ in the inner halo to $\sim 70$ km s$^{-1}$ in the outer halo, corresponding to a pure-thermal temperature of $\log\, [T/ \text{K}] \approx 5.7 - 5.5$. Because $b_\mathrm{eff, HI}$ is $N_\mathrm{HI}$-weighted, it is biased toward cooler, more neutral phases, so these temperatures lie on the lower side of our MW-halo's virial temperatures, which are $ \log\, [T_{\rm vir}/\,\text{K}]\sim 5.6 - 5.9$. However, the large scatter of $1-2\sigma$ at small radii reflects the multi-phase nature of the inner halo, where some sightlines intersect cooler, denser structures with narrower profiles, while others probe hotter phases in the inner halo. At large radii ($R > 0.6R_{\rm 200}$), the smaller median value for $b_\mathrm{eff, HI}$ compared to the inner halo indicates a mild decrease in the median gas temperature of the CBLAs. This is perfectly in line with the expectations from analytic models, in which CBLA absorbers are hotter in the inner halos as they adjust to a hydrostatic equilibrium within a characteristic cooling radius of $0.6R_{200}$ \citep[see][their Table 1]{Richter_2020_hot_halos}, beyond which the gas is expected to be mildly cooler and isothermal.

Taken together, the declining trends for $N_{\mathrm{HI}}$, EW, and effective Doppler parameter with increasing radius appear to reflect the transition from a complex, dynamically active inner hot CGM to a smoother and more isothermal coronal gas layer in the outer halo. In Appendix \ref{appendix:CBLA_sample} we provide some additional spectral CBLA profiles for this particular galaxy halo at different impact parameters.

\begin{figure}
\centering
   \includegraphics[width=0.4\textwidth]{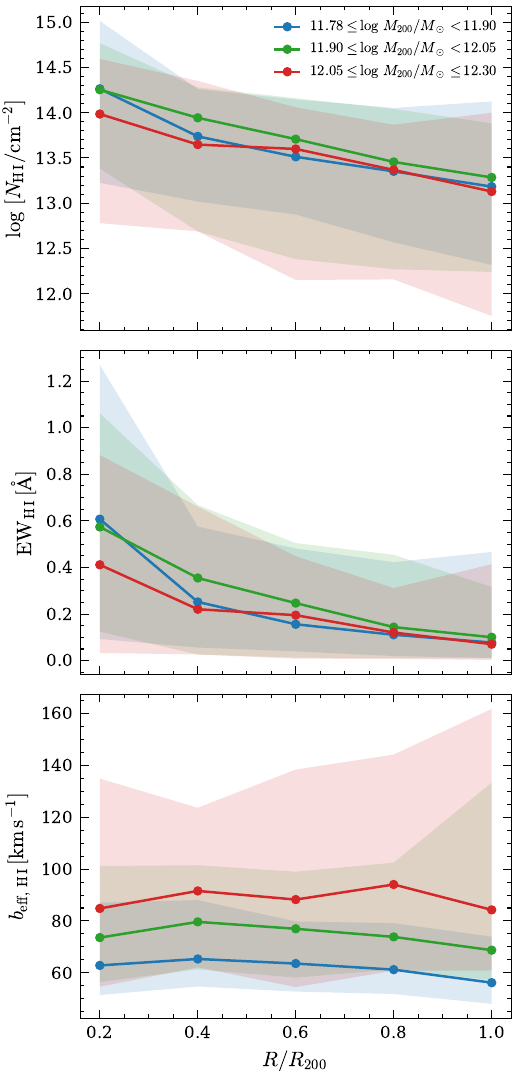}
   \caption{Same as Figure \ref{figure3}, but now for the full sample of 15 MW galaxy halos (675 sightlines) divided into three halo mass bins (colour-coded; see upper panel).
   \label{figure4_mass_bins}}
\end{figure}

\begin{figure*}[h]
\centering
   \includegraphics[width= 0.90\textwidth]{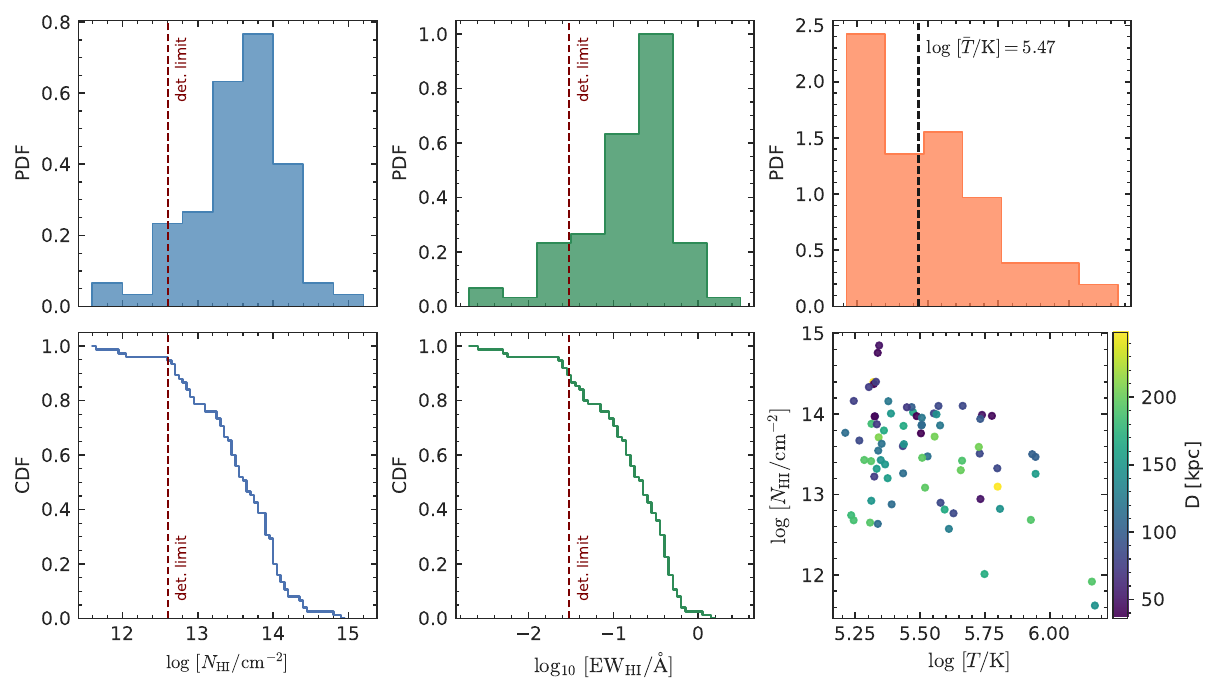}
    \caption{
    Top panels: Distributions of \ion{H}{I} column densities, $\log N_{\text{HI}}$, \ion{H}{I} equivalent widths ($\mathrm{EW}_\mathrm{HI}$), and average gas temperatures, $\log T$, for all CBLAs (i.e. warm-hot gas) identified in the 75 sightlines across our sample of 15 MW-like galaxies from TNG50 at $z=0$.
     Bottom left and middle panels: Cumulative fractions of $\log N_{\text{HI}}$ and $\mathrm{EW}_\mathrm{HI}$, respectively.
    Bottom right panel: Dependence of $\log N_{\text{HI}}$ and $\log T$ colour-coded by the impact parameter $D$.
    \label{cbla_random_distr}}
\end{figure*}

\subsection{CBLAs in halos of different masses}

After exploring one representative MW-mass halo, we expanded our analysis to our full sample of 15 MW-halos with masses between $11.78 \leq \log M_{200}/M_\odot < 12.30$. To systematically explore the CBLA properties as a function of halo mass, we divided our galaxy sample into three different mass bins and plotted in Figure \ref{figure4_mass_bins} the same parameters as in Fig. \ref{figure3}, now colour-coded for the different mass ranges. As for our representative model halo, each galaxy in our sample was surrounded by 45 sightlines in a ring-like pattern with radii spanning from $0.2$ to $1.0 R_{200}$ (see above). The three mass ranges were: low-mass halos ($11.78 \leq \log M_{200}/M_\odot < 11.90$, blue; 7 halos, 315 sightlines), intermediate-mass halos ($11.90 \leq \log M_{200}/M_\odot < 12.05$, green; 4 halos, 180 sightlines), and high-mass halos ($12.05 \leq \log M_{200}/M_\odot \leq 12.30$, red; 4 halos, 180 sightlines). 

The three panels show, from top to bottom, the median \ion{H}{I} column density $\log N_\mathrm{HI}$, the equivalent width ($\mathrm{EW}_\mathrm{HI}$), and the effective Doppler parameter $b_\mathrm{eff, HI}$ as a function of the normalized radius $R/R_{200}$ for the three mass bins using the colour coding listed above. The colour-shaded regions represent the $\sigma$ scatter across sightlines in each bin. Across the full sample, both $N_\mathrm{HI}$ and $\mathrm{EW}_\mathrm{HI}$ show a steady decline with distance with similar trends for all three mass bins. The median \ion{H}{I} column densities of the CBLAs in the inner halos have values of $\log N_\mathrm{HI} \approx 14.0 - 14.3$ ($\mathrm{EW}_\mathrm{HI}\approx 0.4-0.6$~\AA), while near $R_{200}$ these values reduce to $\log N_\mathrm{HI} \approx 13.0-13.2$ ($\mathrm{EW}_\mathrm{HI}\leq 0.1$~\AA). Across the full radial range, galaxy halos in the highest mass bin (red) have smaller median \ion{H}{I} column densities and median EWs than galaxies in the lower mass bins (blue and green). Only at $R \approx R_{200}$, the median values  $\log N_\mathrm{HI}$ and $\mathrm{EW}_\mathrm{HI}$ for low-, mid-, and high-mass halos become almost similar, indicating that the mass-dependence of $\log N_\mathrm{HI}$ and $\mathrm{EW}_\mathrm{HI}$ diminishes in the outer CGM. These trends indicate that, for a given absorber size, the neutral gas fractions in CBLAs in the inner halos around more massive galaxies are smaller compared to lower-mass halos due to the higher gas temperatures
\citep[see also][]{2024MNRAS.528.3745D}.
This is expected in collisionally ionised gas that adjusts to the halo virial temperature within its cooling radius \citep[see discussion in][]{Richter_2020_hot_halos}. The higher gas temperature in CBLAs around more massive galaxies in TNG50 is also evident in the distribution of $b_\mathrm{eff, HI}$, as displayed in the lower panel of Fig. \ref{figure4_mass_bins}, where the median $b_\mathrm{eff, HI} \approx 80-90$ km\,s$^{-1}$ in the highest mass bin (red), while it is systematically smaller in the two lower mass bins ($b_\mathrm{eff, HI} \approx 70-80$ km\,s$^{-1}$ for 
intermediate-mass halos, $11.90 \leq \log M_{200}/M_\odot < 12.05$, and $b_\mathrm{eff, HI}\approx 55-65$ km\,s$^{-1}$ for low-mass halos, $11.78 \leq \log M_{200}/M_\odot < 11.90$. 
For the latter two mass bins (green and blue), $b_\mathrm{eff, HI}$ peaks near $0.4 R_{200}$, while it steadily declines towards larger radii, possibly indicating a mild temperature decline beyond the halos' cooling radii (see discussion above). Only for the highest mass bin (red), such a behaviour is not evident, but here the range in $b_\mathrm{eff, HI}$ is substantial, possibly hiding any systematic trend. 

In general, all three panels in Fig. \ref{figure4_mass_bins} exhibit a substantial scatter in the radial trends for $\log N_\mathrm{HI}$, $\mathrm{EW}_\mathrm{HI}$, and $b_\mathrm{eff, HI}$, reflecting the patchy, multi-phase nature of the warm-hot CGM in the TNG50 simulations. This patchiness marks a striking difference compared to CBLAs in semi-analytic models \citep{Richter_2020_hot_halos}, where CBLAs arise (by definition) from the smooth coronal gas distribution in the hydrostatic model halos.  
A more detailed comparison between the simulated CBLAs in our study and the semi-analytical CBLA models from \citet{Richter_2020_hot_halos} is provided in Sect.\, \ref{comparision_observations}.

\subsection{Statistical properties of CBLAs along randomly placed sightlines around galaxies}\label{CBLA_random}

To better reflect the stochastic nature of quasar sightlines in observational CGM studies, which are randomly distributed around the CGM host galaxies, we took an additional, observationally motivated approach to study CBLAs in TNG50. For this, we selected five sightlines per galaxy at random impact parameters within $R_{200}$ for our sample of 15 galaxies, yielding a total of 75 sightlines for the following analysis. Then, we derived the total CGM \ion{H}{I} column density for each sightline, where we combined the contribution of the warm-hot phase (CBLA phase; $T>10^5$ K) with that of the cooler CGM phases (see also Sect. \ref{hot_cgm} and Fig.\,\ref{figure1}). 

Figure \ref{cbla_random_distr} shows the distribution of the neutral hydrogen column densities, $\log N_{\mathrm{HI}}$, equivalent widths, $\mathrm{EW}_\mathrm{HI}$, and average temperatures, $T$, for all CBLAs identified in our 75 randomly placed sightlines. CBLA absorption is detected along all 75 sightlines, indicating a CBLA covering fraction of 100 $\%$.
The column densities span a range from $\log N_{\text{HI}} \sim 11.6 - 15.0$, with a broad peak centred near $\log N_{\text{HI}} \approx 13.6$
(Figure \ref{cbla_random_distr}, upper left panel). 
This peak is substantially higher than the characteristic
peak value of $12.9$ derived from the semi-analytic model reported in \citep{Richter_2020_hot_halos}, indicating a larger amount and/or enhanced patchiness of the hot circumgalactic gas in TNG50 compared to the idealized quasi-hydrostatic model. We find that about $92\%$ of the CBLAs lie in the range $\log N_{\text{HI}} \sim 12.4 - 14.4$. The corresponding $\mathrm{EW}_\mathrm{HI}$ distribution  
(Figure \ref{cbla_random_distr}, upper middle panel)
is strongly skewed toward lower $\mathrm{EW}_\mathrm{HI}$ values, with $90.7\%$ of the values lying within the range $0.0-0.5$~\AA. On the other hand, the temperature distribution (Figure \ref{cbla_random_distr}, upper right panel) places the majority of the absorbers within the warm-hot, collisionally ionised regime at $\log\ [T/\text{K}]=5.2-6.4$, with a median CBLA temperature of $\log\ [T/\text{K}] = 5.47$. The temperature range $\log\ [T/\text{K}] = 5.0-5.5$ is a regime in which also highly ionised metal ions such as O\,{\sc vi} and N\,{\sc v} are expected to arise in a metal-enriched CGM. The connection between CBLAs and high metal ions is discussed in Sect.\, \ref{connection_CBLA_high_ion}.

\begin{figure*}
\sidecaption
\includegraphics[width=10cm]{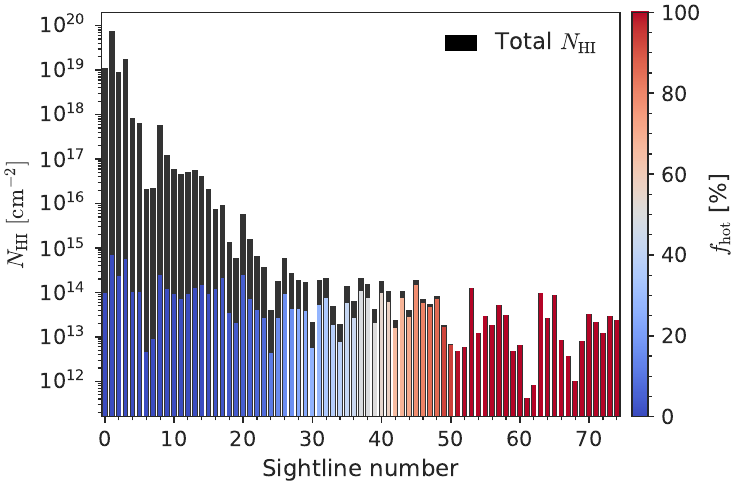}
\caption{Distribution of \ion{H}{I} column densities for the 75 CGM absorbers discussed in Sect.~\ref{CBLA_random}. The black bars represent the total \ion{H}{I} column density (i.e. including all gas phases), while the coloured bars show the \ion{H}{I} column density of the warm-hot component ($T \geq 10^{5}\,\mathrm{K}$), i.e. the CBLA column density (see also Fig.~\ref{figure1}). The colour coding indicates the hot gas fraction $f_{\mathrm{hot}}$, as defined in Sect.~\ref{CBLA_random}.}
\label{NHI_bar_sightlines_colormapped}
\end{figure*}

In real UV spectra (such as obtained with HST/COS), weak CBLAs with $\log N_{\text{HI}} \lesssim 12.6$ (EW $\leq 0.03$\,\AA) would be very hard (or even impossible) to detect, in particular in spectra of moderate-to-low signal-to-noise (S/N) ratios of S/N $<20$  \citep[as it is common in IGM surveys][]{Danforth_2016}
and if the CBLA absorption is overlaid by Ly$\alpha$ absorption from cooler CGM components 
\citep[see examples in][]{Richter_2020_hot_halos}.
This limitation is even more pronounced for very broad absorbers with Doppler widths of $b \sim 100-150$ km s$^{-1}$, which generally require column densities of at least $\log N_{\text{HI}} \gtrsim 12.8-13.0$ to be detectable at S/N $\gtrsim 40$ \citep{Richter_2020_hot_halos}. 
Consequently, many of the CBLAs shown here would likely remain undetected in real observational data. For instance, the observational CBLAs observed with HST/STIS at high S/N and discussed in \citet{Richter_2020_hot_halos} exhibit column densities (EWs) in the range $\log N_{\text{HI}} = 13.1-13.3$ ($0.07 - 0.11$\,\AA). This aspect is further discussed in Section \ref{discussion}. 

At the high end of the column density distribution, a small fraction of sightlines ($\sim 4\%$) exhibit CBLAs with $\log N_{\text{HI}} \gtrsim 14.4$, corresponding to systems with Ly$\alpha$ equivalent widths exceeding the $1\,\AA$. These systems obviously trace unusually dense and/or extended pockets of hot gas in the CGM of their host galaxies. Note that such strong CBLAs are not reproduced in the semi-analytic model \citep{Richter_2020_hot_halos}, where the CBLAs span a column-density range of $\log N_{\text{HI}} = 11.5-13.6$. We further explore the origin and nature of the strong CBLAs in Section \ref{CBLA_deep}. 

In the bottom (left and middle) panels of Figure \ref{cbla_random_distr}, we show the cumulative distributions for $\log N_\text{HI}$ and the $\mathrm{EW}_\mathrm{HI}$. These panels visualize the relative fractions of CBLAs that would be observable in real spectra above a typical detection limit (expressed in values of $\log N_\text{HI}$ and $\mathrm{EW}_\mathrm{HI}$) that is usually defined by the local S/N and by blending effects. As we can see, roughly $50 \%$ of the CBLA absorbers in our sample exceed $\log N_\text{HI}\sim 13.5$, corresponding to $\mathrm{EW}_\mathrm{HI}\sim 0.1 - 0.2$~\AA{}, i.e. near the sensitivity threshold of HST/COS spectra. Only about $10\%$ of absorbers are stronger than $\log N_\text{HI}= 14.0$ or with a $\mathrm{EW}_\mathrm{HI} >0.4$~\AA.

The scatter plot (bottom right), on the other hand, shows the relation of $\log N_\text{HI}$ and $\log\ T$ colour-coded by impact parameter. We see that the strongest absorbers, with $\log N_{\text{HI}} \gtrsim 14.0$, predominantly arise at $\log T \approx 5.3-5.6$. At higher temperatures ($\log T \gtrsim 5.8$), increasing collisional ionisation leads to systematically lower column densities, typically $\log N_{\text{HI}} \lesssim 13$. The colour distribution further shows that strong and weak absorbers occur at all radii ($D \approx 30-160$ kpc), with only a mild trend for hotter systems to appear at larger impact parameters and the highest $\log N_{\text{HI}}$ CBLAs to occur at the smallest. This trend, however, is weak, implying that the presence of CBLAs is not set primarily by distance from the galaxy centre (even though we expect them to arise within the virialized halo), but instead depends on local variations in density, temperature, and the small-scale structure of the warm-hot CGM. To help the reader interpret these trends more clearly, we show in Appendix~\ref{appendix:random_sample_radial} the direct relation between impact parameter, temperature, and column density for this CBLA sample.

\begin{figure}
\centering
   \includegraphics[width= 0.45\textwidth]{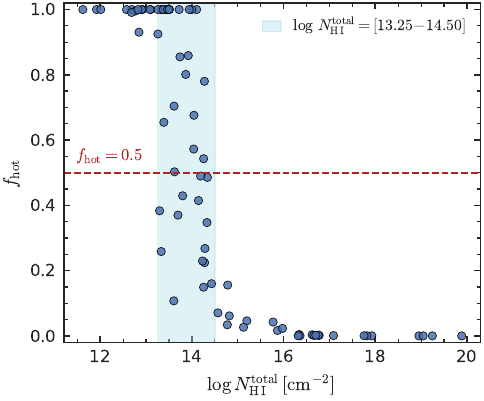}
    \caption{Fraction of hot gas, $f_{\rm hot}$, as a function of the total (i.e. all gas phases) \ion{H}{I} column density for our sample of 75 CGM absorbers. The horizontal dashed line marks  $f_{\rm hot} = 0.5$, separating hot-dominated from cold-dominated absorbers. The shaded region highlights the \ion{H}{I} column density range $\log N_\mathrm{HI}^{\rm total} = 13.25 - 14.50$, where both hot- and cold-dominated absorbers co-exist. 
    \label{fhot_N_total}}
\end{figure}

In Figure \ref{NHI_bar_sightlines_colormapped}, we show (sightline-by-sightline) how the CBLA \ion{H}{I} column density
(i.e. the column density from the warm-hot gas; coloured bars;  $N_\text{HI}^{\text{hot}}$) contributes to the total \ion{H}{I} column density (black bars;  $N_\text{HI}^\text{total}$) for all 75 sightlines. The colour code (see right axis label) indicates the warm-hot gas fraction, $f_{\text{hot}}$, defined as $f_\text{hot}=N_\text{ HI}^\text{hot}/N_\text{HI}^\text{total}$. The sightlines are sorted by increasing values of $f_{\text{hot}}$. There is a large variation in $f_{\text{hot}}$ among the 75 sightlines. In some cases, the warm-hot component contributes only a minor fraction of the total \ion{H}{I} column density, while in many others, the hot phase actually dominates, then accounting for more than 95\% of the total column density, where CBLAs provide the column-density floor of the circumgalactic \ion{H}{I} with its 100$\%$ covering fraction. We find that roughly $35$\% of our sightlines are hot-dominated ($f_\text{hot} > 90\%$), about $32\%$ are cold-dominated ($f_\text{hot} < 10\%$), while the remaining $\sim 35 \%$ are mixed ($ 10 \% < f_\text{hot} < 90\%$). 

In Figure \ref{fhot_N_total}, we show the hot gas fraction $f_{\text{hot}}$ as a function of the total \ion{H}{I} column density $N_\text{HI}^\text{total}$. We find that above $\log \ N_\text{HI}^\text{total} \approx 14.5$ all absorbers are cold-dominated, with $f_{\text{hot}} \lesssim 0.1$, while for $\log \ N_\text{HI}^\text{total} \leq 13.2$, most of them are hot-dominated with $f_{\text{hot}} \ge 0.85$. However, between these two limits, in the interval of  $13.25 \leq \log \ N_\text{HI}^\text{total} \leq 14.5$, absorbers with similar column densities show large variations in $f_{\text{hot}}$, i.e. they can be either hot-dominated or cold-dominated. This means that column density alone does not uniquely trace the thermal phase of the gas: Two absorbers with nearly identical $\log \ N_\text{HI}^\text{total}$ may correspond to either cold, narrow Ly$\alpha$ lines or to warm–hot, broad CBLA-like profiles.

\section{Strong CBLAs}\label{CBLA_deep}

As we have seen in the previous section, a small fraction of the sightlines in our randomized sample exhibit strong CBLAs with high column densities as large as $\log N_{\text{HI}} \approx 14.9$. Such systems stand out from the rest of the sample and are difficult to reconcile with what is typically expected from smoothly distributed, hot coronal gas in a quasi-hydrostatic equilibrium \citet{Richter_2020_hot_halos}. From an observer's point of view, Ly$\alpha$ absorbers at this column-density level represent strong, fully saturated lines that (in the common CGM spectral classification scheme) would typically be interpreted as photoionised Ly$\alpha$ features tracing the cooler ($T<10^5$ K) CGM, which they are not. These strong CBLAs therefore represent a new, previously unnoticed absorber class that traces pockets of warm-hot ($T>10^5$ K) gas in the CGM of their host galaxies.

\begin{figure}[t!]
\centering
   \includegraphics[width= 0.45\textwidth]{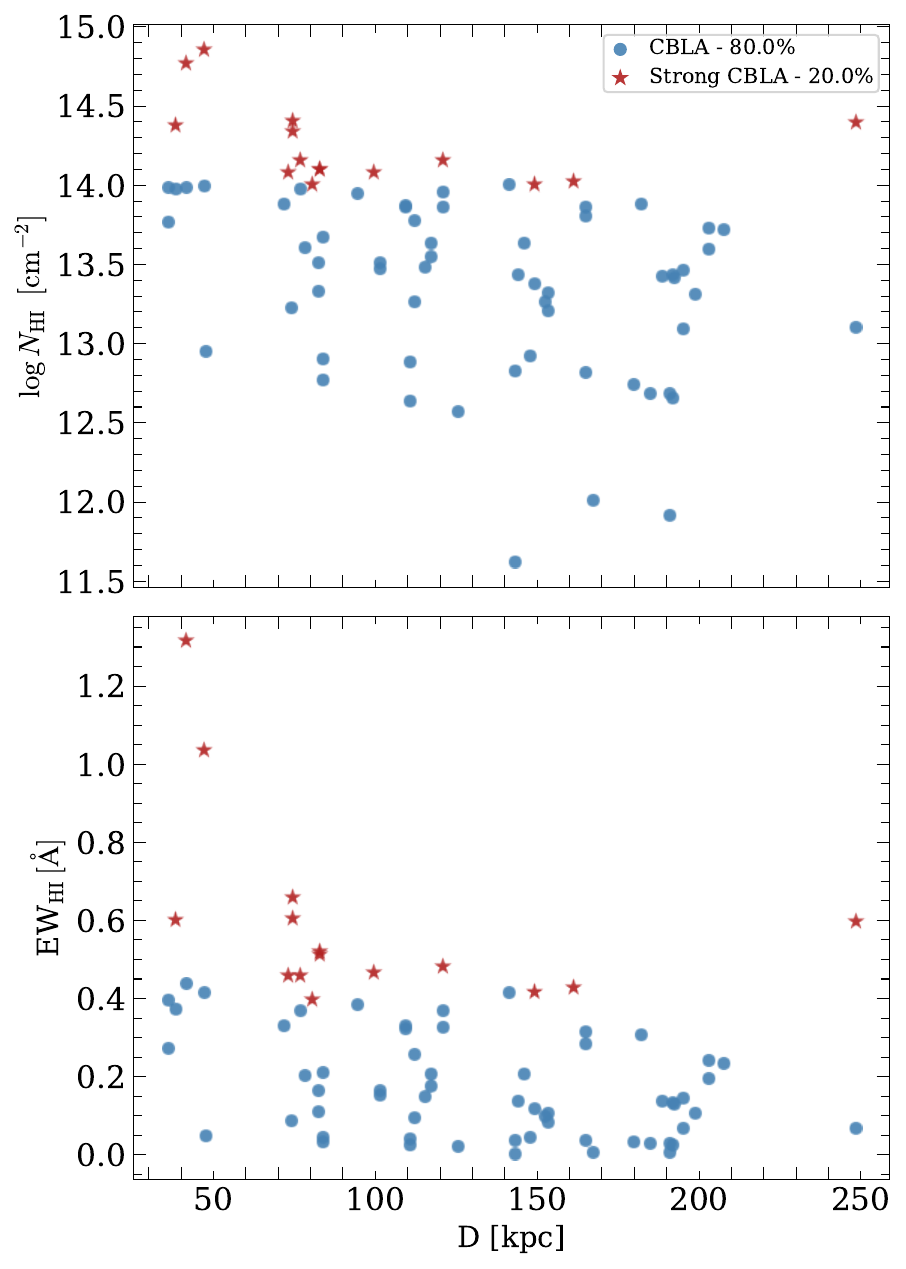}
    \caption{Neutral hydrogen column density ($\log N_{\mathrm{HI}}$, top) and equivalent width ($\mathrm{EW}_\mathrm{HI}$, bottom) as a function of impact parameter $D$ for our sample of 75 simulated CBLAs. Strong CBLAs with $\log N_{\text{HI}} > 14.0$ are marked as red stars (see Section\ \ref{CBLA_deep}).
    \label{NHI_EW_impact}}
\end{figure}

\subsection{Possible origins of strong CBLAs}

One can think of (at least) two important processes in galaxy evolution that may be able to circulate large amounts of warm-hot gas in the circumgalactic environment of MW-like galaxies:
galactic winds/outflows \citep[e.g.][]{Nelson2019,P_roux_2020, Truong_2021, Pillepich_2021,Laktionov_2025} and the hot-gas accretion (potentially in combination with cooling flows) from the IGM \citep{Stern_2024}. While the former process is expected to inject hot gas predominantly into the inner CGM perpendicular to the disk, the latter process could potentially distribute warm-hot accretion streams throughout the entire halo even at larger distances, but preferentially along the disk \citep[due to the conservation of angular momentum; see][]{Stern_2024}. Both of these processes would be able to produce large-scale flows of warm-hot gas with \ion{H}{I} column densities large enough to appear as a strong CBLA when observed in absorption.  

To further explore these scenarios, we investigated 
the radial distribution of strong CBLAs in our sample in Figure \ref{NHI_EW_impact}. For this, we defined absorbers as strong CBLAs if their \ion{H}{I} column density was $\log N_{\text{HI}} \geq 14.0$. This column density was chosen as a cut-off value because it is substantially larger ($0.4$ dex) than the peak \ion{H}{I} column density in our random sample (Fig.\ \ref{cbla_random_distr}) and at this column density, a typical CBLA line with $b=50$ km\,s$^{-1}$ reaches an absorption depth of $\sim 50\%$. Based on this criterion, we found that $80$\% of the absorbers in our sample fall within the ``normal'' CBLA regime (61 systems), while the remaining $20$\% are classified as strong CBLAs (14 systems). In Figure \ref{NHI_EW_impact}, we plot the \ion{H}{I} column density (upper panel) and the $\mathrm{EW}_\mathrm{HI}$ (lower panel) against the impact parameter, $D$ for all 75 CBLAs, where normal CBLAs are indicated in blue and strong CBLAs are indicated in red. 

Although there is a tendency for strong CBLAs with the highest column densities and equivalent widths to cluster at smaller impact parameters ($D \leq 100$ kpc), several strong CBLAs were also found at significantly larger impact parameters, even beyond $200$ kpc. One might interpret the preference for small $D$ as being consistent with a simple spherically symmetric halo, where sightlines at smaller impact parameters probe longer path lengths and higher average warm–hot gas densities. However, a purely geometric effect is not sufficient to explain their high \ion{H}{I} column densities: over the range of $D = 50 -200$ kpc, the path length changes only by a factor of $1.5\ (\sim 0.2\ \text{dex})$, while strong CBLAs exceed the bulk of the sample by up to $\sim 1\ \text{dex}$ in $\log N_\text{HI}$. This indicates that, in addition to geometry, strong CBLAs require locally enhanced gas densities that are not uniquely captured by a smooth spherical halo model.

\begin{figure}[t!]
\centering
   \includegraphics[width= 0.45\textwidth]{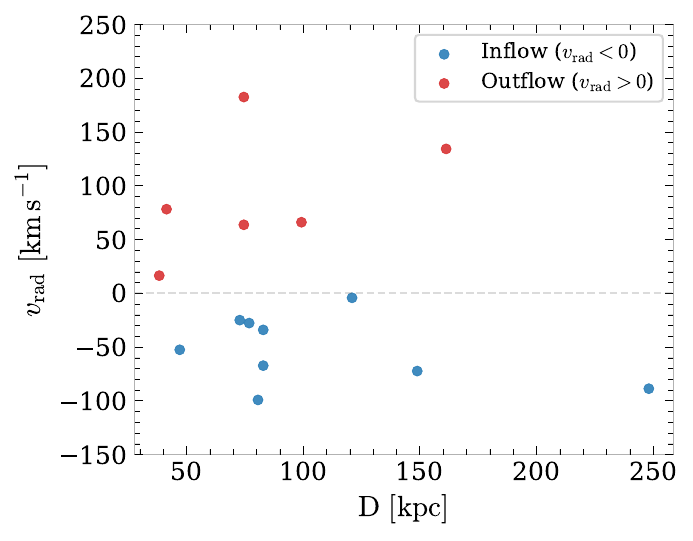}
    \caption{Radial velocities of strong CBLAs as a function of impact parameter $D$. Blue and red points correspond to inflowing ($v_{\rm rad} < 0$) and outflowing ($v_{\rm rad} > 0$) gas, respectively. The dashed horizontal line indicates zero radial velocity, separating inflows from outflows.
\label{vel_rad_strong_CBLAs}}
\end{figure}

\begin{figure*}[t!]
\centering
  \hspace*{-1cm}\includegraphics[width=1.0\textwidth]{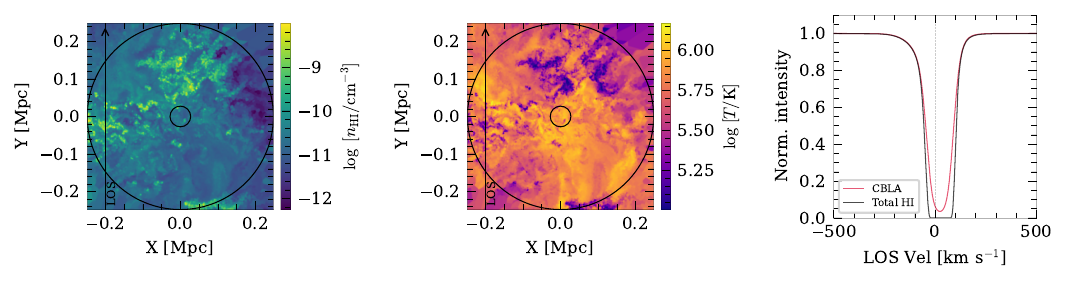}
    \caption{Face-on 2D-slice view of a MW-like halo from TNG50 at $z=0$ with $\log M_{200}/M_\odot = 12.03$ showing the neutral hydrogen density ($\log n_{\text{HI}}$, left) and temperature (middle) for the warm-hot gas ($T \geq 10^{5}$ K). The LOS is indicated by the black arrow; the bigger black circle marks $R_{200}$, and the smaller one denotes the extent of the \ion{H}{I} disk for reference, which in this case $R_{\rm HI-disk} = 27.74$ kpc. Additionally, the right panel represents the normalized Ly$\alpha$ absorption along this particular sightline localised at $D=248.12$ kpc. As can be seen, the LOS intersects a localized, hot \ion{H}{I}-rich clump embedded in the diffuse hot corona, which produces the strong CBLA seen in the spectrum.  
    \label{strong_CBLA_slices}}
\end{figure*}

\subsection{Kinematics of strong CBLAs}

To further investigate the origin and nature of these systems in the context of outflowing and infalling warm–hot gas, we show in Figure \ref{vel_rad_strong_CBLAs} the radial velocities of strong CBLAs in TNG50 plotted against the projected distances from the galaxy centre to assess whether they are predominantly inflowing or outflowing.

As can be seen, the strong CBLAs span both regimes: around $\sim 43 \%$ of the systems (6 out of 14) represent outflowing material with positive velocities in the range $v_{\rm rad} \approx 100-200$ km\,s$^{-1}$, whereas $\sim 57\%$ of the systems represent inflowing material with negative velocities at $v_{\rm rad} \approx -50$ to $-150$  km\,s$^{-1}$. This mixed behaviour is consistent with the two physical processes discussed above. In particular, the outflowing strong CBLAs have radial velocities that fall within the characteristic range of warm–hot winds predicted for MW-mass halos in TNG50 \citep[e.g.][]{Nelson2019}, where warm-hot gas typically reaches $v \approx 150-300$ km\,s$^{-1}$ at radii of $50-150$ kpc. 

While the radial velocity analysis of the \ion{H}{I} component already showed whether the strong CBLAs are inflowing or outflowing, \ion{H}{I} alone cannot capture the full thermal or ionisation state of the warm-hot CGM structure hosting the absorption. These systems arose within warm–hot, multi-phase structures, so their dynamical behaviour is better understood when also considering associated high metal ions. \ion{O}{VI}, in particular, is expected to be also present in the structures that trace these strong CBLAs, since in collisional ionisation equilibrium, the \ion{O}{VI} ion fraction peaks near $T \sim 10^{5.5}\ \text{K}$. Analysing the \ion{O}{VI} kinematics of the same systems would therefore provide a complementary view: it would allow us to assess whether the warm-hot phase sampled by \ion{O}{VI} moves coherently with the overall \ion{H}{I} in CBLAs, or whether the different phases are kinematically separated from each other. This is further discussed in Sect.\, \ref{connection_CBLA_high_ion}.

Figure \ref{strong_CBLA_slices} visualizes the global gas environment of a strong CBLA and provides an example of the visual appearance of a CGM absorption system that hosts a strong CBLA. In this figure, we display a strong CBLA with $\log N_{\text{HI}} = 14.4$, located at a very large impact parameter of $D = 248$ kpc from its host galaxy, thus at the edge of its $R_{\rm 200}$ radius ($248.12$ kpc). In Figure \ref{NHI_EW_impact}, this strong CBLA is the one with the largest value for $D$ in the outermost right part of the two panels. 

In the left and middle panels of Figure \ref{strong_CBLA_slices}, we show the face-on 2D slices of the corresponding galaxy halo up to $\pm R_{\rm 200}$, showing the \ion{H}{I} density (left panel) and temperature (middle panel) distribution of the warm-hot gas with $T \geq 10^{5}$ K. These two maps reveal that neutral hydrogen in the warm-hot coronal gas is not smoothly distributed, but instead located in irregular, clumpy structures. Even at million-degree temperatures, localized high-density regions (i.e., clouds) with higher \ion{H}{I} content are present. The sightline (black arrow) intersects one of these clumpy regions within the otherwise diffuse, ionised CGM, giving rise to the strong Ly$\alpha$ absorption seen in the right panel. The temperature map in the middle panel indicates that the gas that gives rise to the absorption is cooler ($ \log\, [T/\text{K}] \approx 5.0-5.2$) than the surrounding halo ($ \log\, [T/\text{K}] \approx 5.8-6.0$) along the LOS, placing the absorbing gas phase into the so-called transition-temperature regime \citep{wakker2012}.

The resulting Ly$\alpha$ absorption line from this sightline (right panel) is fully saturated. As in Fig.\ \ref{figure1}, we indicate
with the red curve the \ion{H}{I} profile arising from the warm-hot phase gas alone, while the black curve shows the total \ion{H}{I} profile including all gas phases (hot, warm, and cool gas). Although the warm-hot gas adds only a small fraction of the total \ion{H}{I} column density ($f_\text{hot}= 0.042$; see also Fig.\ \ref{NHI_bar_sightlines_colormapped}, sightline number 20), it dominates the shape of the broad Ly$\alpha$ absorption component.
In view of the large impact parameter and the location within the plane of the rotating galaxy disk, this massive warm-hot \ion{H}{I}-bearing structure could have arisen from several physical processes. From Fig.~\ref{vel_rad_strong_CBLAs}, we found that this structure is inflowing. One possibility is a cooling flow from the gaseous halo, as discussed by \citet{Stern_2024}, but it may also have been associated with stripped or mixing material from infalling satellite galaxies \citep[e.g.][]{Ramesh_2023}. 

While several studies \citep[e.g.][]{10.1093/mnras/stad3142, 2024MNRAS.527.3494W, Damle_2025} have shown that satellites can contribute a substantial fraction of the cold gas in the CGM of MW-like halos, their contribution to the warm–hot gas phase remains unexplored. Therefore, to test whether our CBLAs could be associated with satellite gas, we carried out a simple analysis at $z = 0$ to identify sightlines intersecting gas bound to satellites. We found that $\leq 1\%$ of our sightlines intersect this gas. Similarly, for our MW-like halos, we found that satellite gas is not the dominant source of warm-hot \ion{H}{I} gas, with an average contribution of $\sim 0.006\%$. We note, however, that this analysis only includes gas currently associated with satellites at $z = 0$. To fully understand their contribution, we would need to track the origin of this gas over time, which is beyond the scope of this work. In the Appendix \ref{appendix:strong_CBLA_along_LOS}, we provide a more detailed description of the physical conditions along this example sightline that lead to the formation of this strong CBLA.

\section{Discussion}\label{discussion}

One key result of our study is that CBLAs are ubiquitous in the CGM of MW-like halos in TNG50 out to $R_{200}$. CBLA absorption was detected along each of the sightlines in our random LOS sample at $\log N_{\text{HI}}>11.6$ (Sect. \ref{CBLA_random}), confirming that the
warm-hot phase of the CGM at $T>10^5$ K, despite its small neutral gas fraction, provides 100\% cross-section for H\,{\sc i} Ly$\alpha$ absorption. 80\% of the sightlines exhibited CBLA absorption at a column-density level of $\log N_{\text{HI}} > 13.0$ (Fig. \ref{cbla_random_distr}), a level at which BLAs are principally detectable in HST/COS spectra, if the CBLA absorption can be separated from the Ly$\alpha$ absorption of cooler CGM components (see discussion below). In about half of the sightlines in our random sample (Sect. \ref{CBLA_random}), CBLA absorption from warm-hot gas dominated the column density of the CGM absorption (Fig. \ref{NHI_bar_sightlines_colormapped}).
This value increased to 100 \% when considering only 
low-column density H\,{\sc i} absorbers with $\log N_{\text{HI}} < 14.0$. In other words: in TNG50, the Ly$\alpha$ absorption cross-section for 
low-column density H\,{\sc i} absorbers with $\log N_{\text{HI}} < 14.0$ in MW-like galaxy halos
was dominated by the warm-hot component of the CGM.
CBLAs therefore provide the HI column density floor in low redshift CGM around these types of galaxies. 

However, our study demonstrates that there also exists a population of strong CBLAs with $\log N_{\text{HI}}>14.0$, which made up about 20\% of the CBLA absorber 
population in TNG50. These strong CBLAs represent strong, saturated Ly$\alpha$ lines and thus mimic an H\,{\sc i} absorption pattern that an observer would typically assign to the cooler, photoionised CGM gas phase. The occurrence of these strong CBLAs at both small and large impact parameters (Fig. \ref{NHI_EW_impact}) suggests that hot-gas outflows \citep[e.g.][]{Laktionov_2025}, warm-hot accretion streams \citep[e.g.][]{Stern_2024}, and stripping material from satellites \citep[e.g.][]{Ramesh_2023} can contribute to this interesting new sub-population of CBLAs that have not been recognized before as an individual absorber class.

In general, we found that the strength of CBLAs declines with increasing impact parameter (Figs. \ref{figure4_mass_bins} \& \ref{NHI_EW_impact}), possibly related to the decreasing absorption pathlength and small mean density at larger radii. However, large scatter exists in these relations, pointing towards
an inhomogeneous, clumpy distribution of warm-hot gas along typical halo sightlines in the galaxy-mass range traced here. Indeed, it is well known from other studies that
MW-like halos at $\sim10^{12} M_\odot$ exhibit the strongest neutral CGM, consistent with their position near the peak of star formation efficiency and their ability to access 
both cold and hot accretion modes \citep[e.g.][]{Conroy_2006, Behroozi_2013,2024MNRAS.528.3745D}. 
More massive halos ($\gtrsim 10^{12} M_\odot$) begin to develop virially shock-heated atmospheres that suppress the neutral fraction of the halo gas. Therefore, the slightly lower values for $\log N_\text{HI}$ and Ly$\alpha$ EW in the inner regions of more massive galaxies (Fig. \ref{figure4_mass_bins}; upper
two panels) are consistent with these considerations. Taken together, our study demonstrates that CBLAs represent an important absorber class that needs to be taken into account when characterising the absorption properties of the CGM around 
MW-like galaxies.

\subsection{Comparison with analytic models and observations}\label{comparision_observations}

In \cite{Richter_2020_hot_halos}, a first detailed analysis of CBLAs was carried out using a semi-analytical approach, in which analytic equations for the radial density and temperature profiles of a quasi-hydrostatic hot halo gas 
residing in an NFW potential were adopted, accounting for gas cooling and fragmentation under realistic conditions. This study focused on predicting the strength, spectral 
profiles, and cross-section properties of CBLAs as a function of halo mass and LOS impact parameters. For halo masses between $10.6 \leq\log \left(M/M_{\odot} \right) \leq 12.6$, the model predicted characteristic \ion{H}{I} column densities of CBLAs in the range $\log N_{\text{HI}}=12.4-13.4$ with a peak value in the distribution at $\log N_{\text{HI}}=12.9$. 

Compared to the semi-analytic model of \citet{Richter_2020_hot_halos}, the \ion{H}{I} column-density distribution of CBLAs in TNG50 was shifted towards higher values with the majority of CBLAs having $\log N_{\text{HI}}=13.0-14.0$ and a peak at $\log N_{\text{HI}}=13.7$. This substantial offset is probably due to the differences in how the two approaches treat the warm-hot CGM. Unlike TNG50, the semi-analytical model cannot resolve detailed interactions between galaxies and the CGM, such as turbulence, small-scale clustering, or complex kinematic motions and the resulting fragmentation of the gas. The enhanced cooling in such warm-hot gas fragments and the enhanced hydrogen recombination will lead to lower gas temperatures (e.g. in the transition-temperature regime) and higher neutral gas fractions. This, in turn, will increase the H\,{\sc i} column densities along sightlines through such a clumpy medium compared to the (idealized) smooth gas distribution in the semi-analytic approach, where the
gas is (locally) isothermal near the halo's virial temperature. The observed radial decline of the CBLA column density in TNG50 (Figs. \ref{figure4_mass_bins} \& \ref{NHI_EW_impact}), on the other hand, is in agreement with the declining trend seen in the semi-analytic model 
\citet[][; their Fig. 5]{Richter_2020_hot_halos}.

From an observational point of view, CBLAs have been studied systematically only very 
recently using HST/COS spectra. In their paper, \citet{sameer_cbla} have investigated 240 CGM absorption
components around 47 galaxies at redshift $z < 0.7$ using absorption from HST/COS and a ``cloud-by-cloud''
Bayesian modelling approach and non-equilibrium ionisation models, such as time-dependent photoionisation and collisional ionisation models (TDP) from \cite{gnat_2017}. They identified a population of 48 highly ionised, high-temperature absorption components (`TDP-High'), for which they obtained a median temperature of $10^{5.71}$K and an H\,{\sc i} column density range of $\log N_{\text{HI}}=12.0-13.5$ \citep[][their Fig.\,12]{sameer_cbla}. These systems represent CBLA analogues in the high-temperature wing of the CBLA distribution in TNG50. \citep{sameer_cbla} further identified a population of 35 multi-phase-absorbers (`TPD-Low'-systems) with higher H\,{\sc i} column densities ($\log N_{\text{HI}}=12.0-17.5$) and a lower median temperature ($10^{4.62}$ K). Some of these systems also contained hot gas with $T>10^{5}$K and most likely represent cases in which CBLAs and cooler CGM absorbers are aligned in velocity space, making it extremely difficult (or even impossible) to separate the contribution of the hot and cool gas components to the overall H\,{\sc i} optical depth. This aspect is further discussed in Sect. \ref{challenges_CBLAs}. The four initial CBLA examples presented in \citet{Richter_2020_hot_halos} spanned a column density range
$\log N_{\text{HI}}=13.1-13.3$ with $b$-values between $90$ and $150$ km\,s$^{-1}$, but no information on the gas temperature was available for these systems. 

\subsection{Detection challenges for CBLAs}\label{challenges_CBLAs}

In simulations, we have the crucial advantage of being able to disentangle the contributions from the different gas phases so that we can easily characterise the spectral signatures from these phases, such as presented in this study for the warm-hot gas component and the CBLAs. Detecting CBLAs in real spectral data is, in contrast, extremely challenging because their broad, shallow Ly$\alpha$ profiles are difficult to identify in continua of noisy spectral data and they are often masked by narrower absorption from cooler CGM gas (as discussed above).

On the one hand, the detectability of broad spectral features in spectra depends critically on their depth, width, and the S/N ratio in the data. Using high-resolution HST/STIS observations, \citet{2006A&A...451..767R} established an empirical detection criterion for BLAs in the WHIM that was later applied to CBLAs in \citet[][their equation 7]{Richter_2020_hot_halos}. Following this criterion, a typical CBLA with $\log N_{\text{HI}} \approx 13.0$ and $b_{\mathrm{HI}} \approx 130\ \mathrm{km\ s^{-1}}$ (i.e., $\log (N/b) \approx 10.9$) required a S/N ratio of approximately 40 per resolution element for reliable identification. In the semi-analytic CBLA model \citep{Richter_2020_hot_halos}, $80\%$ of CBLAs had $\log (N/b) \geqslant 10.5$, making them especially difficult to detect at large impact parameters where column densities drop. Similar detectability challenges have been seen for BLAs in general \citep[e.g.][]{Savage_2014,Danforth_2016}, where broad, shallow features require a profile fitting to separate thermal from non-thermal broadening. In our study, however, the CBLA column-density distribution was shifted towards larger values of $N_\text{HI}$ compared to the semi-analytic approach. In fact, $\sim60\%$ of our absorbers in TNG50 had on average $\log (N/b) \approx 11.6$ ($\sim 0.7$ dex stronger), so that these should be detectable at a S/N of $\geqslant 6$, which is available in many HST/COS spectra.

On the other hand, however, when a CBLA is embedded within a multi-component Ly$\alpha$ profile, its  detection depends also critically on i) the complexity of the \ion{H}{I} absorption pattern in the cooler gas components and ii) the velocity offset of the CBLA absorption from those cooler components, which in most cases dominate the \ion{H}{I} optical depth. Several observational studies have reported the detection of broad Ly$\alpha$ absorption features along sightlines close to galaxies that mimic the properties of CBLAs as discussed here \citep[e.g.][]{Savage_2014,Stocke_2014,2017ApJ...850L..10J,Werk_2016,Tumlinson_2013,Richter_2020_hot_halos,sameer_cbla}. While in earlier studies, these were reported as individual cases when spectroscopically de-composing multi-component absorption systems \citep[see, e.g.][]{Savage_2014,Stocke_2014}, the more systematic de-composition of CGM into photoionised and collisionally ionised gas components by \citet{sameer_cbla} using a Bayesian modelling approach indicated a detection rate as high as $\sim 64\%$ (about 47 sightlines show at least one `TDP-High' component, see their Fig. 20 and Table 3) for $\log N_{\text{HI}} > 12.0$, which is in good agreement with our prediction ($\sim 80\%$; see Fig.\ref{cbla_random_distr}).

\subsection{CBLAs and their connection to high metal ions}\label{connection_CBLA_high_ion}
The detection of high metal ions (e.g. \ion{O}{VI} and \ion{N}{V}) that are associated with the CBLA absorption provides an important guide to the location of warm-hot gas within a complex multi-phase CGM absorber \citep[see, e.g.][for example cases]{Savage_2014,Stocke_2014}. \ion{O}{VI}, in particular, is widespread in the CGM of the Milky Way \citep{sembach2003} and other low-redshift galaxies \citep[e.g.][]{sameer2024}, predominantly tracing gas in the range $T\approx 10^{5.0}-10^{5.5}$ K. In our TNG50 simulation, $90 \%$ of the sightlines with a CBLA also showed \ion{O}{VI} absorption, where 65 percent of the \ion{O}{VI} absorbers had column densities of $\log N(\ion{O}{VI})\geq 13.6$. This detection rate is in excellent agreement with observational data of the \ion{O}{VI} cross section in the CGM of low-redshift galaxies \citep{sameer2024}. The majority of the CGM \ion{O}{VI} absorption features in our TNG50 sightlines showed, however, a significant velocity offset ($\Delta v$) from the CBLA \ion{H}{I} central optical depth, where $\Delta v$ spanned a range between $-50$ and $+50$ km\,s$^{-1}$, typically. This is not at all surprising, because  
the \ion{O}{VI} absorption predominantly samples metal-enriched warm-hot gas at $T\approx 10^{5.0}-10^{5.5}$ K. The CBLAs, on the other hand, trace also gas at higher temperatures and are independent of the metal abundance, where the \ion{H}{I} optical depth distribution depends on the local gas density and (temperature-dependent) ionisation fraction. These effects result in CBLA absorption profiles and line centroids that differ from those of \ion{O}{VI}. While it would be highly interesting to further explore the complex connection between CBLAs and \ion{O}{VI} (and other high ions) in more detail (also in view of the origin of the gas and its outflow/inflow characteristics), such an analysis clearly is beyond the scope of this paper. We are, however, planning to address these aspects in a forthcoming follow-up study.

\subsection{Implications for constraining the CGM baryon budget}

The detection (or non-detection) of CBLAs also has implications for our understanding 
of the baryon distribution in and around galaxies in the context of the 
long-standing ``missing baryon'' problem. Baryons, constituting approximately  $17\%$ of the Universe's mass \citep{1984Natur.311..517B,Dunkley_2009,refId0}, are the building blocks of stars, galaxies, and other cosmic structures. However, a large portion of the baryons predicted by the $\Lambda$CDM model is not observed in galaxies \citep{2008IAUS..244..136M, McGaugh_2010, Anderson2010, Caudill_2023, Nicastro_2023}. Observational studies have shown that the photoionised, bound CGM can contribute at least 25\% of the total baryonic budget of an $L^{*}$ galaxy, a fraction comparable to or exceeding the baryon mass in stars and interstellar gas in the disk ($14-24$\%, \citealt{10.1093/mnras/stv318, McGaugh_2010}). When the contributions from all CGM phases are combined \citep{Tumlinson_2013, Anderson2013, Peeples_2014, Werk_2014}, the diffuse halo gas accounts for at least $\sim 35$\% of the total baryonic mass, thereby recovering more than half of the ``missing'' baryons ($\sim 60$\%). Including an additional $\sim 20$\% to account for saturation effects in \ion{H}{I} column density measurements brings the total baryon content of these halos close to the cosmological value \citep{Werk_2014}. 

In the context of these results, it is interesting to evaluate the baryon content of the 
warm-hot gas traced by CBLAs in our TNG50 simulations. For each halo we defined the total baryonic mass within $R_\text{200}$ as $M_\text{b}(< R_\text{200}) = M_\text{gas}(< R_\text{200}) + M_\text{stars}(< R_\text{200}) + M_\text{BH}(< R_\text{200})$. We then defined the warm-hot CGM mass, $M_\text{WH-CGM}$, as the mass in gas with $T>10^5$ K within $0.1R_\text{200} \leq r \leq R_\text{200}$. Therefore, the corresponding baryonic mass fraction residing in the warm-hot CGM was $f_\text{b,WH-CGM} = M_\text{WH-CGM}(< R_\text{200})/M_\text{b}(< R_\text{200})$. Using this definition, the median baryon fraction of 
the warm-hot CGM at $T>10^5$ K of the 15 TNG50 galaxy halos described above came out as 
$\langle f_\text{b,WH-CGM}\rangle=0.29$ with values spanning a range from $f_\text{b,WH-CGM}=0.17$
to $0.42$. These numbers underline that the warm-hot circumgalactic gas phase represents 
a major baryon reservoir in Milky-Way type galaxies in TNG50. CBLAs, however, predominantly trace the 
cooler part of this phase at sub-virial temperatures, as indicated in Fig.\, \ref{cbla_random_distr}, a temperature regime
in which the neutral gas fractions are the highest. If we limit our CGM baryon estimate in the 15
TNG50 halos to the range $T=10^{5.0-5.5}$ K, the 
transition-temperature regime, the median baryon 
fraction was reduced to a value $\langle f_{\rm b,TT-CGM} \rangle =0.07$. While this is still a substantial baryon reservoir, the baryon fraction is significantly smaller than the
value of $\langle f_\text{b,WH-CGM} \rangle =0.29$ for entire warm-hot temperature range ($T>10^5$) K. These numbers therefore indicated that the bulk of the circumgalactic baryons 
reside in the hot phase at $T>10^{5.5}$ K near the halo's virial temperature, which is extremely difficult to detect in H\,{\sc i} Ly$\alpha$ due to the very small neutral gas 
fractions and detection rates in such a hot plasma (see Figure \ref{cbla_random_distr}, right panel). 

X-ray observations in emission and absorption can be used to trace the million-degree gas phase in the CGM of the Milky Way and nearby galaxies
\citep[e.g.][]{Miller_2013,miller2015,anderson2016}. With current X-ray instruments, however, 
the sample size and sensitivity of such observations remain quite limited, which introduces
substantial uncertainties for estimating the baryon budget of the hot CGM gas phase from those. 
For the Milky Way, on the other hand, the million-degree gas phase can be also studied 
by using fast radio bursts \citep[FRBS; e.g.][]{cook2023} and (in the near future) by
highly forbidden optical high-ion line transitions in combination with stacking experiments, 
providing independent constraints on the baryonic mass hosted by million-degree CGM. 

\section{Summary and conclusions}\label{conclusions}

In this study, we have used the TNG50 simulation from the IllustrisTNG suite to 
systematically investigate the H\,{\sc i} Ly$\alpha$ absorption signatures 
of the warm-hot ($T>10^5$ K) circumgalactic medium around MW-like galaxies at $z=0$. 
For this, we have decomposed the Ly$\alpha$ absorption features of the CGM into different 
phases (cold/warm-hot) around 15 MW-like galaxies in the mass range
$11.7 \leq \log M_{200}/M_\odot \leq 12.3$. We then studied the strength and spatial distribution of the so-called coronal broad 
Ly$\alpha$ absorbers that trace the $T>10^5$ K CGM, characterising their properties. The main results of our study are summarized as follows:\\
\\
1) The TNG50 simulations imply that CBLAs are ubiquitous around Milky-Way type galaxies 
and have a large cross section. In fact, CBLA absorption is detected along each
of the 75 randomly placed sightlines around the considered 15 galaxy halos in TNG50 at impact 
parameters $D \leq R_{200}$ with H\,{\sc i} column densities covering the range 
$\log N_{\text{HI}}=11.6-15.4$. For a limiting H\,{\sc i} column density of $\log N_{\text{HI}}\ge 13.0$ (as typical for HST/COS spectra), the detection rate 
is still 80\%. These numbers indicate that the warm-hot gas component of the 
CGM as traced by CBLAs is widespread throughout the halos of the TNG50 galaxies.\\ 
\\
2) CBLAs provide the H\,{\sc i} column density floor in the CGM of MW-like galaxies.
Even along sightlines in which the CGM H\,{\sc i} column density is dominated by 
cooler gas components at $T\leq 10^5\, \text{K}$, the CBLA component provides a significant
contribution to the overall H\,{\sc i} optical depth. In $\sim 50 \%$ of the 
sightlines in our random sample of 75 sightlines the CBLA absorption from warm-hot
gas actually dominates the column density of the overall CGM H\,{\sc i} absorption.
Therefore, our study indicates that the H\,{\sc i} absorption from warm-hot gas needs 
to be taken into account when it comes to the correct modelling of CGM absorption
systems based on real data.\\
\\
3) CBLAs span a wide range of physical properties and trace a substantial baryon reservoir
in the CGM. Our study shows that the strength of CBLAs (H\,{\sc i} column density and
equivalent width) declines with increasing impact parameter for a given halo mass by
a factor of a few. While these trends are similar for all halo masses sampled in our
study, the average H\,{\sc i} column densities in higher-mass halos are somewhat lower 
compared to lower-mass halos, particularly in the inner halo regions. We argue that
this mild mass dependence reflects the (on average) higher virial temperatures of
more massive halos and a resulting lower (average) neutral gas fraction. In our
simulation, CBLAs trace warm-hot gas in a temperature range $T=10^{5.2-6.4}$ K,
which accounts for $\sim 7 \%$ (median value) of the overall baryon budget in
the TNG50 galaxies, which is roughly a quarter of the total baryon budget contained
in the CGM.\\
\\
4) There exists a previously observationally unidentified population of strong CBLAs, which may have been present in observational data but not recognized as arising from warm-hot gas, and instead attributed to cooler gas. These strong CBLAs, which exhibit substantial H\,{\sc i} column densities up to $\log N_{\text{HI}}=14.9$, represent a new absorber class that traces massive, 
extended circumgalactic gas structures composed of warm-hot gas. There is a 
mild tendency for particularly strong CBLAs (those with the highest column 
densities and equivalent widths) to cluster at smaller impact parameters 
($D\leq100$ kpc). There are, however, also several strong CBLAs found at 
significantly larger impact parameters, even beyond $D=200$ kpc.
We discuss that galactic outflows, accretion streams of warm-hot gas or stripped material from satellites can contribute to the population of strong CBLAs. However, for the latter, a proper analysis would require tracking the origin of the gas over time, which is beyond the scope of this work.\\
\\
In conclusion, our study demonstrates that CBLAs represent an important
absorber class that needs to be considered when interpreting the
spectral signatures from the multi-phase CGM of MW-like galaxies at 
low redshift and analysing UV spectral data from instruments such as HST/COS
\citep{sameer_cbla}. Our study further underlines the importance 
of magneto-hydrodynamical simulations i) to cover cooling and fragmentation processes in the warm-hot CGM that
lead to locally enhanced gas densities and larger CBLA H\,{\sc i} column densities
compared to semi-analytic models \citep{Richter_2020_hot_halos} and ii) to separate 
the absorption signatures from the warm-hot phase from those of the cooler
CGM phases, which in real observations is often difficult or even impossible
due to blending effects.

In the future, it will be important to explore how robust the occurrence and properties of CBLAs are across different cosmological simulations. Their frequency and strength may vary with the adopted feedback and CGM-physics prescriptions, or they may instead be governed primarily by large-scale gas accretion and halo assembly. Determining how sensitive CBLAs are to these underlying physical processes could make them a valuable probe of the baryon cycle and a means to distinguish between different galaxy-formation models. We also plan to continue our systematic study of CBLAs and associated high-ion metal lines by combining TNG50 with the Synthetic Absorption Line Spectral Almanac tool \citep[SALSA;][]{nelson2025}, and to evaluate the role of CBLAs in shaping CGM metallicity estimates derived from UV observations.

\begin{acknowledgements}

DN acknowledges funding from the Deutsche Forschungsgemeinschaft (DFG) through an Emmy Noether Research Group (grant number NE 2441/1-1). This work is supported by the Deutsche Forschungsgemeinschaft (DFG, German Research Foundation) under Germany's Excellence Strategy EXC 2181/1 – 390900948 (the Heidelberg STRUCTURES Excellence Cluster).
The authors thank an anonymous referee for helpful comments and suggestions.

\end{acknowledgements}

\bibliographystyle{aa}
\bibliography{bibliography}

@ARTICLE{Nelson2021,
       author = {{Nelson}, Dylan and {Springel}, Volker and {Pillepich}, Annalisa and {Rodriguez-Gomez}, Vicente and {Torrey}, Paul and {Genel}, Shy and {Vogelsberger}, Mark and {Pakmor}, Ruediger and {Marinacci}, Federico and {Weinberger}, Rainer and {Kelley}, Luke and {Lovell}, Mark and {Diemer}, Benedikt and {Hernquist}, Lars},
        title = "{The IllustrisTNG Simulations: Public Data Release}",
      journal = {J. Open Source Software},
         year = 2021,
       volume = {7},
          eid = {30},
        pages = {30},
          doi = {10.21105/joss.03030}
}

@inproceedings{10.1007/978-3-030-66792-4_1,
	address = "Cham",
	author = "Pillepich, Annalisa and Nelson, Dylan and Springel, Volker and Pakmor, Rüdiger and Hernquist, Lars and Vogelsberger, Mark and Weinberger, Rainer and Genel, Shy and Marinacci, Federico and Torrey, Paul and Naiman, Jill",
	booktitle = "High Performance Computing in Science and Engineering '19",
	editor = "Nagel, Wolfgang E. and Kröner, Dietmar H. and Resch, Michael M.",
	isbn = "978-3-030-66792-4",
	pages = "5–22",
	publisher = "Springer International Publishing",
	title = "The TNG50 Simulation: Highly-Resolved Galaxies in a Large Cosmological Volume to the Present Day",
	year = "2021"
}

@article{10.1093/mnras/stu1536,
    author = {Vogelsberger, Mark and Genel, Shy and Springel, Volker and Torrey, Paul and Sijacki, Debora and Xu, Dandan and Snyder, Greg and Nelson, Dylan and Hernquist, Lars},
    title = "{Introducing the Illustris Project: simulating the coevolution of dark and visible matter in the Universe}",
    journal = {\mnras},
    volume = {444},
    number = {2},
    pages = {1518-1547},
    year = {2014},
    month = {08},
    issn = {0035-8711},
    doi = {10.1093/mnras/stu1536},
    url = {https://doi.org/10.1093/mnras/stu1536},
}

@article{10.1093/mnras/stu1654,
    author = {Genel, Shy and Vogelsberger, Mark and Springel, Volker and Sijacki, Debora and Nelson, Dylan and Snyder, Greg and Rodriguez-Gomez, Vicente and Torrey, Paul and Hernquist, Lars},
    title = "{Introducing the Illustris project: the evolution of galaxy populations across cosmic time}",
    journal = {\mnras},
    volume = {445},
    number = {1},
    pages = {175-200},
    year = {2014},
    month = {09},
    issn = {0035-8711},
    doi = {10.1093/mnras/stu1654},
    url = {https://doi.org/10.1093/mnras/stu1654},
}

@ARTICLE{2013MNRAS.436.3031V,
       author = {{Vogelsberger}, Mark and {Genel}, Shy and {Sijacki}, Debora and {Torrey}, Paul and {Springel}, Volker and {Hernquist}, Lars},
        title = "{A model for cosmological simulations of galaxy formation physics}",
      journal = {\mnras},
         year = 2013,
        month = dec,
       volume = {436},
       number = {4},
        pages = {3031-3067},
          doi = {10.1093/mnras/stt1789},
}

@ARTICLE{2015MNRAS.452..575S,
       author = {{Sijacki}, Debora and {Vogelsberger}, Mark and {Genel}, Shy and {Springel}, Volker and {Torrey}, Paul and {Snyder}, Gregory F. and {Nelson}, Dylan and {Hernquist}, Lars},
        title = "{The Illustris simulation: the evolving population of black holes across cosmic time}",
      journal = {\mnras},
         year = 2015,
        month = sep,
       volume = {452},
       number = {1},
        pages = {575-596},
          doi = {10.1093/mnras/stv1340},
}

@article{Springel_2010,
   title={E pur si muove:Galilean-invariant cosmological hydrodynamical simulations on a moving mesh},
   volume={401},
   ISSN={1365-2966},
   url={http://dx.doi.org/10.1111/j.1365-2966.2009.15715.x},
   DOI={10.1111/j.1365-2966.2009.15715.x},
   number={2},
journal = {\mnras},
   publisher={Oxford University Press (OUP)},
   author={Springel, Volker},
   year={2010},
   month=jan, pages={791–851} }

@article{10.1111/j.1365-2966.2011.19591.x,
    author = {Pakmor, Ruediger and Bauer, Andreas and Springel, Volker},
    title = "{Magnetohydrodynamics on an unstructured moving grid}",
    journal = {\mnras},
    volume = {418},
    number = {2},
    pages = {1392-1401},
    year = {2011},
    month = {11},
    issn = {0035-8711},
    doi = {10.1111/j.1365-2966.2011.19591.x},
    url = {https://doi.org/10.1111/j.1365-2966.2011.19591.x},
}

@ARTICLE{Bagla_2002,
       author = {{Bagla}, Jasjeet},
        title = "{TreePM: A code for cosmological N-body simulations}",
      journal = {J. Astrophys. Astron.},
         year = 2002,
       volume = {23},
        pages = {185},
          doi = {10.1007/BF02702282}
}

@article{Bode_2003,
   title={Tree Particle‐Mesh: An Adaptive, Efficient, and Parallel Code for Collisionless Cosmological Simulation},
   volume={145},
   ISSN={1538-4365},
   url={http://dx.doi.org/10.1086/345538},
   DOI={10.1086/345538},
   number={1},
   journal = {\apjs},
   publisher={American Astronomical Society},
   author={Bode, Paul and Ostriker, Jeremiah P.},
   year={2003},
   month=mar, pages={1–13} }

@article{Faucher-Giguere_2009,
	author = {Claude-André Faucher-Giguère and Adam Lidz and Matias Zaldarriaga and Lars Hernquist},
	doi = {10.1088/0004-637X/703/2/1416},
	journal = {\apj},
	month = {sep},
	number = {2},
	pages = {1416},
	publisher = {The American Astronomical Society},
	title = {A NEW CALCULATION OF THE IONIZING BACKGROUND SPECTRUM AND THE EFFECTS OF He ii REIONIZATION},
	url = {https://dx.doi.org/10.1088/0004-637X/703/2/1416},
	volume = {703},
	year = {2009}
}

@article{10.1046/j.1365-8711.2003.06206.x,
    author = {Springel, Volker and Hernquist, Lars},
    title = "{Cosmological smoothed particle hydrodynamics simulations: a hybrid multiphase model for star formation}",
    journal = {\mnras},
    volume = {339},
    number = {2},
    pages = {289-311},
    year = {2003},
    month = {02},
    issn = {0035-8711},
    doi = {10.1046/j.1365-8711.2003.06206.x},
    url = {https://doi.org/10.1046/j.1365-8711.2003.06206.x},
}

@article{Emami_2021,
	author = {Razieh Emami and Lars Hernquist and Charles Alcock and Shy Genel and Sownak Bose and Rainer Weinberger and Mark Vogelsberger and Xuejian Shen and Joshua S. Speagle and Federico Marinacci and John C. Forbes and Paul Torrey},
	doi = {10.3847/1538-4357/ac088b},
	journal = {\apj},
	month = {aug},
	number = {1},
	pages = {7},
	publisher = {The American Astronomical Society},
	title = {Inferring the Morphology of Stellar Distribution in TNG50: Twisted and Twisted-stretched Shapes},
	url = {https://dx.doi.org/10.3847/1538-4357/ac088b},
	volume = {918},
	year = {2021}
}

@ARTICLE{2018MNRAS.473.4077P,
       author = {{Pillepich}, Annalisa and {Springel}, Volker and {Nelson}, Dylan and {Genel}, Shy and {Naiman}, Jill and {Pakmor}, R{\"u}diger and {Hernquist}, Lars and {Torrey}, Paul and {Vogelsberger}, Mark and {Weinberger}, Rainer and {Marinacci}, Federico},
        title = "{Simulating galaxy formation with the IllustrisTNG model}",
      journal = {\mnras},
         year = 2018,
        month = jan,
       volume = {473},
       number = {3},
        pages = {4077-4106},
          doi = {10.1093/mnras/stx2656},
}

@article{Nelson2019,
  author = {Nelson, Dylan and Pillepich, Annalisa and Springel, Volker and Pakmor, Rüdiger and Weinberger, Rainer and Genel, Shy and Torrey, Paul and Vogelsberger, Mark and Marinacci, Federico and Hernquist, Lars},
  title = {First results from the TNG50 simulation: galactic outflows driven by supernovae and black hole feedback},
  journal = {\mnras},
  volume = {490},
  number = {3},
  pages = {3234--3261},
  year = {2019},
  month = aug,
  doi = {10.1093/mnras/stz2306}
}

@ARTICLE{2019MNRAS.490.3196P,
       author = {{Pillepich}, Annalisa and {Nelson}, Dylan and {Springel}, Volker and {Pakmor}, R{\"u}diger and {Torrey}, Paul and {Weinberger}, Rainer and {Vogelsberger}, Mark and {Marinacci}, Federico and {Genel}, Shy and {van der Wel}, Arjen and {Hernquist}, Lars},
        title = "{First results from the TNG50 simulation: the evolution of stellar and gaseous discs across cosmic time}",
      journal = {\mnras},
         year = 2019,
        month = dec,
       volume = {490},
       number = {3},
        pages = {3196-3233},
          doi = {10.1093/mnras/stz2338},
}

@article{10.1046/j.1365-8711.2001.04912.x,
    author = {Springel, Volker and White, Simon D. M. and Tormen, Giuseppe and Kauffmann, Guinevere},
    title = "{Populating a cluster of galaxies – I. Results at z = 0}",
    journal = {\mnras},
    volume = {328},
    number = {3},
    pages = {726-750},
    year = {2001},
    month = {12},
    issn = {0035-8711},
    doi = {10.1046/j.1365-8711.2001.04912.x},
    url = {https://doi.org/10.1046/j.1365-8711.2001.04912.x},
}

@ARTICLE{2014ApJ...783L..20P,
       author = {{Pakmor}, R{\"u}diger and {Marinacci}, Federico and {Springel}, Volker},
        title = "{Magnetic Fields in Cosmological Simulations of Disk Galaxies}",
      journal = {\apjl},
         year = 2014,
        month = mar,
       volume = {783},
       number = {1},
          eid = {L20},
        pages = {L20},
          doi = {10.1088/2041-8205/783/1/L20},
}

@article{refId0,
	author = {{Planck Collaboration} and {Ade, P. A. R.} and {Aghanim, N.} and {Arnaud, M.} and {Ashdown, M.} and {Aumont, J.} and {Baccigalupi, C.} and {Banday, A. J.} and {Barreiro, R. B.} and {Bartlett, J. G.} and {Bartolo, N.} and {Battaner, E.} and {Battye, R.} and {Benabed, K.} and {Benoît, A.} and {Benoit-Lévy, A.} and {Bernard, J.-P.} and {Bersanelli, M.} and {Bielewicz, P.} and {Bock, J. J.} and {Bonaldi, A.} and {Bonavera, L.} and {Bond, J. R.} and {Borrill, J.} and {Bouchet, F. R.} and {Boulanger, F.} and {Bucher, M.} and {Burigana, C.} and {Butler, R. C.} and {Calabrese, E.} and {Cardoso, J.-F.} and {Catalano, A.} and {Challinor, A.} and {Chamballu, A.} and {Chary, R.-R.} and {Chiang, H. C.} and {Chluba, J.} and {Christensen, P. R.} and {Church, S.} and {Clements, D. L.} and {Colombi, S.} and {Colombo, L. P. L.} and {Combet, C.} and {Coulais, A.} and {Crill, B. P.} and {Curto, A.} and {Cuttaia, F.} and {Danese, L.} and {Davies, R. D.} and {Davis, R. J.} and {de Bernardis, P.} and {de Rosa, A.} and {de Zotti, G.} and {Delabrouille, J.} and {Désert, F.-X.} and {Di Valentino, E.} and {Dickinson, C.} and {Diego, J. M.} and {Dolag, K.} and {Dole, H.} and {Donzelli, S.} and {Doré, O.} and {Douspis, M.} and {Ducout, A.} and {Dunkley, J.} and {Dupac, X.} and {Efstathiou, G.} and {Elsner, F.} and {Enßlin, T. A.} and {Eriksen, H. K.} and {Farhang, M.} and {Fergusson, J.} and {Finelli, F.} and {Forni, O.} and {Frailis, M.} and {Fraisse, A. A.} and {Franceschi, E.} and {Frejsel, A.} and {Galeotta, S.} and {Galli, S.} and {Ganga, K.} and {Gauthier, C.} and {Gerbino, M.} and {Ghosh, T.} and {Giard, M.} and {Giraud-Héraud, Y.} and {Giusarma, E.} and {Gjerløw, E.} and {González-Nuevo, J.} and {Górski, K. M.} and {Gratton, S.} and {Gregorio, A.} and {Gruppuso, A.} and {Gudmundsson, J. E.} and {Hamann, J.} and {Hansen, F. K.} and {Hanson, D.} and {Harrison, D. L.} and {Helou, G.} and {Henrot-Versillé, S.} and {Hernández-Monteagudo, C.} and {Herranz, D.} and {Hildebrandt, S. R.} and {Hivon, E.} and {Hobson, M.} and {Holmes, W. A.} and {Hornstrup, A.} and {Hovest, W.} and {Huang, Z.} and {Huffenberger, K. M.} and {Hurier, G.} and {Jaffe, A. H.} and {Jaffe, T. R.} and {Jones, W. C.} and {Juvela, M.} and {Keihänen, E.} and {Keskitalo, R.} and {Kisner, T. S.} and {Kneissl, R.} and {Knoche, J.} and {Knox, L.} and {Kunz, M.} and {Kurki-Suonio, H.} and {Lagache, G.} and {Lähteenmäki, A.} and {Lamarre, J.-M.} and {Lasenby, A.} and {Lattanzi, M.} and {Lawrence, C. R.} and {Leahy, J. P.} and {Leonardi, R.} and {Lesgourgues, J.} and {Levrier, F.} and {Lewis, A.} and {Liguori, M.} and {Lilje, P. B.} and {Linden-Vørnle, M.} and {López-Caniego, M.} and {Lubin, P. M.} and {Macías-Pérez, J. F.} and {Maggio, G.} and {Maino, D.} and {Mandolesi, N.} and {Mangilli, A.} and {Marchini, A.} and {Maris, M.} and {Martin, P. G.} and {Martinelli, M.} and {Martínez-González, E.} and {Masi, S.} and {Matarrese, S.} and {McGehee, P.} and {Meinhold, P. R.} and {Melchiorri, A.} and {Melin, J.-B.} and {Mendes, L.} and {Mennella, A.} and {Migliaccio, M.} and {Millea, M.} and {Mitra, S.} and {Miville-Deschênes, M.-A.} and {Moneti, A.} and {Montier, L.} and {Morgante, G.} and {Mortlock, D.} and {Moss, A.} and {Munshi, D.} and {Murphy, J. A.} and {Naselsky, P.} and {Nati, F.} and {Natoli, P.} and {Netterfield, C. B.} and {Nørgaard-Nielsen, H. U.} and {Noviello, F.} and {Novikov, D.} and {Novikov, I.} and {Oxborrow, C. A.} and {Paci, F.} and {Pagano, L.} and {Pajot, F.} and {Paladini, R.} and {Paoletti, D.} and {Partridge, B.} and {Pasian, F.} and {Patanchon, G.} and {Pearson, T. J.} and {Perdereau, O.} and {Perotto, L.} and {Perrotta, F.} and {Pettorino, V.} and {Piacentini, F.} and {Piat, M.} and {Pierpaoli, E.} and {Pietrobon, D.} and {Plaszczynski, S.} and {Pointecouteau, E.} and {Polenta, G.} and {Popa, L.} and {Pratt, G. W.} and {Prézeau, G.} and {Prunet, S.} and {Puget, J.-L.} and {Rachen, J. P.} and {Reach, W. T.} and {Rebolo, R.} and {Reinecke, M.} and {Remazeilles, M.} and {Renault, C.} and {Renzi, A.} and {Ristorcelli, I.} and {Rocha, G.} and {Rosset, C.} and {Rossetti, M.} and {Roudier, G.} and {Rouillé d’Orfeuil, B.} and {Rowan-Robinson, M.} and {Rubiño-Martín, J. A.} and {Rusholme, B.} and {Said, N.} and {Salvatelli, V.} and {Salvati, L.} and {Sandri, M.} and {Santos, D.} and {Savelainen, M.} and {Savini, G.} and {Scott, D.} and {Seiffert, M. D.} and {Serra, P.} and {Shellard, E. P. S.} and {Spencer, L. D.} and {Spinelli, M.} and {Stolyarov, V.} and {Stompor, R.} and {Sudiwala, R.} and {Sunyaev, R.} and {Sutton, D.} and {Suur-Uski, A.-S.} and {Sygnet, J.-F.} and {Tauber, J. A.} and {Terenzi, L.} and {Toffolatti, L.} and {Tomasi, M.} and {Tristram, M.} and {Trombetti, T.} and {Tucci, M.} and {Tuovinen, J.} and {Türler, M.} and {Umana, G.} and {Valenziano, L.} and {Valiviita, J.} and {Van Tent, F.} and {Vielva, P.} and {Villa, F.} and {Wade, L. A.} and {Wandelt, B. D.} and {Wehus, I. K.} and {White, M.} and {White, S. D. M.} and {Wilkinson, A.} and {Yvon, D.} and {Zacchei, A.} and {Zonca, A.}},
	doi = {10.1051/0004-6361/201525830},
	journal = {A\&A},
	pages = {A13},
	title = {Planck 2015 results - XIII. Cosmological parameters},
	url = {https://doi.org/10.1051/0004-6361/201525830},
	volume = 594,
	year = 2016
}

@ARTICLE{1985ApJ...292..371D,
       author = {{Davis}, M. and {Efstathiou}, G. and {Frenk}, C.~S. and {White}, S.~D.~M.},
        title = "{The evolution of large-scale structure in a universe dominated by cold dark matter}",
      journal = {\apj},
         year = 1985,
        month = may,
       volume = {292},
        pages = {371-394},
          doi = {10.1086/163168},
}

@ARTICLE{2014MNRAS.437.1750M,
       author = {{Marinacci}, Federico and {Pakmor}, R{\"u}diger and {Springel}, Volker},
        title = "{The formation of disc galaxies in high-resolution moving-mesh cosmological simulations}",
      journal = {\mnras},
         year = 2014,
        month = jan,
       volume = {437},
       number = {2},
        pages = {1750-1775},
          doi = {10.1093/mnras/stt2003},
}

@ARTICLE{2003ApJ...591..499A,
       author = {{Abadi}, Mario G. and {Navarro}, Julio F. and {Steinmetz}, Matthias and {Eke}, Vincent R.},
        title = "{Simulations of Galaxy Formation in a {\ensuremath{\Lambda}} Cold Dark Matter Universe. I. Dynamical and Photometric Properties of a Simulated Disk Galaxy}",
      journal = {\apj},
         year = 2003,
        month = jul,
       volume = {591},
       number = {2},
        pages = {499-514},
          doi = {10.1086/375512},
}

@ARTICLE{Aumer2013,
       author = {{Aumer}, Michael and {White}, Simon D. M. and {Naab}, Thorsten and {Scannapieco}, Cecilia},
        title = "{Towards a more realistic population of bright spiral galaxies in cosmological simulations}",
      journal = {\mnras},
         year = 2013,
       volume = {434},
       number = {4},
        pages = {3142},
          doi = {10.1093/mnras/stt1230}
}

@article{Richter_2006,
   title={Tracing baryons in the warm-hot intergalactic medium  with broad Ly α 
 absorption},
   volume={445},
   ISSN={1432-0746},
   url={http://dx.doi.org/10.1051/0004-6361:20053636},
   DOI={10.1051/0004-6361:20053636},
   number={3},
   journal={\aap},
   publisher={EDP Sciences},
   author={Richter, P. and Savage, B. D. and Sembach, K. R. and Tripp, T. M.},
   year={2006},
   month=jan, pages={827–842} }

@article{Richter_2020_hot_halos,
	author = {Richter, Philipp},
	doi = {10.3847/1538-4357/ab7937},
	journal = {\apj},
	month = {mar},
	number = {1},
	pages = {33},
	publisher = {The American Astronomical Society},
	title = {Hot Gas in Galaxy Halos Traced by Coronal Broad Ly\ensuremath{\alpha} Absorbers},
	url = {https://dx.doi.org/10.3847/1538-4357/ab7937},
	volume = {892},
	year = {2020}
}

@ARTICLE{1991ApJ...379...52W,
       author = {{White}, Simon D.~M. and {Frenk}, Carlos S.},
        title = "{Galaxy Formation through Hierarchical Clustering}",
      journal = {\apj},
         year = 1991,
        month = sep,
       volume = {379},
        pages = {52},
          doi = {10.1086/170483},
}

@article{10.1093/mnras/stz2894,
    author = {Wilkins, Stephen M and Lovell, Christopher C and Stanway, Elizabeth R},
    title = "{Recalibrating the cosmic star formation history}",
    journal = {\mnras},
    volume = {490},
    number = {4},
    pages = {5359-5365},
    year = {2019},
    month = {10},
    issn = {0035-8711},
    doi = {10.1093/mnras/stz2894},
    url = {https://doi.org/10.1093/mnras/stz2894},
}

@ARTICLE{1994ApJ...435...22K,
       author = {{Kennicutt}, Robert C., Jr. and {Tamblyn}, Peter and {Congdon}, Charles E.},
        title = "{Past and Future Star Formation in Disk Galaxies}",
      journal = {\apj},
         year = 1994,
        month = nov,
       volume = {435},
        pages = {22},
          doi = {10.1086/174790},
}

@article{10.1093/mnras/stv318,
    author = {Popping, Gergö and Behroozi, Peter S. and Peeples, Molly S.},
    title = "{Evolution of the atomic and molecular gas content of galaxies in dark matter haloes}",
    journal = {\mnras},
    volume = {449},
    number = {1},
    pages = {477-493},
    year = {2015},
    month = {03},
    issn = {0035-8711},
    doi = {10.1093/mnras/stv318},
    url = {https://doi.org/10.1093/mnras/stv318},
}

@article{Peeples_2014,
	author = {Molly S. Peeples and Jessica K. Werk and Jason Tumlinson and Benjamin D. Oppenheimer and J. Xavier Prochaska and Neal Katz and David H. Weinberg},
	doi = {10.1088/0004-637X/786/1/54},
	journal = {\apj},
	month = {apr},
	number = {1},
	pages = {54},
	publisher = {The American Astronomical Society},
	title = {A BUDGET AND ACCOUNTING OF METALS AT z ∼ 0: RESULTS FROM THE COS-HALOS SURVEY*},
	url = {https://dx.doi.org/10.1088/0004-637X/786/1/54},
	volume = {786},
	year = {2014}
}

@article{Tumlinson_2013,
	author = {Jason Tumlinson and Christopher Thom and Jessica K. Werk and J. Xavier Prochaska and Todd M. Tripp and Neal Katz and Romeel Davé and Benjamin D. Oppenheimer and Joseph D. Meiring and Amanda Brady Ford and John M. O’Meara and Molly S. Peeples and Kenneth R. Sembach and David H. Weinberg},
	doi = {10.1088/0004-637X/777/1/59},
	journal = {\apj},
	month = {oct},
	number = {1},
	pages = {59},
	publisher = {The American Astronomical Society},
	title = {THE COS-HALOS SURVEY: RATIONALE, DESIGN, AND A CENSUS OF CIRCUMGALACTIC NEUTRAL HYDROGEN*},
	url = {https://dx.doi.org/10.1088/0004-637X/777/1/59},
	volume = {777},
	year = {2013}
}

@ARTICLE{1984Natur.311..517B,
       author = {{Blumenthal}, G.~R. and {Faber}, S.~M. and {Primack}, J.~R. and {Rees}, M.~J.},
        title = "{Formation of galaxies and large-scale structure with cold dark matter.}",
      journal = {\nat},
         year = 1984,
        month = oct,
       volume = {311},
        pages = {517-525},
          doi = {10.1038/311517a0},
}

@article{Dunkley_2009,
	author = {J. Dunkley and D. N. Spergel and E. Komatsu and G. Hinshaw and D. Larson and M. R. Nolta and N. Odegard and L. Page and C. L. Bennett and B. Gold and R. S. Hill and N. Jarosik and J. L. Weiland and M. Halpern and A. Kogut and M. Limon and S. S. Meyer and G. S. Tucker and E. Wollack and E. L. Wright},
	doi = {10.1088/0004-637X/701/2/1804},
	journal = {\apj},
	month = {aug},
	number = {2},
	pages = {1804},
	publisher = {The American Astronomical Society},
	title = {FIVE-YEAR WILKINSON MICROWAVE ANISOTROPY PROBE (WMAP*) OBSERVATIONS: BAYESIAN ESTIMATION OF COSMIC MICROWAVE BACKGROUND POLARIZATION MAPS},
	url = {https://dx.doi.org/10.1088/0004-637X/701/2/1804},
	volume = {701},
	year = {2009}
}

@INPROCEEDINGS{2008IAUS..244..136M,
       author = {{McGaugh}, Stacy S.},
        title = "{The Halo by Halo Missing Baryon Problem}",
    booktitle = {Dark Galaxies and Lost Baryons},
         year = 2008,
       editor = {{Davies}, Jonathan I. and {Disney}, Michael J.},
       series = {IAU Symposium},
       volume = {244},
        month = may,
        pages = {136-145},
          doi = {10.1017/S1743921307013920},
}

@article{McGaugh_2010,
	author = {Stacy S. McGaugh and James M. Schombert and W. J. G. de Blok and Matthew J. Zagursky},
	doi = {10.1088/2041-8205/708/1/L14},
	journal = {\apjl},
	month = {dec},
	number = {1},
	pages = {L14},
	publisher = {The American Astronomical Society},
	title = {THE BARYON CONTENT OF COSMIC STRUCTURES},
	url = {https://dx.doi.org/10.1088/2041-8205/708/1/L14},
	volume = {708},
	year = {2009}
}

@ARTICLE{1986A&A...155L...8B,
       author = {{Bergeron}, J.},
        title = "{The MG II absorption system in the QSO PKS 2128-12 : a galaxy disc/halo with a radius of 65 kpc.}",
      journal = {\aap},
         year = 1986,
        month = jan,
       volume = {155},
        pages = {L8-L11},
}

@ARTICLE{1995ApJ...442..538L,
       author = {{Lanzetta}, Kenneth M. and {Bowen}, David V. and {Tytler}, David and {Webb}, John K.},
        title = "{The Gaseous Extent of Galaxies and the Origin of Lyman-Alpha Absorption Systems: A Survey of Galaxies in the Fields of Hubble Space Telescope Spectroscopic Target QSOs}",
      journal = {\apj},
         year = 1995,
        month = apr,
       volume = {442},
        pages = {538},
          doi = {10.1086/175459},
}

@ARTICLE{Caudill_2023,
       author = {{Caudill}, Lucas},
        title = "{The Missing Baryon Problem via Cosmological Zoom-in Simulations}",
      journal = {arXiv e-prints},
         year = 2023,
        month = jul,
          eid = {arXiv:2307.10498},
        pages = {arXiv:2307.10498},
          doi = {10.48550/arXiv.2307.10498},
archivePrefix = {arXiv},
       eprint = {2307.10498},
 primaryClass = {astro-ph.CO},
       adsurl = {https://ui.adsabs.harvard.edu/abs/2023arXiv230710498C}
}

@ARTICLE{Nicastro_2023,
       author = {{Nicastro}, Fabrizio and {Krongold}, Y. and {Fang}, T. and {Fraternali}, F. and {Mathur}, S. and {Bianchi}, S. and {De Rosa}, A. and {Piconcelli}, E. and {Zappacosta}, L. and {Bischetti}, M. and {Feruglio}, C. and {Gupta}, A. and {Zhou}, Z.},
        title = "{X-Ray Detection of the Galaxy's Missing Baryons in the Circumgalactic Medium of L* Galaxies}",
      journal = {\apjl},
         year = 2023,
        month = sep,
       volume = {955},
       number = {1},
          eid = {L21},
        pages = {L21},
          doi = {10.3847/2041-8213/acec70},
archivePrefix = {arXiv},
       eprint = {2302.04247},
 primaryClass = {astro-ph.GA},
       adsurl = {https://ui.adsabs.harvard.edu/abs/2023ApJ...955L..21N}}

@ARTICLE{Tumlinson_2017,
   author = {{Tumlinson}, Jason and {Peeples}, Molly S. and {Werk}, Jessica K.},
   title = "{The Circumgalactic Medium}",
   journal = {ARA\&A},
   year = 2017,
   volume = 55,
   pages = {389--432},
   doi = {10.1146/annurev-astro-091916-055240}}

@article{refId0metalcontent,
	author = {{Méndez-Hernández, H.} and {Cassata, P.} and {Ibar, E.} and {Amorín, R.} and {Aravena, M.} and {Bardelli, S.} and {Cucciati, O.} and {Garilli, B.} and {Giavalisco, M.} and {Guaita, L.} and {Hathi, N.} and {Koekemoer, A.} and {Le Brun, V.} and {Lemaux, B. C.} and {Maccagni, D.} and {Ribeiro, B.} and {Tasca, L.} and {Tejos, N.} and {Thomas, R.} and {Tresse, L.} and {Vergani, D.} and {Zamorani, G.} and {Zucca, E.}},
	doi = {10.1051/0004-6361/202142553},
	journal = {A\&A},
	pages = {A56},
	title = {Metal content of the circumgalactic medium around star-forming galaxies at z ∼ 2.6 as revealed by the VIMOS Ultra-Deep Survey},
	url = {https://doi.org/10.1051/0004-6361/202142553},
	volume = 666,
	year = 2022
}

@article{Carr_2023,
	author = {Christopher Carr and Greg L. Bryan and Drummond B. Fielding and Viraj Pandya and Rachel S. Somerville},
	doi = {10.3847/1538-4357/acc4c7},
	journal = {\apj},
	month = {may},
	number = {1},
	pages = {21},
	publisher = {The American Astronomical Society},
	title = {Regulation of Star Formation by a Hot Circumgalactic Medium},
	url = {https://dx.doi.org/10.3847/1538-4357/acc4c7},
	volume = {949},
	year = {2023}
}

@ARTICLE{Anderson2010,
  author = {{Anderson}, Michael E. and {Bregman}, Joel N.},
  title = {Do Hot Halos Around Galaxies Contain the Missing Baryons?},
  journal = {\apj},
  year = {2010},
  month = may,
  volume = {714},
  number = {1},
  pages = {320--331},
  doi = {10.1088/0004-637X/714/1/320}
}

@ARTICLE{Anderson2013,
  author = {{Anderson}, Michael E. and {Bregman}, Joel N. and {Dai}, Xinyu},
  title = {Extended Hot Halos around Isolated Galaxies Observed in the ROSAT All-Sky Survey},
  journal = {\apj},
  year = {2013},
  month = jan,
  volume = {762},
  number = {2},
  pages = {106},
  doi = {10.1088/0004-637X/762/2/106}
}

@ARTICLE{2012MNRAS.425.1270S,
       author = {{Stinson}, G.~S. and {Brook}, C. and {Prochaska}, J. Xavier and {Hennawi}, Joe and {Shen}, Sijing and {Wadsley}, J. and {Pontzen}, Andrew and {Couchman}, H.~M.~P. and {Quinn}, T. and {Macci{\`o}}, Andrea V. and {Gibson}, Brad K.},
        title = "{MAGICC haloes: confronting simulations with observations of the circumgalactic medium at z=0}",
      journal = {\mnras},
         year = 2012,
        month = sep,
       volume = {425},
       number = {2},
        pages = {1270-1277},
          doi = {10.1111/j.1365-2966.2012.21522.x},
}

@ARTICLE{2013MNRAS.432...89F,
       author = {{Ford}, Amanda Brady and {Oppenheimer}, Benjamin D. and {Dav{\'e}}, Romeel and {Katz}, Neal and {Kollmeier}, Juna A. and {Weinberg}, David H.},
        title = "{Hydrogen and metal line absorption around low-redshift galaxies in cosmological hydrodynamic simulations}",
      journal = {\mnras},
         year = 2013,
        month = jun,
       volume = {432},
       number = {1},
        pages = {89-112},
          doi = {10.1093/mnras/stt393},
}

@ARTICLE{2017MNRAS.465.2966S,
       author = {{Suresh}, Joshua and {Rubin}, Kate H.~R. and {Kannan}, Rahul and {Werk}, Jessica K. and {Hernquist}, Lars and {Vogelsberger}, Mark},
        title = "{On the OVI abundance in the circumgalactic medium of low-redshift galaxies}",
      journal = {\mnras},
         year = 2017,
        month = mar,
       volume = {465},
       number = {3},
        pages = {2966-2982},
          doi = {10.1093/mnras/stw2499},
}

@article{Ho_2020,
	author = {Stephanie H. Ho and Crystal L. Martin and Joop Schaye},
	doi = {10.3847/1538-4357/abbe88},
	journal = {\apj},
	month = {nov},
	number = {1},
	pages = {76},
	publisher = {The American Astronomical Society},
	title = {Morphological and Rotation Structures of Circumgalactic Mg ii Gas in the EAGLE Simulation and the Dependence on Galaxy Properties},
	url = {https://dx.doi.org/10.3847/1538-4357/abbe88},
	volume = {904},
	year = {2020}
}

@ARTICLE{2016MNRAS.460.2157O,
       author = {{Oppenheimer}, Benjamin D. and {Crain}, Robert A. and {Schaye}, Joop and {Rahmati}, Alireza and {Richings}, Alexander J. and {Trayford}, James W. and {Tumlinson}, Jason and {Bower}, Richard G. and {Schaller}, Matthieu and {Theuns}, Tom},
        title = "{Bimodality of low-redshift circumgalactic O VI in non-equilibrium EAGLE zoom simulations}",
      journal = {\mnras},
         year = 2016,
        month = aug,
       volume = {460},
       number = {2},
        pages = {2157-2179},
          doi = {10.1093/mnras/stw1066},
}

@article{DeFelippis_2021,
	author = {Daniel DeFelippis and Nicolas F. Bouché and Shy Genel and Greg L. Bryan and Dylan Nelson and Federico Marinacci and Lars Hernquist},
	doi = {10.3847/1538-4357/ac2cbf},
	journal = {\apj},
	month = {dec},
	number = {1},
	pages = {56},
	publisher = {The American Astronomical Society},
	title = {A Comparison of Circumgalactic Mg ii Absorption between the TNG50 Simulation and the MEGAFLOW Survey},
	url = {https://dx.doi.org/10.3847/1538-4357/ac2cbf},
	volume = {923},
	year = {2021}
}

@article{Zheng_2020,
	author = {Yong Zheng and Molly S. Peeples and Brian W. O’Shea and Raymond C. Simons and Cassandra Lochhaas and Lauren Corlies and Jason Tumlinson and Britton D. Smith and Ramona Augustin},
	doi = {10.3847/1538-4357/ab960a},
	journal = {\apj},
	month = {jun},
	number = {2},
	pages = {143},
	publisher = {The American Astronomical Society},
	title = {Figuring Out Gas \& Galaxies in Enzo (FOGGIE). III. The Mocky Way: Investigating Biases in Observing the Milky Way’s Circumgalactic Medium},
	url = {https://dx.doi.org/10.3847/1538-4357/ab960a},
	volume = {896},
	year = {2020}
}

@article{10.1093/mnras/stab1434,
    author = {Szakacs, Roland and Péroux, Céline and Zwaan, Martin and Hamanowicz, Aleksandra and Klitsch, Anne and Fresco, Alejandra Y and Augustin, Ramona and Biggs, Andrew and Kulkarni, Varsha and Rahmani, Hadi},
    title = "{MUSE-ALMA haloes VI: coupling atomic, ionized, and molecular gas kinematics of galaxies}",
    journal = {\mnras},
    volume = {505},
    number = {4},
    pages = {4746-4761},
    year = {2021},
    month = {06},
    issn = {0035-8711},
    doi = {10.1093/mnras/stab1434},
    url = {https://doi.org/10.1093/mnras/stab1434},
}

@ARTICLE{2022MNRAS.512.3717D,
       author = {{Damle}, Mitali and {Sparre}, Martin and {Richter}, Philipp and {Hani}, Maan H. and {Nuza}, Sebasti{\'a}n E. and {Pfrommer}, Christoph and {Grand}, Robert J.~J. and {Hoffman}, Yehuda and {Libeskind}, Noam and {Sorce}, Jenny G. and {Steinmetz}, Matthias and {Tempel}, Elmo and {Vogelsberger}, Mark and {Wang}, Peng},
        title = "{Cold and hot gas distribution around the Milky-Way - M31 system in the HESTIA simulations}",
      journal = {\mnras},
         year = 2022,
        month = may,
       volume = {512},
       number = {3},
        pages = {3717-3737},
          doi = {10.1093/mnras/stac663},
}

@article{10.1093/mnras/stab1673,
    author = {Augustin, Ramona and Péroux, Céline and Hamanowicz, Aleksandra and Kulkarni, Varsha and Rahmani, Hadi and Zanella, Anita},
    title = "{Clumpiness of observed and simulated cold circumgalactic gas}",
    journal = {\mnras},
    volume = {505},
    number = {4},
    pages = {6195-6205},
    year = {2021},
    month = {06},
    issn = {0035-8711},
    doi = {10.1093/mnras/stab1673},
    url = {https://doi.org/10.1093/mnras/stab1673},
}

@article{Tumlinson_2011,
	author = {J. Tumlinson and J. K. Werk and C. Thom and J. D. Meiring and J. X. Prochaska and T. M. Tripp and J. M. O'Meara and M. Okrochkov and K. R. Sembach},
	doi = {10.1088/0004-637X/733/2/111},
	journal = {\apj},
	month = {may},
	number = {2},
	pages = {111},
	publisher = {The American Astronomical Society},
	title = {MULTIPHASE GAS IN GALAXY HALOS: THE O vi LYMAN-LIMIT SYSTEM TOWARD J1009+0713*},
	url = {https://dx.doi.org/10.1088/0004-637X/733/2/111},
	volume = {733},
	year = {2011}
}

@INPROCEEDINGS{2018AAS...23144002D,
       author = {{Dunn}, Brianne and {Smith}, Beverly J.},
        title = "{Hot, Cold, and Warm Interstellar Gas in Galaxy Mergers}",
    booktitle = {American Astronomical Society Meeting Abstracts \#231},
         year = 2018,
       series = {American Astronomical Society Meeting Abstracts},
       volume = {231},
        month = jan,
          eid = {440.02},
        pages = {440.02},
}

@article{10.1093/mnras/stx3252,
    author = {Hani, Maan H and Sparre, Martin and Ellison, Sara L and Torrey, Paul and Vogelsberger, Mark},
    title = "{Galaxy mergers moulding the circum-galactic medium – I. The impact of a major merger}",
    journal = {\mnras},
    volume = {475},
    number = {1},
    pages = {1160-1176},
    year = {2017},
    month = {12},
    issn = {0035-8711},
    doi = {10.1093/mnras/stx3252},
    url = {https://doi.org/10.1093/mnras/stx3252},
}

@article{refId0coldgas,
	author = {{Decataldo, Davide} and {Shen, Sijing} and {Mayer, Lucio} and {Baumschlager, Bernhard} and {Madau, Piero}},
	doi = {10.1051/0004-6361/202346972},
	journal = {A\&A},
	pages = {A8},
	title = {The origin of cold gas in the circumgalactic medium},
	url = {https://doi.org/10.1051/0004-6361/202346972},
	volume = 685,
	year = 2024
}

@article{10.1093/mnras/stae1248,
    author = {Jana, Ranita and Sarkar, Kartick C and Stern, Jonathan and Sternberg, Amiel},
    title = "{X-ray signatures of galactic outflows into the circumgalactic medium}",
    journal = {\mnras},
    volume = {531},
    number = {2},
    pages = {2757-2774},
    year = {2024},
    month = {05},
    issn = {0035-8711},
    doi = {10.1093/mnras/stae1248},
    url = {https://doi.org/10.1093/mnras/stae1248},
}

@article{Schellenberger_2024,
	author = {Gerrit Schellenberger and Ákos Bogdán and John A. ZuHone and Benjamin D. Oppenheimer and Nhut Truong and Ildar Khabibullin and Fred Jennings and Annalisa Pillepich and Joseph Burchett and Christopher Carr and Priyanka Chakraborty and Robert Crain and William Forman and Christine Jones and Caroline A. Kilbourne and Ralph P. Kraft and Maxim Markevitch and Daisuke Nagai and Dylan Nelson and Anna Ogorzalek and Scott Randall and Arnab Sarkar and Joop Schaye and Sylvain Veilleux and Mark Vogelsberger and Q. Daniel Wang and Irina Zhuravleva},
	doi = {10.3847/1538-4357/ad4548},
	journal = {\apj},
	month = {jul},
	number = {2},
	pages = {85},
	publisher = {The American Astronomical Society},
	title = {Mapping the Imprints of Stellar and Active Galactic Nucleus Feedback in the Circumgalactic Medium with X-Ray Microcalorimeters},
	url = {https://dx.doi.org/10.3847/1538-4357/ad4548},
	volume = {969},
	year = {2024}
}

@ARTICLE{2019MNRAS.485.1595P,
       author = {{P{\'e}roux}, C{\'e}line and {Zwaan}, Martin A. and {Klitsch}, Anne and {Augustin}, Ramona and {Hamanowicz}, Aleksandra and {Rahmani}, Hadi and {Pettini}, Max and {Kulkarni}, Varsha and {Straka}, Lorrie A. and {Biggs}, Andy D. and {York}, Donald G. and {Milliard}, Bruno},
        title = "{Multiphase circumgalactic medium probed with MUSE and ALMA}",
      journal = {\mnras},
         year = 2019,
        month = may,
       volume = {485},
       number = {2},
        pages = {1595-1613},
          doi = {10.1093/mnras/stz202},
}

@article{10.1093/mnras/stz2238,
    author = {Augustin, R and Quiret, S and Milliard, B and Péroux, C and Vibert, D and Blaizot, J and Rasera, Y and Teyssier, R and Frank, S and Deharveng, J-M and Picouet, V and Martin, D C and Hamden, E T and Thatte, N and Pereira Santaella, M and Routledge, L and Zieleniewski, S},
    title = "{Emission from the circumgalactic medium: from cosmological zoom-in simulations to multiwavelength observables}",
    journal = {\mnras},
    volume = {489},
    number = {2},
    pages = {2417-2438},
    year = {2019},
    month = {08},
    issn = {0035-8711},
    doi = {10.1093/mnras/stz2238},
    url = {https://doi.org/10.1093/mnras/stz2238},
}

@article{Corlies_2020,
	author = {Lauren Corlies and Molly S. Peeples and Jason Tumlinson and Brian W. O’Shea and Nicolas Lehner and J. Christopher Howk and John M. O’Meara and Britton D. Smith},
	doi = {10.3847/1538-4357/ab9310},
	journal = {\apj},
	month = {jun},
	number = {2},
	pages = {125},
	publisher = {The American Astronomical Society},
	title = {Figuring Out Gas \& Galaxies in Enzo (FOGGIE). II. Emission from the z = 3 Circumgalactic Medium},
	url = {https://dx.doi.org/10.3847/1538-4357/ab9310},
	volume = {896},
	year = {2020}
}

@ARTICLE{2012MNRAS.420.1731F,
       author = {{Frank}, S. and {Rasera}, Y. and {Vibert}, D. and {Milliard}, B. and {Popping}, A. and {Blaizot}, J. and {Courty}, S. and {Deharveng}, J. -M. and {P{\'e}roux}, C. and {Teyssier}, R. and {Martin}, C.~D.},
        title = "{Observable signatures of the low-z circumgalactic and intergalactic media: ultraviolet line emission in simulations}",
      journal = {\mnras},
         year = 2012,
        month = feb,
       volume = {420},
       number = {2},
        pages = {1731-1753},
          doi = {10.1111/j.1365-2966.2011.20172.x},
}

@ARTICLE{2016ApJ...827..148C,
       author = {{Corlies}, Lauren and {Schiminovich}, David},
        title = "{Empirically Constrained Predictions for Metal-line Emission from the Circumgalactic Medium}",
      journal = {\apj},
         year = 2016,
        month = aug,
       volume = {827},
       number = {2},
          eid = {148},
        pages = {148},
          doi = {10.3847/0004-637X/827/2/148},
}

@ARTICLE{2019astro2020T.403T,
       author = {{Tuttle}, Sarah and {Corlies}, Lauren and {Hamden}, Erika and {Lehner}, Nicolas and {O'Meara}, John M. and {Peeples}, Molly S. and {Rafelski}, Marc and {Tumlinson}, Jason and {Wang}, Q. Daniel and {Werk}, Jessica},
        title = "{Mapping the CGM in Emission}",
      journal = {Astro2020: Decadal Survey on Astronomy and Astrophysics},
         year = 2019,
        month = may,
       volume = {54},
       number = {1},
          eid = {041},
        pages = {041},
          doi = {10.3847/25c2cfeb.a00742b4},
}

@article{Mathur_2023,
   title={Probing the hot circumgalactic medium of external galaxies in X-ray absorption II: a luminous spiral galaxy at z ≈ 0.225},
   volume={525},
   ISSN={1745-3933},
   url={http://dx.doi.org/10.1093/mnrasl/slad085},
   DOI={10.1093/mnrasl/slad085},
   number={1},
journal = {\mnras},
   publisher={Oxford University Press (OUP)},
   author={Mathur, Smita and Das, Sanskriti and Gupta, Anjali and Krongold, Yair},
   year={2023},
   month=jun, pages={L11–L16} }

@article{10.1093/mnras/stab871,
    author = {Anand, Abhijeet and Nelson, Dylan and Kauffmann, Guinevere},
    title = "{Characterizing the abundance, properties, and kinematics of the cool circumgalactic medium of galaxies in absorption with SDSS DR16}",
    journal = {\mnras},
    volume = {504},
    number = {1},
    pages = {65-88},
    year = {2021},
    month = {03},
    issn = {0035-8711},
    doi = {10.1093/mnras/stab871},
    url = {https://doi.org/10.1093/mnras/stab871},
}

@ARTICLE{2014MNRAS.445..794T,
       author = {{Turner}, Monica L. and {Schaye}, Joop and {Steidel}, Charles C. and {Rudie}, Gwen C. and {Strom}, Allison L.},
        title = "{Metal-line absorption around z {\ensuremath{\approx}} 2.4 star-forming galaxies in the Keck Baryonic Structure Survey}",
      journal = {\mnras},
         year = 2014,
        month = nov,
       volume = {445},
       number = {1},
        pages = {794-822},
          doi = {10.1093/mnras/stu1801},
}

@article{annurev:/content/journals/10.1146/annurev-astro-021820-120014,
	author = "Péroux, Céline and Howk, J. Christopher",
	doi = "10.1146/annurev-astro-021820-120014",
	issn = "1545-4282",
    journal = {\araa},
	number = "Volume 58, 2020",
	pages = "363–406",
	publisher = "Annual Reviews",
	title = "The Cosmic Baryon and Metal Cycles",
	type = "Journal Article",
	url = "https://www.annualreviews.org/content/journals/10.1146/annurev-astro-021820-120014",
	volume = "58",
	year = "2020"
}

@ARTICLE{2015ApJ...813...46B,
       author = {{Borthakur}, Sanchayeeta and {Heckman}, Timothy and {Tumlinson}, Jason and {Bordoloi}, Rongmon and {Thom}, Christopher and {Catinella}, Barbara and {Schiminovich}, David and {Dav{\'e}}, Romeel and {Kauffmann}, Guinevere and {Moran}, Sean M. and {Saintonge}, Amelie},
        title = "{Connection between the Circumgalactic Medium and the Interstellar Medium of Galaxies: Results from the COS-GASS Survey}",
      journal = {\apj},
         year = 2015,
        month = nov,
       volume = {813},
       number = {1},
          eid = {46},
        pages = {46},
          doi = {10.1088/0004-637X/813/1/46},
}

@article{Lopez_2008,
	author = {S. Lopez and L. F. Barrientos and P. Lira and N. Padilla and D. G. Gilbank and M. D. Gladders and J. Maza and N. Tejos and M. Vidal and H. K. C. Yee},
	doi = {10.1086/587678},
	journal = {\apj},
	month = {jun},
	number = {2},
	pages = {1144},
	publisher = {},
	title = {Galaxy Clusters in the Line of Sight to Background Quasars. I. Survey Design and Incidence of Mg II Absorbers at Cluster Redshifts*},
	url = {https://dx.doi.org/10.1086/587678},
	volume = {679},
	year = {2008}
}

@ARTICLE{2016MNRAS.455.2662T,
       author = {{Tejos}, Nicolas and {Prochaska}, J. Xavier and {Crighton}, Neil H.~M. and {Morris}, Simon L. and {Werk}, Jessica K. and {Theuns}, Tom and {Padilla}, Nelson and {Bielby}, Rich M. and {Finn}, Charles W.},
        title = "{Towards the statistical detection of the warm-hot intergalactic medium in intercluster filaments of the cosmic web}",
      journal = {\mnras},
         year = 2016,
        month = jan,
       volume = {455},
       number = {3},
        pages = {2662-2697},
          doi = {10.1093/mnras/stv2376},
}

@article{Werk_2016,
   title={THE COS-HALOS SURVEY: ORIGINS OF THE HIGHLY IONIZED CIRCUMGALACTIC MEDIUM OF STAR-FORMING GALAXIES},
   volume={833},
   ISSN={1538-4357},
   url={http://dx.doi.org/10.3847/1538-4357/833/1/54},
   DOI={10.3847/1538-4357/833/1/54},
   number={1},
   journal = {\apj},
   publisher={American Astronomical Society},
   author={Werk, Jessica K. and Prochaska, J. Xavier and Cantalupo, Sebastiano and Fox, Andrew J. and Oppenheimer, Benjamin and Tumlinson, Jason and Tripp, Todd M. and Lehner, Nicolas and McQuinn, Matthew},
   year={2016},
   month=dec, pages={54} }

@article{Prochaska_2017,
   title={The COS-Halos Survey: Metallicities in the Low-redshift Circumgalactic Medium∗},
   volume={837},
   ISSN={1538-4357},
   url={http://dx.doi.org/10.3847/1538-4357/aa6007},
   DOI={10.3847/1538-4357/aa6007},
   number={2},
   journal = {\apj},
   publisher={American Astronomical Society},
   author={Prochaska, J. Xavier and Werk, Jessica K. and Worseck, Gábor and Tripp, Todd M. and Tumlinson, Jason and Burchett, Joseph N. and Fox, Andrew J. and Fumagalli, Michele and Lehner, Nicolas and Peeples, Molly S. and Tejos, Nicolas},
   year={2017},
   month=mar, pages={169} }

@article{Keeney_2017,
doi = {10.3847/1538-4365/aa6b59},
url = {https://dx.doi.org/10.3847/1538-4365/aa6b59},
year = {2017},
month = {may},
publisher = {The American Astronomical Society},
volume = {230},
number = {1},
pages = {6},
author = {Brian A. Keeney and John T. Stocke and Charles W. Danforth and J. Michael Shull and Cameron T. Pratt and Cynthia S. Froning and James C. Green and Steven V. Penton and Blair D. Savage},
title = {Characterizing the Circumgalactic Medium of Nearby Galaxies with HST/COS and HST/STIS Absorption-line Spectroscopy. II. Methods and Models∗},
journal = {\apjs}
}

@article{Bowen_2016,
doi = {10.3847/0004-637X/826/1/50},
url = {https://dx.doi.org/10.3847/0004-637X/826/1/50},
year = {2016},
month = {jul},
publisher = {The American Astronomical Society},
volume = {826},
number = {1},
pages = {50},
author = {David V. Bowen and Doron Chelouche and Edward B. Jenkins and Todd M. Tripp and Max Pettini and Donald G. York and Brenda L. Frye},
title = {THE STRUCTURE OF THE CIRCUMGALACTIC MEDIUM OF GALAXIES: COOL ACCRETION INFLOW AROUND NGC 1097*},
journal = {\apj}
}

@article{Rudie_2013,
doi = {10.1088/0004-637X/769/2/146},
url = {https://dx.doi.org/10.1088/0004-637X/769/2/146},
year = {2013},
month = {may},
publisher = {The American Astronomical Society},
volume = {769},
number = {2},
pages = {146},
author = {Gwen C. Rudie and Charles C. Steidel and Alice E. Shapley and Max Pettini},
title = {THE COLUMN DENSITY DISTRIBUTION AND CONTINUUM OPACITY OF THE INTERGALACTIC AND CIRCUMGALACTIC MEDIUM AT REDSHIFT 〈z〉 = 2.4*},
journal = {\apj}
}

@article{Penton_2004,
doi = {10.1086/382877},
url = {https://dx.doi.org/10.1086/382877},
year = {2004},
month = {may},
publisher = {},
volume = {152},
number = {1},
pages = {29},
author = {Steven V. Penton and John T. Stocke and J. Michael Shull},
title = {The Local Lyα Forest. IV. Space Telescope Imaging Spectrograph G140M Spectra and Results on the Distribution and Baryon Content of H I Absorbers*},
journal = {\apjs}
}

@article{Prochaska_2011,
doi = {10.1088/0004-637X/740/2/91},
url = {https://dx.doi.org/10.1088/0004-637X/740/2/91},
year = {2011},
month = {oct},
publisher = {The American Astronomical Society},
volume = {740},
number = {2},
pages = {91},
author = {J. Xavier Prochaska and B. Weiner and H.-W. Chen and J. Mulchaey and K. Cooksey},
title = {PROBING THE INTERGALACTIC MEDIUM/GALAXY CONNECTION. V. ON THE ORIGIN OF Lyα AND O vi ABSORPTION AT z &lt; 0.2},
journal = {\apj}
}

@article{Miller_2013,
doi = {10.1088/0004-637X/770/2/118},
url = {https://dx.doi.org/10.1088/0004-637X/770/2/118},
year = {2013},
month = {jun},
publisher = {The American Astronomical Society},
volume = {770},
number = {2},
pages = {118},
author = {Matthew J. Miller and Joel N. Bregman},
title = {THE STRUCTURE OF THE MILKY WAY'S HOT GAS HALO},
journal = {\apj}
}

@article{McQuinn_2018,
doi = {10.3847/1538-4357/aa9d3f},
url = {https://dx.doi.org/10.3847/1538-4357/aa9d3f},
year = {2018},
month = {jan},
publisher = {The American Astronomical Society},
volume = {852},
number = {1},
pages = {33},
author = {Matthew McQuinn and Jessica K. Werk},
title = {Implications of the Large O vi Columns around Low-redshift L∗ Galaxies},
journal = {\apj}
}

@article{10.1093/mnras/stz1859,
    author = {Stern, Jonathan and Fielding, Drummond and Faucher-Giguère, Claude-André and Quataert, Eliot},
    title = "{Cooling flow solutions for the circumgalactic medium}",
    journal = {\mnras},
    volume = {488},
    number = {2},
    pages = {2549-2572},
    year = {2019},
    month = {07},
    issn = {0035-8711},
    doi = {10.1093/mnras/stz1859},
    url = {https://doi.org/10.1093/mnras/stz1859},
}

@ARTICLE{2020ApJ...893...82F,
       author = {{Faerman}, Yakov and {Sternberg}, Amiel and {McKee}, Christopher F.},
        title = "{Massive Warm/Hot Galaxy Coronae. II. Isentropic Model}",
      journal = {\apj},
         year = 2020,
        month = apr,
       volume = {893},
       number = {1},
          eid = {82},
        pages = {82},
          doi = {10.3847/1538-4357/ab7ffc},
}

@article{10.1093/mnras/183.3.341,
    author = {White, S. D. M. and Rees, M. J.},
    title = "{Core condensation in heavy halos: a two-stage theory for galaxy formation and clustering}",
    journal = {\mnras},
    volume = {183},
    number = {3},
    pages = {341-358},
    year = {1978},
    month = {07},
    issn = {0035-8711},
    doi = {10.1093/mnras/183.3.341},
    url = {https://doi.org/10.1093/mnras/183.3.341},
}

@ARTICLE{1977ApJ...215..483B,
       author = {{Binney}, J.},
        title = "{The physics of dissipational galaxy formation.}",
      journal = {\apj},
         year = 1977,
        month = jul,
       volume = {215},
        pages = {483-491},
          doi = {10.1086/155378},
}

@ARTICLE{1956ApJ...124...20S,
       author = {{Spitzer}, Lyman, Jr.},
        title = "{On a Possible Interstellar Galactic Corona.}",
      journal = {\apj},
         year = 1956,
        month = jul,
       volume = {124},
        pages = {20},
          doi = {10.1086/146200},
}

@ARTICLE{2017A&A...607A..48R,
       author = {{Richter}, P. and {Nuza}, S.~E. and {Fox}, A.~J. and {Wakker}, B.~P. and {Lehner}, N. and {Ben Bekhti}, N. and {Fechner}, C. and {Wendt}, M. and {Howk}, J.~C. and {Muzahid}, S. and {Ganguly}, R. and {Charlton}, J.~C.},
        title = "{An HST/COS legacy survey of high-velocity ultraviolet absorption in the Milky Way's circumgalactic medium and the Local Group}",
      journal = {\aap},
         year = 2017,
        month = nov,
       volume = {607},
          eid = {A48},
        pages = {A48},
          doi = {10.1051/0004-6361/201630081},
}

@article{Khabibullin_2018,
   title={X-ray emission from warm-hot intergalactic medium: the role of resonantly scattered cosmic X-ray background},
   volume={482},
   ISSN={1365-2966},
   url={http://dx.doi.org/10.1093/mnras/sty2992},
   DOI={10.1093/mnras/sty2992},
   number={4},
journal = {\mnras},
   publisher={Oxford University Press (OUP)},
   author={Khabibullin, I and Churazov, E},
   year={2018},
   month=nov, pages={4972–4984} }

@ARTICLE{2006A&A...451..767R,
       author = {{Richter}, P. and {Fang}, T. and {Bryan}, G.~L.},
        title = "{Simulations of thermally broadened H I Ly {\ensuremath{\alpha}} absorption arising in the warm-hot intergalactic medium}",
      journal = {\aap},
         year = 2006,
        month = jun,
       volume = {451},
       number = {3},
        pages = {767-776},
          doi = {10.1051/0004-6361:20054556},
}

@article{ refId0whim,
	author = {{Prause, N.} and {Reimers, D.} and {Fechner, C.} and {Janknecht, E.}},
	title = {The baryon density at z = 0.9–1.9 - Tracing the warm-hot intergalactic medium with broad Lyman α absorption},
	DOI= "10.1051/0004-6361:20077283",
	url= "https://doi.org/10.1051/0004-6361:20077283",
	journal = {\aap},
	year = 2007,
	volume = 470,
	number = 1,
	pages = "67-72",
}

@ARTICLE{2010ApJ...712.1443N,
       author = {{Narayanan}, Anand and {Savage}, Blair D. and {Wakker}, Bart P.},
        title = "{Highly Ionized Plasma in the Halo of a Luminous Spiral Galaxy Near z = 0.225}",
      journal = {\apj},
         year = 2010,
        month = apr,
       volume = {712},
       number = {2},
        pages = {1443-1460},
          doi = {10.1088/0004-637X/712/2/1443},
}

@article{Danforth_2016,
doi = {10.3847/0004-637X/817/2/111},
url = {https://dx.doi.org/10.3847/0004-637X/817/2/111},
year = {2016},
month = {jan},
publisher = {The American Astronomical Society},
volume = {817},
number = {2},
pages = {111},
author = {Charles W. Danforth and Brian A. Keeney and Evan M. Tilton and J. Michael Shull and John T. Stocke and Matthew Stevans and Matthew M. Pieri and Blair D. Savage and Kevin France and David Syphers and Britton D. Smith and James C. Green and Cynthia Froning and Steven V. Penton and Steven N. Osterman},
title = {AN HST/COS SURVEY OF THE LOW-REDSHIFT INTERGALACTIC MEDIUM. I. SURVEY, METHODOLOGY, AND OVERALL RESULTS*},
journal = {\apj}
}

@article{Stocke_2014,
doi = {10.1088/0004-637X/791/2/128},
url = {https://dx.doi.org/10.1088/0004-637X/791/2/128},
year = {2014},
month = {aug},
publisher = {The American Astronomical Society},
volume = {791},
number = {2},
pages = {128},
author = {John T. Stocke and Brian A. Keeney and Charles W. Danforth and David Syphers and H. Yamamoto and J. Michael Shull and James C. Green and Cynthia Froning and Blair D. Savage and Bart Wakker and Tae-Sun Kim and Emma V. Ryan-Weber and Glenn G. Kacprzak},
title = {ABSORPTION-LINE DETECTIONS OF 105–106 K GAS IN SPIRAL-RICH GROUPS OF GALAXIES*},
journal = {\apj}
}

@article{Savage_2014,
doi = {10.1088/0067-0049/212/1/8},
url = {https://dx.doi.org/10.1088/0067-0049/212/1/8},
year = {2014},
month = {apr},
publisher = {The American Astronomical Society},
volume = {212},
number = {1},
pages = {8},
author = {B. D. Savage and T.-S. Kim and B. P. Wakker and B. Keeney and J. M. Shull and J. T. Stocke and J. C. Green},
title = {THE PROPERTIES OF LOW REDSHIFT INTERGALACTIC O vi ABSORBERS DETERMINED FROM HIGH S/N OBSERVATIONS OF 14 QSOs WITH THE COSMIC ORIGINS SPECTROGRAPH*},
journal = {\apjs}
}

@ARTICLE{2017ApJ...850L..10J,
       author = {{Johnson}, Sean D. and {Chen}, Hsiao-Wen and {Mulchaey}, John S. and {Schaye}, Joop and {Straka}, Lorrie A.},
        title = "{The Extent of Chemically Enriched Gas around Star-forming Dwarf Galaxies}",
      journal = {\apjl},
         year = 2017,
        month = nov,
       volume = {850},
       number = {1},
          eid = {L10},
        pages = {L10},
          doi = {10.3847/2041-8213/aa9370},
}

@ARTICLE{Richter_2008,
       author = {{Richter}, P. and {Paerels}, F. B. S. and {Kaastra}, J. S.},
        title = "{FUV and X-Ray Absorption in the Warm-Hot Intergalactic Medium}",
      journal = {Space Sci. Rev.},
         year = 2008,
       volume = {134},
        pages = {25},
          doi = {10.1007/s11214-008-9325-4}
}

@article{ refId0primera,
	author = {{Tüllmann, R.} and {Pietsch, W.} and {Rossa, J.} and {Breitschwerdt, D.} and {Dettmar, R.-J.}},
	title = {The multi-phase gaseous halos of star forming late-type galaxies - I. XMM-Newton observations of the hot ionized medium},
	DOI= "10.1051/0004-6361:20052936",
	url= "https://doi.org/10.1051/0004-6361:20052936",
	journal = {\aap},
	year = 2006,
	volume = 448,
	number = 1,
	pages = "43-75",
}

@ARTICLE{2023MNRAS.519.3154H,
       author = {{Hopkins}, Philip F. and {Wetzel}, Andrew and {Wheeler}, Coral and {Sanderson}, Robyn and {Grudi{\'c}}, Michael Y. and {Sameie}, Omid and {Boylan-Kolchin}, Michael and {Orr}, Matthew and {Ma}, Xiangcheng and {Faucher-Gigu{\`e}re}, Claude-Andr{\'e} and {Kere{\v{s}}}, Du{\v{s}}an and {Quataert}, Eliot and {Su}, Kung-Yi and {Moreno}, Jorge and {Feldmann}, Robert and {Bullock}, James S. and {Loebman}, Sarah R. and {Angl{\'e}s-Alc{\'a}zar}, Daniel and {Stern}, Jonathan and {Necib}, Lina and {Choban}, Caleb R. and {Hayward}, Christopher C.},
        title = "{FIRE-3: updated stellar evolution models, yields, and microphysics and fitting functions for applications in galaxy simulations}",
      journal = {\mnras},
         year = 2023,
        month = feb,
       volume = {519},
       number = {2},
        pages = {3154-3181},
          doi = {10.1093/mnras/stac3489},
}

@ARTICLE{2022MNRAS.509.2720S,
       author = {{Sparre}, Martin and {Whittingham}, Joseph and {Damle}, Mitali and {Hani}, Maan H. and {Richter}, Philipp and {Ellison}, Sara L. and {Pfrommer}, Christoph and {Vogelsberger}, Mark},
        title = "{Gas flows in galaxy mergers: supersonic turbulence in bridges, accretion from the circumgalactic medium, and metallicity dilution}",
      journal = {\mnras},
         year = 2022,
        month = jan,
       volume = {509},
       number = {2},
        pages = {2720-2735},
          doi = {10.1093/mnras/stab3171},
}

@ARTICLE{Marinacci_2018,
       author = {{Marinacci}, Federico and {Vogelsberger}, Mark and {Pakmor}, R{\"u}diger and {Torrey}, Paul and {Springel}, Volker and {Hernquist}, Lars and {Nelson}, Dylan and {Weinberger}, Rainer and {Pillepich}, Annalisa and {Naiman}, Jill and {Genel}, Shy},
        title = "{First results from the IllustrisTNG simulations: radio haloes and magnetic fields}",
      journal = {\mnras},
         year = 2018,
       volume = {480},
       number = {4},
        pages = {5113},
          doi = {10.1093/mnras/sty2206}
}

@article{Nelson2017,
  author = {Nelson, Dylan and Pillepich, Annalisa and Springel, Volker and Weinberger, Rainer and Hernquist, Lars and Pakmor, Rüdiger and Genel, Shy and Torrey, Paul and Vogelsberger, Mark and Kauffmann, Guinevere and Marinacci, Federico and Naiman, Jill},
  title = {First results from the IllustrisTNG simulations: the galaxy colour bimodality},
  journal = {\mnras},
  volume = {475},
  number = {1},
  pages = {624--647},
  year = {2017},
  month = nov,
  doi = {10.1093/mnras/stx3040}
}

@article{Pillepich_2017,
   title={First results from the IllustrisTNG simulations: the stellar mass content of groups and clusters of galaxies},
   volume={475},
   ISSN={1365-2966},
   url={http://dx.doi.org/10.1093/mnras/stx3112},
   DOI={10.1093/mnras/stx3112},
   number={1},
journal = {\mnras},
   publisher={Oxford University Press (OUP)},
   author={Pillepich, Annalisa and Nelson, Dylan and Hernquist, Lars and Springel, Volker and Pakmor, Rüdiger and Torrey, Paul and Weinberger, Rainer and Genel, Shy and Naiman, Jill P and Marinacci, Federico and Vogelsberger, Mark},
   year={2017},
   month=dec, pages={648–675} }

@article{Springel_2017,
   title={First results from the IllustrisTNG simulations: matter and galaxy clustering},
   volume={475},
   ISSN={1365-2966},
   url={http://dx.doi.org/10.1093/mnras/stx3304},
   DOI={10.1093/mnras/stx3304},
   number={1},
journal = {\mnras},
   publisher={Oxford University Press (OUP)},
   author={Springel, Volker and Pakmor, Rüdiger and Pillepich, Annalisa and Weinberger, Rainer and Nelson, Dylan and Hernquist, Lars and Vogelsberger, Mark and Genel, Shy and Torrey, Paul and Marinacci, Federico and Naiman, Jill},
   year={2017},
   month=dec, pages={676–698} }

@article{Werk_2014,
   title={THE COS-HALOS SURVEY: PHYSICAL CONDITIONS AND BARYONIC MASS IN THE LOW-REDSHIFT CIRCUMGALACTIC MEDIUM},
   volume={792},
   ISSN={1538-4357},
   url={http://dx.doi.org/10.1088/0004-637X/792/1/8},
   DOI={10.1088/0004-637x/792/1/8},
   number={1},
   journal = {\apj},
   publisher={American Astronomical Society},
   author={Werk, Jessica K. and Prochaska, J. Xavier and Tumlinson, Jason and Peeples, Molly S. and Tripp, Todd M. and Fox, Andrew J. and Lehner, Nicolas and Thom, Christopher and O’Meara, John M. and Ford, Amanda Brady and Bordoloi, Rongmon and Katz, Neal and Tejos, Nicolas and Oppenheimer, Benjamin D. and Davé, Romeel and Weinberg, David H.},
   year={2014},
   month=aug, pages={8} }

@ARTICLE{tepper2012,
       author = {{Tepper-Garc{\'\i}a}, Thorsten and {Richter}, Philipp and {Schaye}, Joop and {Booth}, C.~M. and {Dalla Vecchia}, Claudio and {Theuns}, Tom},
        title = "{Absorption signatures of warm-hot gas at low redshift: broad H I Ly{\ensuremath{\alpha}} absorbers}",
      journal = {\mnras},
         year = 2012,
        month = sep,
       volume = {425},
       number = {3},
        pages = {1640-1663},
          doi = {10.1111/j.1365-2966.2012.21545.x},
}

@ARTICLE{benbekhti2012,
       author = {{Ben Bekhti}, N. and {Winkel}, B. and {Richter}, P. and {Kerp}, J. and {Klein}, U. and {Murphy}, M.~T.},
        title = "{An absorption-selected survey of neutral gas in the Milky Way halo. New results based on a large sample of Ca ii, Na i, and H i spectra towards QSOs}",
      journal = {\aap},
         year = 2012,
        month = jun,
       volume = {542},
          eid = {A110},
        pages = {A110},
          doi = {10.1051/0004-6361/201118673},
}

@ARTICLE{richter2004,
       author = {{Richter}, Philipp and {Savage}, Blair D. and {Tripp}, Todd M. and {Sembach}, Kenneth R.},
        title = "{FUSE and STIS Observations of the Warm-hot Intergalactic Medium toward PG 1259+593}",
      journal = {\apjs},
         year = 2004,
        month = jul,
       volume = {153},
       number = {1},
        pages = {165-204},
          doi = {10.1086/421297},
}

@ARTICLE{liang2018,
       author = {{Liang}, Cameron J. and {Kravtsov}, Andrey V. and {Agertz}, Oscar},
        title = "{Observing the circumgalactic medium of simulated galaxies through synthetic absorption spectra}",
      journal = {\mnras},
         year = 2018,
        month = sep,
       volume = {479},
       number = {2},
        pages = {1822-1835},
          doi = {10.1093/mnras/sty1668},
}

@ARTICLE{richter2016,
       author = {{Richter}, P. and {Wakker}, B.~P. and {Fechner}, C. and {Herenz}, P. and {Tepper-Garc{\'\i}a}, T. and {Fox}, A.~J.},
        title = "{An HST/COS legacy survey of intervening Si III absorption in the extended gaseous halos of low-redshift galaxies}",
      journal = {\aap},
         year = 2016,
        month = may,
       volume = {590},
          eid = {A68},
        pages = {A68},
          doi = {10.1051/0004-6361/201527038},
}

@ARTICLE{sembach2003,
       author = {{Sembach}, K.~R. and {Wakker}, B.~P. and {Savage}, B.~D. and {Richter}, P. and {Meade}, M. and {Shull}, J.~M. and {Jenkins}, E.~B. and {Sonneborn}, G. and {Moos}, H.~W.},
        title = "{Highly Ionized High-Velocity Gas in the Vicinity of the Galaxy}",
      journal = {\apjs},
         year = 2003,
        month = may,
       volume = {146},
       number = {1},
        pages = {165-208},
          doi = {10.1086/346231},
}

@ARTICLE{lehner2007,
       author = {{Lehner}, N. and {Savage}, B.~D. and {Richter}, P. and {Sembach}, K.~R. and {Tripp}, T.~M. and {Wakker}, B.~P.},
        title = "{Physical Properties, Baryon Content, and Evolution of the Ly{\ensuremath{\alpha}} Forest: New Insights from High-Resolution Observations at z <\raisebox{-0.5ex}\textasciitilde 0.4}",
      journal = {\apj},
         year = 2007,
        month = apr,
       volume = {658},
       number = {2},
        pages = {680-709},
          doi = {10.1086/511749},
}

@ARTICLE{1993ApJS...88..253S,
       author = {{Sutherland}, Ralph S. and {Dopita}, M.~A.},
        title = "{Cooling Functions for Low-Density Astrophysical Plasmas}",
      journal = {\apjs},
         year = 1993,
        month = sep,
       volume = {88},
        pages = {253},
          doi = {10.1086/191823},
}

@ARTICLE{2024MNRAS.528.3745D,
       author = {{Dutta}, Sayak and {Muzahid}, Sowgat and {Schaye}, Joop and {Mishra}, Sapna and {Chen}, Hsiao-Wen and {Johnson}, Sean and {Wisotzki}, Lutz and {Cantalupo}, Sebastiano},
        title = "{MUSEQuBES: mapping the distribution of neutral hydrogen around low-redshift galaxies}",
      journal = {\mnras},
         year = 2024,
        month = feb,
       volume = {528},
       number = {2},
        pages = {3745-3766},
          doi = {10.1093/mnras/stae206},
archivePrefix = {arXiv},
       eprint = {2303.16933},
 primaryClass = {astro-ph.GA},
       adsurl = {https://ui.adsabs.harvard.edu/abs/2024MNRAS.528.3745D}
}

@article{Conroy_2006,
doi = {10.1086/503602},
url = {https://dx.doi.org/10.1086/503602},
year = {2006},
month = {aug},
publisher = {},
volume = {647},
number = {1},
pages = {201},
author = {Conroy, Charlie and Wechsler, Risa H. and Kravtsov, Andrey V.},
title = {Modeling Luminosity-dependent Galaxy Clustering through Cosmic Time},
journal = {\apj}
}

@article{Behroozi_2013,
doi = {10.1088/0004-637X/770/1/57},
url = {https://dx.doi.org/10.1088/0004-637X/770/1/57},
year = {2013},
month = {may},
publisher = {The American Astronomical Society},
volume = {770},
number = {1},
pages = {57},
author = {Behroozi, Peter S. and Wechsler, Risa H. and Conroy, Charlie},
title = {THE AVERAGE STAR FORMATION HISTORIES OF GALAXIES IN DARK MATTER HALOS FROM z = 0–8},
journal = {\apj}
}

@ARTICLE{sameer_cbla,
       author = {{Sameer} and {Charlton}, Jane C. and {Wakker}, Bart P. and {Kacprzak}, Glenn G. and {Nielsen}, Nikole M. and {Churchill}, Christopher W. and {Richter}, Philipp and {Muzahid}, Sowgat and {Ho}, Stephanie H. and {Nateghi}, Hasti and {Rosenwasser}, Benjamin and {Narayanan}, Anand and {Ganguly}, Rajib},
        title = "{Cloud-by-cloud multiphase investigation of the circumgalactic medium of low-redshift galaxies}",
      journal = {\mnras},
         year = 2024,
        month = jun,
       volume = {530},
       number = {4},
        pages = {3827-3854},
          doi = {10.1093/mnras/stae962},
}

@ARTICLE{Laktionov_2025,
       author = {{Laktionov}, Roman and {Sasaki}, Manami and {Wilms}, J{\"o}rn},
        title = "{Deep XMM-Newton observation reveals hot gaseous outflow in NGC 5746}",
      journal = {arXiv e-prints},
         year = 2025,
        month = oct,
          eid = {arXiv:2510.00868},
        pages = {arXiv:2510.00868},
          doi = {10.48550/arXiv.2510.00868},
archivePrefix = {arXiv},
       eprint = {2510.00868},
 primaryClass = {astro-ph.GA},
       adsurl = {https://ui.adsabs.harvard.edu/abs/2025arXiv251000868L}
}

@ARTICLE{Stern_2024,
       author = {{Stern}, Jonathan and {Fielding}, Drummond and {Hafen}, Zachary and {Su}, Kung-Yi and {Naor}, Nadav and {Faucher-Gigu{\`e}re}, Claude-Andr{\'e} and {Quataert}, Eliot and {Bullock}, James},
        title = "{Accretion onto disc galaxies via hot and rotating CGM inflows}",
      journal = {\mnras},
         year = 2024,
        month = may,
       volume = {530},
       number = {2},
        pages = {1711-1731},
          doi = {10.1093/mnras/stae824},
}

@ARTICLE{gnat_2017,
       author = {{Gnat}, Orly},
        title = "{Time-dependent Cooling in Photoionized Plasma}",
      journal = {\apjs},
         year = 2017,
        month = feb,
       volume = {228},
       number = {2},
          eid = {11},
        pages = {11},
          doi = {10.3847/1538-4365/228/2/11},
}

@ARTICLE{cook2023,
       author = {{Cook}, Amanda M. and {Bhardwaj}, Mohit and {Gaensler}, B.~M. and {Scholz}, Paul and {Eadie}, Gwendolyn M. and {Hill}, Alex S. and {Kaspi}, Victoria M. and {Masui}, Kiyoshi W. and {Curtin}, Alice P. and {Dong}, Fengqiu Adam and {Fonseca}, Emmanuel and {Herrera-Martin}, Antonio and {Kaczmarek}, Jane and {Lanman}, Adam E. and {Lazda}, Mattias and {Leung}, Calvin and {Meyers}, Bradley W. and {Michilli}, Daniele and {Pandhi}, Ayush and {Pearlman}, Aaron B. and {Pleunis}, Ziggy and {Ransom}, Scott and {Rahman}, Mubdi and {Sand}, Ketan R. and {Shin}, Kaitlyn and {Smith}, Kendrick and {Stairs}, Ingrid and {Stenning}, David C.},
        title = "{An FRB Sent Me a DM: Constraining the Electron Column of the Milky Way Halo with Fast Radio Burst Dispersion Measures from CHIME/FRB}",
      journal = {\apj},
         year = 2023,
        month = apr,
       volume = {946},
       number = {2},
          eid = {58},
        pages = {58},
          doi = {10.3847/1538-4357/acbbd0}
}

@ARTICLE{miller2015,
       author = {{Miller}, Matthew J. and {Bregman}, Joel N.},
        title = "{Constraining the Milky Way's Hot Gas Halo with O VII and O VIII Emission Lines}",
      journal = {\apj},
         year = 2015,
        month = feb,
       volume = {800},
       number = {1},
          eid = {14},
        pages = {14},
          doi = {10.1088/0004-637X/800/1/14},
}

@ARTICLE{anderson2016,
       author = {{Anderson}, Michael E. and {Churazov}, Eugene and {Bregman}, Joel N.},
        title = "{A deep XMM-Newton study of the hot gaseous halo around NGC 1961}",
      journal = {\mnras},
         year = 2016,
        month = jan,
       volume = {455},
       number = {1},
        pages = {227-243},
          doi = {10.1093/mnras/stv2314},
}

@ARTICLE{wakker2012,
       author = {{Wakker}, Bart P. and {Savage}, Blair D. and {Fox}, Andrew J. and {Benjamin}, Robert A. and {Shapiro}, Paul R.},
        title = "{Characterizing Transition Temperature Gas in the Galactic Corona}",
      journal = {\apj},
         year = 2012,
        month = apr,
       volume = {749},
       number = {2},
          eid = {157},
        pages = {157},
          doi = {10.1088/0004-637X/749/2/157},
}

@ARTICLE{bisht2025,
       author = {{Singh Bisht}, Mukesh and {Das}, Sanskriti and {Mathur}, Smita and {Roy}, Manami and {Krongold}, Yair and {Gupta}, Anjali},
        title = "{Detection of 'super-virial' gas in the Circumgalactic medium of the Milky Way towards PKS 2155-304}",
      journal = {arXiv e-prints},
         year = 2025,
        month = sep,
          eid = {arXiv:2509.02019},
        pages = {arXiv:2509.02019},
          doi = {10.48550/arXiv.2509.02019},
archivePrefix = {arXiv},
       eprint = {2509.02019},
 primaryClass = {astro-ph.GA},
         note = {submitted to ApJ},
       adsurl = {https://ui.adsabs.harvard.edu/abs/2025arXiv250902019S}
}

@ARTICLE{nelson2025,
       author = {{Nelson}, Dylan and {Peroux}, Celine and {Richter}, Philipp and {Pieri}, Matthew M. and {Lopez}, Sebastian and {Bordoloi}, Rongmon and {Zou}, Siwei and {Burchett}, Joseph N. and {Davies}, Rebecca L. and {Ramesh}, Rahul and {Smith}, Matthew C. and {Borthakur}, Sanchayeeta and {Churchill}, Christopher W.},
        title = "{The Synthetic Absorption Line Spectral Almanac (SALSA)}",
      journal = {arXiv e-prints},
         year = 2025,
        month = oct,
          eid = {arXiv:2510.19904},
        pages = {arXiv:2510.19904},
          doi = {10.48550/arXiv.2510.19904},
archivePrefix = {arXiv},
       eprint = {2510.19904},
 primaryClass = {astro-ph.GA},
       adsurl = {https://ui.adsabs.harvard.edu/abs/2025arXiv251019904N}
}

@article{Naiman_2018,
   title={First results from the IllustrisTNG simulations: a tale of two elements – chemical evolution of magnesium and europium},
   volume={477},
   ISSN={1365-2966},
   url={http://dx.doi.org/10.1093/mnras/sty618},
   DOI={10.1093/mnras/sty618},
   number={1},
journal = {\mnras},
   publisher={Oxford University Press (OUP)},
   author={Naiman, Jill P and Pillepich, Annalisa and Springel, Volker and Ramirez-Ruiz, Enrico and Torrey, Paul and Vogelsberger, Mark and Pakmor, Rüdiger and Nelson, Dylan and Marinacci, Federico and Hernquist, Lars and Weinberger, Rainer and Genel, Shy},
   year={2018},
   month=mar, pages={1206–1224} }

@article{Weinberger_2016,
   title={Simulating galaxy formation with black hole driven thermal and kinetic feedback},
   volume={465},
   ISSN={1365-2966},
   url={http://dx.doi.org/10.1093/mnras/stw2944},
   DOI={10.1093/mnras/stw2944},
   number={3},
journal = {\mnras},
   publisher={Oxford University Press (OUP)},
   author={Weinberger, Rainer and Springel, Volker and Hernquist, Lars and Pillepich, Annalisa and Marinacci, Federico and Pakmor, Rüdiger and Nelson, Dylan and Genel, Shy and Vogelsberger, Mark and Naiman, Jill and Torrey, Paul},
   year={2016},
   month=nov, pages={3291–3308} }

@ARTICLE{1996ApJS..105...19K,
       author = {{Katz}, Neal and {Weinberg}, David H. and {Hernquist}, Lars},
        title = "{Cosmological Simulations with TreeSPH}",
      journal = {\apjs},
         year = 1996,
        month = jul,
       volume = {105},
        pages = {19},
          doi = {10.1086/192305},
}

@article{P_roux_2020,
   title={Predictions for the angular dependence of gas mass flow rate and metallicity in the circumgalactic medium},
   volume={499},
   ISSN={1365-2966},
   url={http://dx.doi.org/10.1093/mnras/staa2888},
   DOI={10.1093/mnras/staa2888},
   number={2},
journal = {\mnras},
   publisher={Oxford University Press (OUP)},
   author={Péroux, Céline and Nelson, Dylan and van de Voort, Freeke and Pillepich, Annalisa and Marinacci, Federico and Vogelsberger, Mark and Hernquist, Lars},
   year={2020},
   month=sep, pages={2462–2473} }

@ARTICLE{Truong_2021,
       author = {{Truong}, Nhut and {Pillepich}, Annalisa and {Nelson}, Dylan and {Werner}, Norbert and {Hernquist}, Lars},
        title = "{Predictions for anisotropic X-ray signatures in the circumgalactic medium: imprints of supermassive black hole driven outflows}",
      journal = {\mnras},
         year = 2021,
       volume = {508},
        pages = {1563},
          doi = {10.1093/mnras/stab2638}
}

@article{Pillepich_2021,
   title={X-ray bubbles in the circumgalactic medium of TNG50 Milky Way- and M31-like galaxies: signposts of supermassive black hole activity},
   volume={508},
   ISSN={1365-2966},
   url={http://dx.doi.org/10.1093/mnras/stab2779},
   DOI={10.1093/mnras/stab2779},
   number={4},
journal = {\mnras},
   publisher={Oxford University Press (OUP)},
   author={Pillepich, Annalisa and Nelson, Dylan and Truong, Nhut and Weinberger, Rainer and Martin-Navarro, Ignacio and Springel, Volker and Faber, Sandy M and Hernquist, Lars},
   year={2021},
   month=oct, pages={4667–4695} }

@article{Ramesh_2023,
   title={The circumgalactic medium of Milky Way-like galaxies in the TNG50 simulation – II. Cold, dense gas clouds and high-velocity cloud analogs},
   volume={522},
   ISSN={1365-2966},
   url={http://dx.doi.org/10.1093/mnras/stad951},
   DOI={10.1093/mnras/stad951},
   number={1},
journal = {\mnras},
   publisher={Oxford University Press (OUP)},
   author={Ramesh, Rahul and Nelson, Dylan and Pillepich, Annalisa},
   year={2023},
   month=mar, pages={1535–1555} }

@ARTICLE{sameer2024,
       author = {{Sameer} and {Lehner}, Nicolas and {Howk}, J. Christopher and {Fox}, Andrew J. and {O'Meara}, John M. and {Oppenheimer}, Benjamin D.},
        title = "{The COS CGM Compendium. V. The Dichotomy of O VI Associated with Low- and High-metallicity Cool Gas at z < 1}",
      journal = {\apj},
         year = 2024,
        month = nov,
       volume = {975},
       number = {2},
          eid = {264},
        pages = {264},
          doi = {10.3847/1538-4357/ad7af2},
}

@article{10.1093/mnras/stae2165,
    author = {Pillepich, Annalisa and Sotillo-Ramos, Diego and Ramesh, Rahul and Nelson, Dylan and Engler, Christoph and Rodriguez-Gomez, Vicente and Fournier, Martin and Donnari, Martina and Springel, Volker and Hernquist, Lars},
    title = {Milky Way and Andromeda analogues from the TNG50 simulation},
    journal = {\mnras},
    volume = {535},
    number = {2},
    pages = {1721-1762},
    year = {2024},
    month = {09},
    issn = {0035-8711},
    doi = {10.1093/mnras/stae2165},
    url = {https://doi.org/10.1093/mnras/stae2165},
}

@article{10.1093/mnras/stad3142,
    author = {Roy, Manami and Su, Kung-Yi and Tonnesen, Stephanie and Fielding, Drummond B and Faucher-Giguère, Claude-André},
    title = {Seeding the CGM: how satellites populate the cold phase of milky way haloes},
    journal = {\mnras},
    volume = {527},
    number = {1},
    pages = {265-280},
    year = {2024},
    month = {01},
    issn = {0035-8711},
    doi = {10.1093/mnras/stad3142},
    url = {https://doi.org/10.1093/mnras/stad3142},
}

@ARTICLE{2024MNRAS.527.3494W,
       author = {{Weng}, Simon and {P{\'e}roux}, C{\'e}line and {Ramesh}, Rahul and {Nelson}, Dylan and {Sadler}, Elaine M. and {Zwaan}, Martin and {Bollo}, Victoria and {Casavecchia}, Benedetta},
        title = "{The physical origins of gas in the circumgalactic medium using observationally motivated TNG50 mocks}",
      journal = {\mnras},
         year = 2024,
        month = jan,
       volume = {527},
       number = {2},
        pages = {3494-3516},
          doi = {10.1093/mnras/stad3426},
}

@article{Damle_2025,
	author = {Damle, Mitali and Tonnesen, Stephanie and Sparre, Martin and Richter, Philipp},
	doi = {10.3847/1538-4357/adcee9},
	journal = {\apj},
	month = {jun},
	number = {1},
	pages = {69},
	publisher = {The American Astronomical Society},
	title = {Searching for Correlations between Satellite Galaxy Populations and the Cold Circumgalactic Medium around TNG50 Galaxies},
	url = {https://doi.org/10.3847/1538-4357/adcee9},
	volume = {986},
	year = {2025}
}

\appendix

\section{CBLA sample}
\label{appendix:CBLA_sample}

\begin{figure}[H]
\centering
\includegraphics[width=\columnwidth]{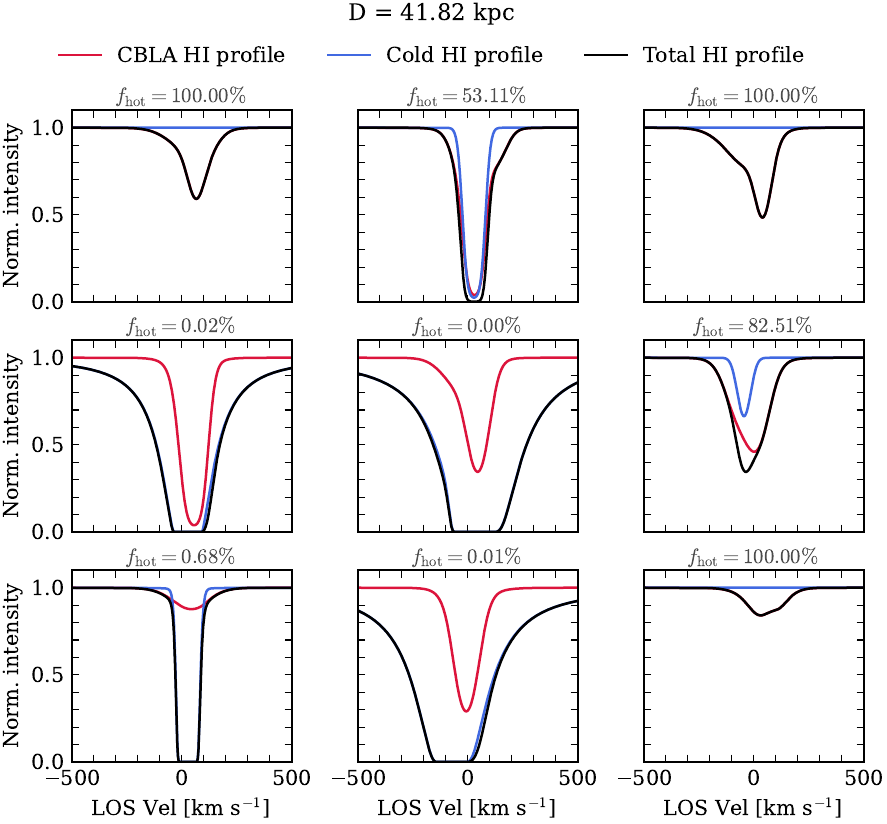}
\caption{Coronal broad Lyman-$\alpha$ absorbers (CBLAs) sample for our ring-pattern distribution at an impact parameter of $D = 0.2R_{\mathrm{200}}$.}
\label{CBLA_0.2}
\end{figure}

\begin{figure}[H]
\centering
\includegraphics[width=\columnwidth]{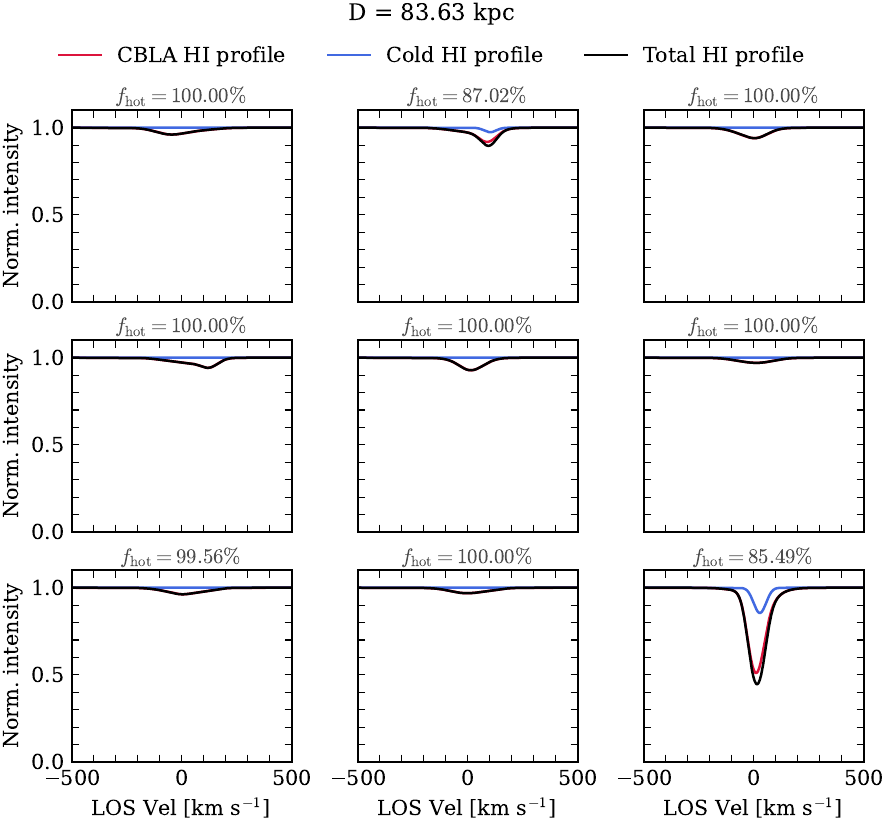}
\caption{Coronal broad Lyman-$\alpha$ absorbers (CBLAs) sample for our ring-pattern distribution at an impact parameter of $D = 0.4R_{\mathrm{200}}$.}
\label{CBLA_0.4}
\end{figure}

\begin{figure}[H]
\centering
\includegraphics[width=\columnwidth]{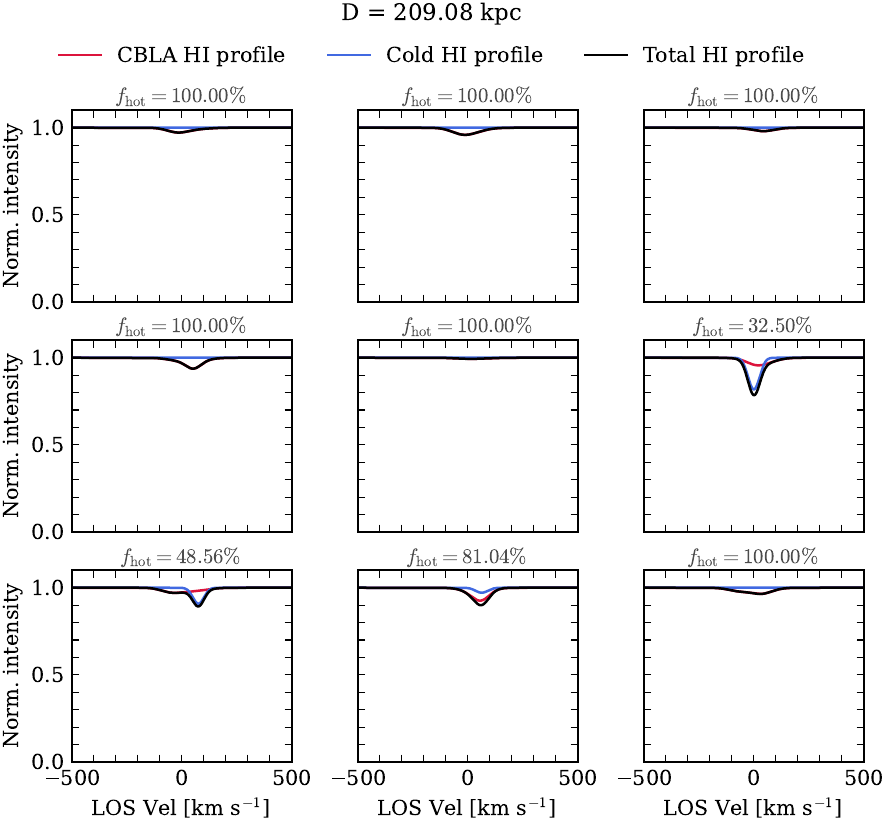}
\caption{Coronal broad Lyman-$\alpha$ absorbers (CBLAs) sample for our ring-pattern distribution at an impact parameter of $D = 1.0R_{\mathrm{200}}$.}
\label{CBLA_1.0}
\end{figure}

In Figures \ref{CBLA_0.2}, \ref{CBLA_0.4}, and \ref{CBLA_1.0} we show some additional CBLA profiles at three representative impact parameters: $R = 0.2R_{\mathrm{200}}$, $0.4R_{\mathrm{200}}$, and $1.0R_{\mathrm{200}}$ of one of our MW-like galaxy halos. This halo in particular has a $R_{\mathrm{200}} = 209.08$ kpc and a halo mass of $M_{\mathrm{200}} = 9.75 \times 10^{11} \mathrm{M}_{\odot}$. Each panel shows the \ion{H}{I} CBLA (red), the cold \ion{H}{I} (blue), and the total \ion{H}{I} profile (black) respectively, together with the hot fraction $f_{hot}$ (defined in Sect.\, \ref{CBLA_random}) along each sightline.

\section{Radial trends of the random CBLA sample}
\label{appendix:random_sample_radial}
In this appendix, we show more directly how the properties of our random sample of CBLAs, namely their temperatures and column densities, vary with impact parameter. This representation helps to better interpret the bottom right panel of Fig.~\ref{cbla_random_distr}, discussed in Sect.~\ref{CBLA_random}.
\begin{figure}
\centering
   \includegraphics[width= 0.40\textwidth]{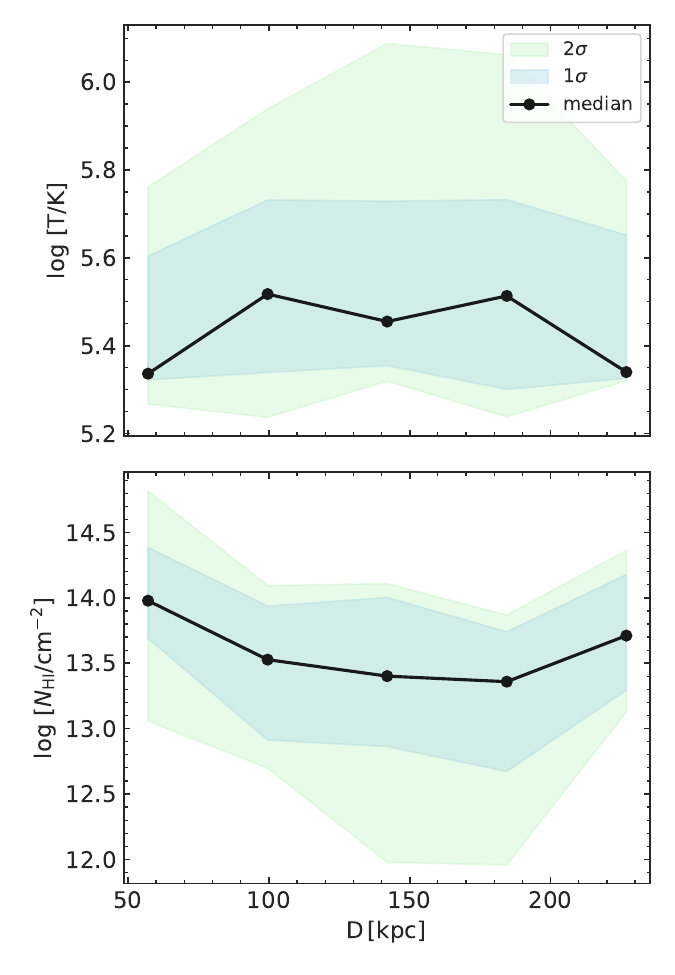}
    \caption{Temperatures (top) and \ion{H}{I} column densities (bottom) of our CBLA random sample as a function of impact parameter. The solid line shows the median at each radial bin, while the shaded regions indicate the $1\sigma$ and $2\sigma$ scatter, respectively.}\label{radial_trend_random_sample}
\end{figure}

Figure~\ref{radial_trend_random_sample} shows the median trends of these quantities as a function of impact parameter, together with their scatter. There is a slight tendency for hotter systems to appear at intermediate impact parameters ($\sim 100-170\, \text{kpc}$) and for the highest column density absorbers to be found at smaller impact parameters ($\sim 50-100\, \text{kpc}$). However, the uncertainties are large and the scatter is significant, making these trends insignificant overall. 
We also see that most CBLAs trace gas in the range $\log T \sim 5.3 - 5.6$, i.e. the warm-hot phase. At larger impact parameters, however, they also probe hotter, more virialized gas, reaching temperatures up to $\log T \sim 6.0$. 
This supports the picture discussed in Sect.~\ref{CBLA_random}, where the presence and properties of CBLAs are not primarily set by distance from the galaxy centre, but instead reflect local variations in the physical conditions and structure of the warm-hot CGM.

\section{LOS properties of a strong CBLA}
\label{appendix:strong_CBLA_along_LOS}

As shown in Fig. \ref{cbla_random_distr}, most of the simulated CBLAs in our sample show properties broadly consistent with theoretical expectations: weak, shallow Ly$\alpha$ absorption associated with hot, highly ionised gas (see also Appendix \ref{appendix:CBLA_sample}). However, our sample also reveals a small number of outlier absorbers that depart significantly from this typical CBLA profile as discussed in Sec. \ref{CBLA_deep}. In these cases, the Ly$\alpha$ absorption is not only broad but also unusually deep, sometimes approaching saturation levels (see Fig. \ref{strong_CBLA_slices}). Such systems stand out from the rest of the sample and are difficult to reconcile with what is typically expected from purely hot coronal gas. 

\begin{figure}
\centering
   \includegraphics[width= 0.55\textwidth]{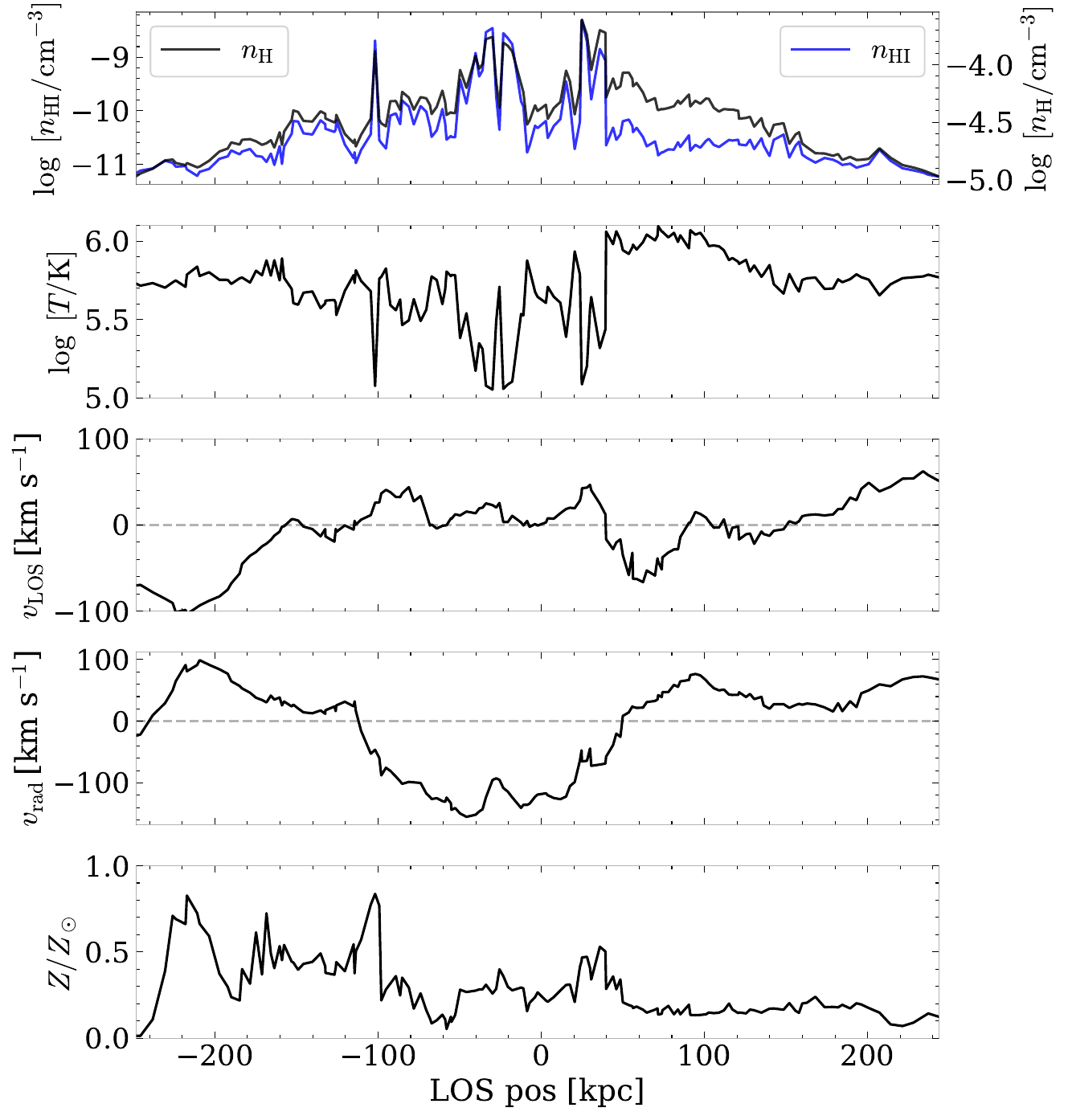}
    \caption{Multipanel view of the variation of some physical properties along the LOS producing the strong CBLA at $D=248.12$ kpc. From top to bottom: (1) total hydrogen density ($n_{\mathrm{H}}$, black) and neutral hydrogen density ($n_{\mathrm{HI}}$, blue); (2) temperature $T$; (3) LOS velocity $v_{\rm LOS}$; (4) radial velocity $v_{\rm rad}$ relative to the galaxy centre; (5) metallicity in units of solar metallicity $Z/Z_{\odot}$.\label{Multipanel_strong_CBLA}}
\end{figure}

To further explore the origin of these rare, strong absorbers, we take a closer look at the gas distribution and physical conditions along the LOS for the strong CBLA identified in Sect.~\ref{CBLA_deep} and shown in Fig.~\ref{strong_CBLA_slices}. The corresponding LOS properties are presented in Fig.~\ref{Multipanel_strong_CBLA}. The absorber is located at an impact parameter of $D = 248.12$ kpc.

As seen in Fig.~\ref{Multipanel_strong_CBLA}, most of the LOS traces diffuse, highly ionised CGM gas with low total and neutral hydrogen densities ($\log n_{\mathrm{H}}\sim-4.5$ and $\log n_{\mathrm{HI}}\sim-11.0$). Near the absorber, however, both densities increase sharply, revealing the presence of a compact overdense cloud. In this region, the neutral fraction increases from $f_{\rm HI}\sim10^{-6}$ in the surrounding halo to $f_{\rm HI}\sim10^{-5}$ inside the cloud. Combined with the higher gas density, this enhancement is enough to produce the unusually strong Ly$\alpha$ absorption seen in Fig.~\ref{strong_CBLA_slices}. The temperature remains within the warm-hot regime ($\log\, T\sim5.0-6.0$), although it decreases to $\log\, T\sim5.0$ inside the cloud. The LOS velocity remains relatively coherent across the absorber, suggesting that the gas moves as a single structure, while the negative radial velocity indicates inward motion toward the galaxy. Finally, the metallicity at the cloud position reaches $\sim0.25-0.50\,Z_{\odot}$, implying that the absorber is enriched and likely associated with recycled or pre-enriched halo gas rather than pristine IGM accretion.

\end{document}